\documentclass[traditabstract]{aa} % for the abstract without structuration 
\usepackage{graphicx}
\usepackage{txfonts}
\usepackage{lscape}
\newcommand{\mic}{$\mu$m}

\newcommand{\Lsun}{L$_{\odot}$}
\newcommand{\Msun}{M$_{\odot}$}

\newcommand{\km}{$\rm{km\,s^{-1}}$}
\newcommand{\water}[0]{H$_2$O}

\newcommand{\hh}[0]{H$_2$}

\newcommand{\CII}{[C\,{\sc ii}]}
\newcommand{\Oone}{[O\,{\sc i}]}
\newcommand{\Odrie}{[O\,{\sc i}]\,63}
\newcommand{\Ohon}{[O\,{\sc i}]\,145}
\newcommand{\Oopt}{[O\,{\sc i}]\,6300\,\AA}
\newcommand{\teff}{$T_{\mathrm{eff}}$}
\newcommand{\trot}{$T_{\mathrm{rot}}$}
\newcommand{\tvib}{$T_{\mathrm{vib}}$}
\newcommand{\irratio}{F$_{\mathrm{IR}}$/F$_*$}

\newcommand{\lUV}{L$_{\mathrm{UV}}$}

  \def\prodimo{{\sc ProDiMo}}

\begin{document}

   \title{GASPS observations of Herbig Ae/Be stars with PACS/Herschel\thanks{{\it Herschel} is an ESA space 
   observatory with science instruments provided by European-led Principal Investigator consortia and with 
   important participation from NASA.}}

   \subtitle{The atomic and molecular content of their protoplanetary discs}

   \author{G. Meeus\inst{1}, B. Montesinos\inst{2}, I. Mendigut\'ia\inst{1}, I. Kamp\inst{3},  W.F. Thi\inst{4}, C. Eiroa\inst{1}, 
    C.A. Grady\inst{5,6,7}, G. Mathews\inst{8}, G. Sandell\inst{9}, C. Martin-Za\"idi\inst{4}, S. Brittain\inst{10}, W.R.F. Dent\inst{11}, 
    C. Howard\inst{9}, F. M\'enard\inst{4}, C. Pinte\inst{4}, A. Roberge\inst{12},  B. Vandenbussche\inst{13}
           \and
          J.P. Williams \inst{8}           }

   \institute{Universidad Autonoma de Madrid, Dpt. Fisica Teorica, Campus Cantoblanco, Spain
              \email{gwendolyn.meeus@uam.es}
         \and
            Dept. of Astrophysics, CAB (CSIC-INTA), ESAC Campus, P.O. Box 78, 28691 Villanueva de la Ca\~nada, Spain %2
          \and
            Kapteyn Astronomical Institute, Postbus 800, 9700 AV Groningen, The Netherlands %3
          \and
            UJF-Grenoble 1 / CNRS-INSU, Institut de Plan\'{e}tologie et d'Astrophysique (IPAG) UMR 5274, Grenoble, 38041, France %4
          \and
            Eureka Scientific, 2452 Delmer, Suite 100, Oakland CA 96002, USA%5
          \and
            ExoPlanets and Stellar Astrophysics Laboratory, Code 667, Goddard Space Flight Center, Greenbelt, MD 20771 USA%6
          \and
          Goddard Center for Astrobiology%7
          \and
             Institute for Astronomy (IfA), University of Hawaii, 2680 Woodlawn Dr., Honolulu, HI 96822, USA %8
          \and
             SOFIA-USRA, NASA Ames Research Center, Building N232, Rm 146, P. O. Box 1, Moffett Field, CA 94035, USA. %9
          \and
            Department of Physics \& Astronomy, 118 Kinard Laboratory, Clemson University, Clemson, SC 29634, USA%10
          \and
            ALMA SCO, Alonso de C\'ordova 3107, Vitacura, Santiago, Chile%11
          \and
            Exoplanets and Stellar Astrophysics Lab, NASA Goddard Space Flight Center, Code 667, Greenbelt, MD, 20771, USA %12
          \and
            Instituut voor Sterrenkunde, Katholieke Universiteit Leuven, Celestijnenlaan 200D, 3001 Heverlee, Belgium %13 
             }

   \date{Received ; accepted }

% \abstract{}{}{}{}{} 
% 5 {} token are mandatory
 
  \abstract{We observed a sample of 20 representative Herbig Ae/Be stars and five A-type debris 
  discs with PACS onboard of {\em Herschel}. The observations were done in spectroscopic mode, 
  and cover far-IR lines of [O\,{\sc i}], \CII, CO, CH$^+$, \water \ and OH. 
  We have a \Odrie \ \mic \ detection rate of 100\% for the Herbig Ae/Be and 0\% for the debris discs.
  \Ohon  \ \mic \ is only detected in 25 \%, CO $J$=18-17 in 45 \% (and less for higher $J$ transitions)
  of the Herbig Ae/Be stars and for \CII \ 157  \ \mic, we 
  often found spatially variable background contamination. We show the first detection 
  of water in a Herbig Ae disc, HD 163296, which has a settled disc. Hydroxyl is detected as well in this disc. 
  CH$^+$, first seen in HD 100546, is now detected for the second time in a Herbig Ae star, HD 97048. 
  We report fluxes for each line and use the observations as line diagnostics of the gas properties. 
  Furthermore, we look for correlations between the strength of the emission lines and stellar or disc 
  parameters, such as stellar luminosity, UV and X-ray flux, accretion rate, PAH band strength, and flaring. 
  We find that the stellar UV flux is the dominant excitation mechanism of \Odrie, with the highest line 
  fluxes found in those objects with a large amount of flaring and greatest PAH strength. Neither the 
  amount of accretion nor the X-ray luminosity has an influence on the line strength. 
  We find correlations between the line flux of \Odrie \ and \Ohon, CO  $J$ = 18-17 
  and \Oopt, and between the continuum flux at 63 \mic \ and at 1.3 mm, while we find weak correlations 
  between the line flux of \Odrie \ and the PAH luminosity, the line flux of CO $J$ = 3-2, the continuum flux at 63 \mic, 
  the stellar effective temperature and the Br$\gamma$ luminosity. Finally, we use a combination of 
  the \Odrie \ and $^{12}$CO $J$ = 2-1 line fluxes to obtain order of magnitude estimates of the disc gas masses, 
  in agreement of the values we found from detailed modelling of 2 HAEBEs, HD 163296 and HD 169142.
  }

  \titlerunning{Herschel GASPS observations of HAEBEs}
  \authorrunning{G. Meeus et al.}
  
   \maketitle

%++++++++++++++++++++++++++++++++++++++++++++++++++++++++++++++++++++++
%++++++++++++++++++++++++++++++++++++++++++++++++++++++++++++++++++++++
\section{Introduction}
\label{s_intro}

Circumstellar discs around young stars are the sites of planet formation (e.g. Pollack et al. 
\cite{pollack1996}, Alibert et al. \cite{alibert2005}). During the first 10 Myr, the initially gas-rich 
disc will evolve into first a transitional and then a debris disc, while dispersing its gas content. 
The understanding of this dispersal process and what favours/hinders it is a crucial part of the 
planet formation puzzle, as the amount of gas present in a disc is crucial to determine whether 
gas giant planets can still be formed. Furthermore, the disc mass controls the migration of 
planetary bodies of all sizes, from gas giants to meter-sized planetesimals. Three components 
need to be characterised well: the disc geometry, the dust, and the gas content.

The disc geometry of young intermediate-mass stars, the Herbig Ae/Be stars (HAEBEs), is 
constrained through multi-wavelength imaging, interferometry, and radiative transfer 
modelling (e.g. Benisty et al. \cite{benisty2010}). Meeus et al. (\cite{meeus2001})
empirically divided the HAEBE discs into group I (flared) and group II (flat). A general 
consensus exists that discs become flatter as dust grains grow and settle towards the 
midplane (Dullemond \& Dominik \cite{dullemond2004}). Lately, several of the group I sources 
have been found to have an inner opacity hole in the disc (e.g. Grady et al. \cite{grady2007}, 
\cite{grady2009}), possibly due to a lack of small dust grains in the inner disc. In HD 100546, 
the gap may be caused by a planet (e.g. Bouwman et al. \cite{bouwman2003}, Tatulli et al. \cite{tatulli2011}). 

In a study of 53 HAEBEs, 85\% show a silicate emission feature at 10 \mic \, with a variety 
in grain size and crystallinity, proving the presence of {\em warm} small grains in these discs
(Juh\'asz et al. \cite{juhasz2010}). Polycyclic Aromatic Hydrocarbons (PAH) features were clearly 
detected in 70\% of the sample, with a clear preference towards flared discs (Acke et al. 
\cite{acke2010}). PAHs located in the disc atmosphere are transiently excited by UV photons 
and are an important heating source for the gas in the disc surface through the photo-electric effect. 

The study of gas properties is difficult as, in general, emission lines are rather weak. Different gas 
species and transitions probe different regions in the disc: lines in the near- and mid-IR generally 
trace the inner disc ($<$10-20 AU), while lines in the far-IR and mm mainly trace the outer disc. 
We refer to Carmona et al. (\cite{carmona2010}) for a discussion of different gas tracers, their location 
in the disc and observational characteristics. To understand the disc radial and vertical structure, it is 
necessary to observe several transitions of different species, as they arise under different conditions 
(density, temperature, radiation field). \hh \, and CO lines are most often used, since they are the most 
abundant species present, with the canonical \hh \ to CO number ratio in the ISM being 10$^4$. However, 
the detection of \hh \  in the IR has proven to be difficult because of its weak rotational and ro-vibrational 
transitions - it has only been detected in 3 HAEBEs. 
In a survey of 15 Herbig Ae/Be stars with CRIRES, Carmona et al.\  (\cite{carmona2011}) detected 
ro-vibrational transitions of \hh \ at  2.1218 \mic \ in only two objects: HD 97048 and HD 100546. 
Earlier, Bitner et al.\ (\cite{bitner2008}), Carmona et al.\ (\cite{carmona2008}), Martin-Za\"idi et al.\ 
(\cite{claire2009,claire2010}) searched for mid-IR pure rotational lines of \hh \ at 17.035~\mic \ in 
a sample of in total 20 HAEBEs; only two detections were made, in AB Aur and HD 97048. 
In sharp contrast, the detection of CO, although much less abundant, is easier as its rotational/ro-vibrational 
lines are much stronger. CO is routinely detected in HAEBEs (e.g. Thi et al. \cite{thi2001}, Dent et al. 
\cite{dent2005}). Lorenzetti et al. (\cite{lorenzetti2002}) showed {\em ISO}/LWS observations of atomic and 
molecular lines in the far-IR for a sample of  HAEBEs. They detected the fine-structure lines of \Odrie \ \& 
145.5~\mic \ and \CII \ at 157.7~\mic.

Despite the wealth of observations, it is still not clear how HAEBE discs dissipate with time. In the less
massive T Tauri stars, disc dispersal is thought to be initiated by photo-evaporation, mainly due to ionising 
EUV ($h \nu >$ 13.6 eV) photons that first create a gap in the inner disc, which is subsequently rapidly 
viscously accreted. In a next step, the outer disc is efficiently removed through a photo-evaporative disc 
wind (Alexander et al. \cite{alexander2006}). However, Gorti et al. (\cite{gorti2009}) showed that UV can 
rapidly disperse the outer disc, where the bulk of the disc mass is located, thus setting the disc lifetime. Also 
X-rays are thought to play an important role in those discs (e.g. Ercolano et al. \cite{ercolana2008}, Owen 
et al., \cite{owen2012}). And finally, also the accretion of a planet with a mass of a few Jupiter can play
an important role in the dissipation of the disc. Which mechanism is ultimately dominating the dispersion 
process is not yet determined.
 
We present ESA {\it Herschel} Space Observatory (Pilbratt et al. \cite{pilbratt2010}) spectroscopy of 20 
HAEBEs and 5 A-type debris discs, covering several transitions of abundant atoms and molecules that can 
be used as crucial tests of our understanding of disc physics and chemistry in the upper layers of the disc. 
The observations cover a significant part of the disc surface that was not accessible before. Our observations 
are part of the {\it Herschel} Open Time Key Programme (OTKP) "GAS in Protoplanetary Systems" (GASPS; 
P.I. Dent; see Dent et al. \cite{dent2012}). With this paper, we want to obtain a better understanding of HAEBE 
discs by relating several gas tracers and excitation mechanisms with stellar and disc properties. What gas 
species are present in a HAEBE disc and at what temperatures? What is the physical and chemical structure 
of the disc chemistry? What is the dominant excitation mechanism of gas in HAEBE discs? 

In Sect.\,\ref{s_targets}, we describe the sample and our methods to derive the stellar and disc parameters. 
In Sect.\,\ref{s_spec} we present the spectroscopy and the line detections. We discuss gas lines as a 
diagnostic tool in Sect.\,\ref{s_conditions} and look for correlations between the observed line fluxes. We 
relate our detections and upper limits to stellar and disc parameters in Sect.\,\ref{s_ana}. Finally, we round 
off with conclusions in Sect.\,\ref{s_conc}. 

\begin{table*}
\caption{Main stellar parameters of the sample.}
\begin{tabular}{lllrlllllll}
\hline\hline
Star  &\multicolumn{1}{c}{ Alternative}& Sp. Type  & $T_{\rm eff}$ &  $\log g_*$ &     [M/H]   & Refs. &\multicolumn{1}{c}{$d$}  &  $L_*/L_\odot$    &A$_{V}$&\multicolumn{1}{c}{Age}    \\
           &\multicolumn{1}{c}{Name}&           &   (K)         &             &             &       &\multicolumn{1}{c}{(pc)} &                    &&\multicolumn{1}{c}{(Myr)}  \\
\hline
AB Aur       &  HD 31293   & A0 Ve        &   9280    &    4.00     & {\it 0.00}                      & 1     & $139.3\pm 19.0$ & $33.0\pm 9.2$     &0.25&  $5.0\pm 1.0$   \\
HD 31648     &  MWC 480  & A3-5 Ve      &   8250    &    4.00     &     0.00                        & 2     & $137.0\pm 26.2$ & $13.7\pm 5.5$     &0.16&  $8.5\pm 2.0$   \\
HD 35187     &           & A2 Ve        &   8915    & {\it 4.00}  & {\it 0.00}                      & 3     & $114.2\pm 32.4$ & $17.4\pm 10.6$    &0.78&  $9.0\pm 2.0$   \\
HD 36112     &  MWC 758  & A5 IVe       &   7750    &    3.50     & --0.08$^a$                & 4     & $279.3\pm 75.0$ & $33.7\pm 19.3$    &0.16&  $3.7\pm 2.0$   \\
                       &           &              &           &             & --0.14$^b$         & 4     &                 &                   &&  $3.5\pm 2.0$   \\
 CQ Tau         &HD 36910    & F3 Ve        &   6900    & {\it 4.35}  & {\it 0.00}                      & 5     & $113.0\pm 24.0$ & $3.4\pm 1.5$      &1.40&  $4.0\pm 2.0$   \\
HD 97048     &  CU Cha   & A0 Ve        &  10000    &    4.00     & --0.05$^a$                & 4     & $158.5\pm 15.7$ & $30.7\pm 6.1$     &1.15&  $6.5\pm 1.0$   \\
                        &           &              &           &             &--0.75$^c$ & 4     &                 &                   &&  $4.0\pm 0.5$   \\
HD 100453    &           & A9 Ve        &   7400    &    4.20     &  +0.30                          & 6     & $121.5\pm 9.7$  & $8.8\pm1.4$       &0.00&  $>10$          \\
                         &           &              &           &             & --0.02$^a$                & 4     &                 &                   &&  $10.0\pm 2.0$  \\
                         &           &              &           &             & --0.09$^b$         & 4     &                 &                   &&  $9.0\pm 2.0$   \\
HD 100546    &           & B9 Ve         &  10470    &    3.50     & --0.08$^a$                &3, 4, 7& $96.9\pm 4.0$   & $22.7\pm 1.9$     &0.09&  --             \\
                         &           &              &           &             & --1.30$^d$& 4     &                 &                   &&  $3.8\pm 0.5$   \\
HD 104237    & DX Cha    & A4-5 Ve      &   8550    &    3.90     &  +0.16                          & 8     & $114.7\pm 4.7$  & $28.8\pm 2.4$     &0.16&  $5.5\pm 0.5$   \\
HD 135344 B &SAO 206462& F3-4 Ve      &   6810    &    4.40     & +0.14$^a$                 & 10, 4 & $142.0\pm 27.0$ & $8.1\pm 3.1$      &0.37&  $10.0\pm 2.0$  \\
                         &                     &              &           &             & --0.08$^b$         & 4     &                 &                   &&  $9.5\pm 2.0$   \\
HD 139614    &                    & A7 Ve        &   7400    &    4.00     & --0.50                          & 6     & $140\pm 42$     & $7.6\pm 4.6$      &0.00&  $9.2\pm 2.0$   \\
             &           &                   &           &             & --0.19$^a$                & 4     &                 &                   &&  $11.5\pm 2.0$  \\
             &           &                   &           &             & --0.27$^b$         & 4     &                 &                   &&  $10.5\pm 2.0$  \\
HD 141569 A  &                   & B9.5 Ve      &  10000    &    4.28     & --0.50                          & 11    & $116.1\pm 8.1$  & $29.6\pm 4.2$     &0.37&  $4.7\pm 0.3$   \\
HD 142527    &                     & F6 IIIe      &   6250    &    3.62     & {\it 0.00}                      & 12    & $233.1\pm 56.2$ & $33.2\pm 16.9$    &0.59&  $2.0\pm 0.5$   \\
HD 142666    &                     & A8 Ve        &   7500    &    4.30     & +0.20                           & 6     & $145\pm 43$     & $13.5\pm 8.0$     &0.93&  $9.0\pm 2.0$   \\
HD 144668    & HR 5999   & A7 IVe       &   7925    & {\it 4.00}  & {\it 0.00}                      & 13    & $162.9\pm 15.3$ & $50.8\pm 9.5$     &0.47&  $2.8\pm 1.0$   \\
HD 150193    & MWC 863   & A2 IVe       &   8970    &    3.99     & {\it 0.00}                      & 2     & $216.5\pm 76.0$ & $48.7\pm 38.0$    &1.55&  $3.8\pm 2.0$   \\
KK Oph A     &                        & A6 Ve        &   8000    & {\it 4.00}  & {\it 0.00}                      & 16    & {\it 260}       & 13.7              &1.60&  $8.0\pm 2.0$   \\
KK Oph B     &                        & G5 Ve        &   5750    & {\it 4.50}  & {\it 0.00}                      & 17    & {\it 260}       & 2.1               &2.80&    $8.0\pm 2.0$   \\
 51 Oph        &HD 158643    & B9.5 IIIe    &  10250    &    3.57     & +0.10                           & 2     & $124.4\pm 3.7$  & $285.0\pm 17.0$   &0.37&  $0.7\pm 0.5$   \\
HD 163296    & MWC 275   & A1 Ve        &   9250    &    4.07     & +0.20                           & 14    & $118.6\pm 11.1$ & $33.1\pm 6.2$     &0.47&  $5.5\pm 0.5$   \\
HD 169142    & MWC 925   & A7-8 Ve      &   7500    &    4.00     & --0.50                          & 15    & $145\pm 43$     & $9.4\pm 5.6$      &0.00&  $7.7\pm 2.0$   \\
\hline
 49 Cet              &  HD  9672     & A4 V         &   9500    &    4.30     & +0.10             & 2     & $59.4\pm 1.0$   & $21.0\pm 0.7$     &0.22&$8.9_{-2.4}^{+6.1}$, $61_{-46}^{+119}$ \\
HD 32297     &           & A0 V         &   9520    & {\it 4.15}  & {\it 0.00}                      & 18    & $112.4\pm 10.8$ & $10.9\pm 2.1$     &0.62&  --            \\
HR 1998 &HD 38678, $\zeta$ Lep& A2 IV-V      &   8500    &    4.27     & --0.76                          & 9, 19 & $21.6\pm 0.1$   & $14.0\pm 0.1$     &0.00&  $1250\pm250$  \\
HR 4796 A    &HD 109573 A   & A0 Ve        &   9750    &    4.32     & {\it 0.00}                      & 9     & $72.8\pm 1.8$   & $23.4\pm1.1$      &0.00&  $10.0\pm2.0$  \\
HD 158352    & HR 6507   & A7 V         &   7500    &    3.85     & {\it 0.00}                      & 9     & $59.6\pm 0.9$   & $17.7\pm 0.6$     &0.00&  $1000\pm200$  \\
\hline
\end{tabular}
\label{t_para}
{\bf Notes}: Quantities in {\it italics} are assigned. ($a$) These metallicities are weighted averages of all the elemental 
abundances listed in Tables 2, 3 and 4 of  Acke \& Waelkens (2004), see Appendix A of Montesinos  et al. (2009) for 
details; ($b/c/d$) Metallicities assumed to be in the same proportion as the species [Fe {\sc i}]/([Si {\sc ii}])/[Fe {\sc ii}], 
respectively (see Table 3 of Acke \& Waelkens, 2004). Refs.: (1) Woitke et al. (GASPS) (in prep), (2) Montesinos 
et al. (2009), (3) Manoj et al. (2006), (4) Acke \& Waelkens (2004), (5) Testi et al. (2003), (6) Guimar\~aes et al 
(2006), (7) Levenhagen et al. (2006), (8) Fumel \& B\"ohm (2011), (9) Allende-Prieto et al. (1999), (10) M\"uller 
et al. (2011), (11) Mer\'{\i}n et al. (2004), (12) Verhoeff et al. (2011), (13) van Boekel et al.  (2005), (14) Tilling 
et al. (GASPS) (2012), (15) Meeus et al. (GASPS) (2010), (16) Herbig (2005), (17) Carmona et al. (2007), (18) 
Torres et al. (2006), (19) Gray (2006). Distances are from the revised parallaxes by van Leeuwen (2007) 
except for HD 135344 B (M\"uller et al. 2011) and HD 139614, HD 142666 and HD 169142 (van Boekel et al. 
2005).
\end{table*}

%++++++++++++++++++++++++++++++++++++++++++++++++++++++++++++++++++++++
%++++++++++++++++++++++++++++++++++++++++++++++++++++++++++++++++++++++
\section{Targets}
\label{s_targets}

The sample consists of 20 Herbig Ae/Be stars with spectral types ranging between B9.5 and F4, to 
which we will refer as Herbig Ae (HAe) stars. We do not include the more massive Herbig Be stars 
which are, in general, younger and have smaller discs and often an additional remnant envelope 
(e.g. Natta et al. \cite{natta2000}, Verhoeff et al. \cite{verhoeff2012}). We also include HD 141569A, 
an object that completely lacks a near-IR excess for $\lambda < 4.5$ \mic,  attributed to inner disc 
clearing, but still has a substantial amount of primordial gas;  in this paper, we will call this a transitional 
disc. We are aware that several of our sources are also called (pre-)transitional discs in the literature, 
such as HD 100546 (Grady et al. \cite{grady2005}) and HD 135344B (Andrews et al. \cite{andrews2011}), 
which are observed to have a gap in their disc, but for the purpose of this paper, we include them in the 
HAe sample, since they still have a substantial near-IR excess and their total IR excess is much larger 
than that of HD 141569A.
Besides the Herbig Ae sample, we include five debris discs around A-type stars with ages between 
$\sim$10 and 1200 Myr for comparison, as the Herbig Ae stars are seen as their precursors. We list the 
main stellar parameters in Table~\ref{t_para}. The sample is representative for the known Herbig 
Ae stars: there are nine objects in Meeus group I and ten in group II (to which we will refer to
as flaring and flat discs). Furthermore, we have a good coverage of \teff, age, stellar luminosity L$_{*}$ 
and accretion rate. 49 Cet is the only debris disc in our sample for which gas was detected through CO 
observations with the JCMT (Dent et al. \cite{dent2005}). Hughes et al. (\cite{hughes2008}) later resolved 
the CO gas emission using the SMA.

For 3 stars in our sample, members of the GASPS team have performed detailed modeling of their discs with 
the radiative transfer code MCFOST (Pinte et al. \cite{pinte2006}, \cite{pinte2009}) and the thermo-chemical 
code \prodimo \ (Woitke et al. \cite{woitke2009a}): HD 169142 (Meeus et al. \cite{meeus2010}), HD 100546 
(Thi et al. \cite{thi2010}) and HD 163296 (Tilling et al. \cite{tilling2012}).

\begin{table}
\caption{Derived properties of the sample. Disc group classification from Meeus et al. (\cite{meeus2001}) 
and Acke et al. (\cite{acke2010}). For the derivation of $L_{\rm IR}/L_*$, $L_{\rm UV}/L_\odot$ and 
$\log L_{\rm acc}/L_\odot$, see the description in Sect.\ref{s_prop}.  }
\begin{tabular}{lclrr}
\hline\hline
Star                & Disc  & $L_{\rm IR}/L_*$  & $L_{\rm UV}/L_\odot$ & $\log L_{\rm acc}/L_\odot$ \\
                    & group &                   &                      &                            \\
\hline
AB Aur                  & I     & 0.76              & 4.63              & $<0.55$  \\
HD 31648            & II    & 0.46              & 0.75              & $<0.19$  \\
HD 35187            & II    & 0.12              & 2.23              & 1.13     \\
HD 36112            & I     & 0.66              & 1.32              & $<-0.81$ \\
CQ Tau                 & II    & 1.01              & 0.19--0.3         & $<-1.16$ \\
HD 97048            & I     & 0.39              & 7.69              & $<0.95$  \\
HD 100453           & I     & 0.62              & 0.29              & $<-0.91$ \\
HD 100546           & I     & 0.56              & 7.22              & 1.62     \\
HD 104237           & II    & 0.32              & 1.54              & 0.87     \\
HD 135344 B         & I     & 0.56              & 0.11$^{\rm a}$    & $-0.22$  \\
HD 139614           & I     & 0.39              & 0.39              & $<-1.12$ \\
HD 141569 A         & II/TO & 0.009             & 6.83              & $<0.70$  \\
HD 142527           & I     & 0.98              & 0.15$^{\rm a}$    & $<-1.04$ \\
HD 142666           & II    & 0.33              & 0.37--0.68        & 0.85     \\
HD 144668           & II    & 0.51              & 1.55--2.94        & $<0.23$  \\
HD 150193           & II    & 0.48              & 8.53              & 1.32     \\
KK Oph A+B          & II    & 2.01              & 2.35              & 1.70     \\
51 Oph                  & II    & 0.03              & 71.32             & $<0.37$  \\
HD 163296           & II    & 0.29              & 3.21--5.58        & $-0.04$  \\
HD 169142           & I     & 0.42              & 0.45              & $<0.05$  \\
\hline
49 Cet                 & Debris& 6.0 $\times$10$^{-4}$          & 2.96              & $<-0.11$ \\
HD 32297            & Debris& 0.003             & 1.91$^{\rm a}$    & --       \\
HR 1998              & Debris& 6.6 $\times$ 10$^{-5}$          & 1.25              & --       \\
HR 4796 A         & Debris& 0.003             & 5.59              & --       \\
HD 158352          & Debris& 1.6 $\times$ 10$^{-4}$        & 0.64              & --       \\
\hline
\end{tabular}
\label{t_excess}
{\bf Notes}: Disc groups I and II refer to Meeus et al. (2001) classification, 
"TO" stands for "Transition Object". (a) UV fluxes measured on the model photosphere 
(no UV observations available).
\end{table}

\subsection{Stellar and disc properties}
\label{s_prop}

To characterize the sample in a consistent way, we first aimed at determining the stellar component of the 
spectral energy distribution (SED). In the next step, several parameters that can be important in the 
context of gas excitation in the disc were computed, namely, ultraviolet luminosities, infrared excesses, 
and accretion luminosities.

We compiled a set of literature and catalogue stellar parameters (effective temperatures, gravities and 
metallicities), and critically selected what we considered the best. Multi-wavelength 
photometry from different sources and ultraviolet spectra obtained by the International Ultraviolet
Explorer\footnote{http://sdc.cab.inta-csic.es/ines/} were used to construct the spectral energy distribution 
of the star-plus-disc systems.  A specific model 
photosphere for each star was extracted, or computed by interpolation, from the grid of PHOENIX/GAIA 
models (Brott \& Hauschildt \cite{brott2005}). A Rayleigh-Jeans extension up to mm wavelengths was
applied to the original models. The model photosphere was reddened with several values of $E(B-V)$ 
($R_V=3.1$) and normalized to the flux at $V$ (0.55 $\mu$m), until the best fit to the optical photometry 
was achieved. In some cases, the photospheric model fits the ultraviolet spectrum fairly well, implying only 
stellar photospheric contribution to that spectral range, whereas in other cases an excess, attributed to 
accretion processes, was apparent.

The photospheric, infrared, and ultraviolet fluxes (required to estimate the stellar luminosity, $L_{\rm IR}/L_*$ 
and $L_{\rm UV}/L_\odot$), were all computed from the {\it dereddened} data or model. The photospheric flux 
was estimated by integrating the model photosphere; the infrared flux was computed by fitting a spline to the 
observed data set, integrating from the wavelength $\lambda_0$ where the fit started separating from the 
photospheric model up to 200 \mic; the UV flux between 1150--2430 \AA{ }was used to compute $L_{\rm UV}$.
Accretion rates were estimated from the observed photometric excesses in the Balmer discontinuity. These 
were modelled assuming a magnetospheric accretion geometry, following the procedure by Mendigut\'{\i}a 
et al. (\cite{mendi2011b}). Upper limits on the accretion rates are provided for most stars, given that the Balmer 
excesses are negligible.

In Table~\ref{t_para} we give some of the basic parameters of the stars, namely, identifications, spectral types, 
effective temperatures, gravities, metallicities and the corresponding references (cols. 1--7), distances (col. 8), 
and two parameters derived from those and the SEDs, namely, stellar luminosities and ages (cols. 9 and 10). 
In~Table \ref{t_excess} we list the disc groups according to Meeus et al. (\cite{meeus2001}) or the
evolutionary status of the stars, fractional infrared luminosities, ultraviolet luminosities and accretion luminosities
(cols. 2--5). Stars with two values of $L_{\rm UV}/L_\odot$ in col. 4 show two different emissivity levels in IUE 
spectra obtained in different epochs; the numbers corresponding to the low and high states are given.

The age estimations were done by placing the values of $\log T_{\rm eff}$ - $\log L_*/L_\odot$ for each star 
on an HR diagram containing tracks and isochrones computed for its particular metallicity. For some stars this parameter 
is unknown, therefore a solar abundance was assumed, which introduces an uncertainty which is difficult to estimate. 
Metallicity is an important parameter to take into account when determining ages using this procedure since the position 
of a set of tracks and isochrones in the HR diagram changes substantially with metallicity.
The evolutionary tracks and isochrones for a scaled solar mixture from the Yonsei-Yale group (Yi et al.  \cite{yi2001}) -- 
Y$^2$ in their notation -- have been used in this work. From the Y$^2$ set, the isochrones with $Z\!=\!0.02$ (solar) have 
been used for those stars with measured metallicities [M/H] between $-0.10$ and $+0.10$ and for the stars for which no 
abundance determinations are available. The remaining metallicities are treated with the isochrones in the grid whose 
value of $Z$ is closer to $0.02\times 10^{\rm [M/H]}$.

There is one star (HD 32297) for which a determination of the age was impossible; its position falls below the main sequence 
in a $Z\!=\!0.02$ HR diagram, therefore, it is quite likely that this star has subsolar abundance. The evolutionary stage of 
49 Cet is interesting and was studied in detail in Montesinos et al. (\cite{montesinos2009}). Two ages, corresponding
to PMS and MS isochrones, are listed in Table \ref{t_para}.

\subsubsection{KK Oph}

Special attention had to be paid to KK Oph. This object is a close binary separated by 1.61 arcsec (300 AU at 
our adopted distance, see discussion below; Leinert et al., \cite{leinert1997}), the hot component (A) is a 
Herbig Ae star and the cool component (B) is a T Tauri star, both of them actively accreting (Herbig, \cite{herbig2005}, 
and references therein). The SED shows an infrared excess from $\sim 1$ $\mu$m onwards. All the available 
photometry corresponds to both components, therefore it is not straighforward to estimate parameters for each 
star from these data alone. In addition, the system is highly variable (Hillenbrand et al. \cite{hillenbrand1992}; Herbst 
\& Shevchenko, \cite{herbst1999}; Eiroa et al. \cite{eiroa2001}; Oudmaijer et al. \cite{oudmaijer2001}), adding 
further complications to the analysis.

Herbig (\cite{herbig2005}) determined an spectral type A6 V for KK Oph A, and Carmona et al. (\cite{carmona2007}) 
- who only studied the secondary component - found a spectral type G5 V for KK Oph B. In both cases the authors 
made use of high resolution spectroscopy which allowed them to separate the spectra of each component. Effective 
temperatures of 8000 and 5750 K have been assigned to the stars according to their spectral types.  To estimate 
luminosities from the available photometry we have followed the approach by Carmona et al. (\cite{carmona2007}) 
where each star is affected by a different extinction, adopting $A_V\!=\!1.6$ and 2.8 mag for components A and B, 
respectively.  We have worked on the set of optical and near-IR photometry provided by Hillenbrand et al. 
(\cite{hillenbrand1992}), and assumed that the contribution to the flux at $U$ and $B$ from the cool component 
was negligible. The distance to this system is uncertain, with published values of 170--200 pc. We find these distances 
too low to obtain values for the luminosities of the stars that are in agreement with their spectral type and luminosity 
class; a lower limit of $\sim\!260$ pc provides sensible results and has been adopted in this work, which is the best 
we could derive with the available data set. 

%++++++++++++++++++++++++++++++++++++++++++++++++++++++++++++++++++++++
%++++++++++++++++++++++++++++++++++++++++++++++++++++++++++++++++++++++
\section{{\it Herschel} PACS  spectroscopy}
\label{s_spec}

\begin{table}
\caption{Ranges and lines targeted with the PACS spectrometer. }
\begin{tabular}{cclrl}
\hline
\hline
Set         & Observed Range      & Species & Transition                                             & Wavelength\\
               & $\lambda$ (\mic)      &                  &                                                               &\multicolumn{1}{c}{(\mic)}  \\
\hline
A            &  62.68 -- 63.68          &\Oone        &$^3P_1 \rightarrow \ ^3P_2$           & 63.184         \\
               &                                     &o-\water     &8$_{18} \rightarrow 7_{07}$            &63.324        \\
                & 188.77 -- 190.30      &DCO$^+$&$J$ = 22-21                                         & 189.570       \\
               \hline
B             & 71.90   -- 73.05       &o-\water   &7$_{07} \rightarrow 6_{16}$              &71.946       \\
               &                                    &CH$^+$   &$J$ = 5-4                                               & 72.14        \\
               &                                    &CO            &$J$ = 36-35                                          & 72.843         \\
               &  144.0 -- 146.1         &p-\water   &4$_{13} \rightarrow 3_{22}$              & 144.518          \\
               &                                    & CO           & $J$ = 18-17                                          & 144.784       \\ 
               &                                    &\Oone       & $^3P_0 \rightarrow \  ^3P_1$           & 145.525       \\ 
               \hline
C            &  78.55   -- 79.45        &o-\water  &4$_{23} \rightarrow 3_{12}$               &78.741          \\
               &                                     &OH           &1/2 - - 3/2                                                & 79.11     \\
               &                                     &OH           &1/2 +- 3/2                                                & 79.18       \\
               &                                     &CO           &$J$ = 33-32                                            & 79.360         \\
               & 157.1 -- 158.9           & \CII          &$^2P_{3/2} \rightarrow \ ^2P_{1/2}$ &157.741       \\
               &                                     &p-\water   &3$_{31} \rightarrow 4_{04}$               &158.309          \\
               \hline
D            & 89.45  --  90.50        &p-\water   & 3$_{22} \rightarrow 2_{11}$             &89.988           \\
               &                                    &CH$+$     &$J$ = 4-3                                               &90.02         \\
               &                                    &CO            &$J$ = 29-28                                           &90.163          \\
               &178.9 -- 181.0           &o-\water   &2$_{12} \rightarrow 1_{01}$              & 179.527         \\
               &                                    &CH$^+$   &$J$ = 2-1                                               & 179.610        \\
               &                                    &o-\water   &2$_{21} \rightarrow 2_{12}$              & 180.488        \\
\hline
\end{tabular}
\label{t_setting}
\end{table}

We obtained PACS (Poglitsch et al. 2010) spectroscopy in both line and range modes 
(PacsLineSpec, 1669\,s and PacsRangeSpec, 5150\,s). The observation identifiers (obsids) 
can be found in Table~\ref{t_obsid}.1 of Appendix A. In a later stage, we obtained deeper 
range scans to confirm tentative detections by doubling the integration time for 7 sources. 
All the observations were carried out in ChopNod mode, in order to remove the emission of 
the telescope and background. PACS is an IFU with 25 spaxels, 9.\arcsec4 on each side. 
Due to the characteristics of the PSF at 60\,\mic, only $\sim$70 \% of a point source flux falls 
in the central spaxel, with a decrease towards longer wavelengths, down to 45\% at 180\,\mic.

The spectroscopic data were reduced with the official release version 8.0.1 of the 
Herschel Interactive Processing Environment (HIPE; Ott \cite{ott2010}), using standard tasks provided 
in HIPE. These include bad pixel flagging; chop on/off subtraction; spectral response function division; 
rebinning with oversample = 2 and upsample =1, corresponding to the native resolution of the instrument; 
spectral flatfielding and finally averaging of the two nod positions. In order to conserve the best signal and 
not to introduce additional noise, we only extracted the central spaxel, and corrected for the flux 
loss with an aperture correction provided by the PACS instrument team ('pointSourceLossCorrection.py'). 
This is only possible when the source is well-centered. If that was not the case (e.g. for HD 142666), several 
spaxels needed to be taken into consideration to obtain the full target's emission.

The main lines targeted are the fine structure lines of \Oone, \CII \ and the molecular lines of CO, OH, 
CH$^+$ and \water. In total we observed eight spectral regions; the details of the transitions can be found 
in Table\,\ref{t_setting}. The spectral resolution varies between 3400 (shortest wavelengths) and 1100 
(longest wavelengths), which are equivalent to $\sim$ 88 \km \ at 60\,\mic, and 177 \km \ at 190\,\mic. In our 
sample, we do not resolve the emission lines as our objects do not have such high-velocity components. The 
line flux sensitivity is of the order a few $10^{-18}$ W/m$^2$. The lines were identified by manual inspection 
of automated gaussian fits at the expected line position, taking the instrumental resolution as the expected 
FWHM of the line. We extracted the fluxes of the detected lines using a gaussian fit to the emission lines with 
a first-order polynomial to the continuum. We used the RMS on the continuum (excluding the line) to derive a 
1$\sigma$ error on the line by integrating a gaussian with height equal to the continuum RMS and width of the 
instrumental FWHM. This approach is necessary as HIPE currently does not deliver errors for the spectra. In case 
of a non-detection, we give a 3$\sigma$ upper limit, also calculated from the continuum RMS. The absolute flux 
calibration error given by the PACS instrument team is currently $<$15 \%. The measured line fluxes - in case 
of detection - or their upper limits are listed in Tables~\ref{t_atomflux} and \ref{t_molflux}. Several atomic and
molecular lines were observed, as described in the following paragraphs.

\begin{table}
\caption{Atomic line strengths in units of 10$^{-18}$ W/m$^2$ with 1$\sigma$ continuum 
RMS between brackets in case of a detection or 3$\sigma$ upper limits.  For HD 158352, 
not all settings were observed, therefore we write 'n.a.' (not available) when the spectrum 
is lacking. $^a$: an absorption feature is observed in the spectrum, that is caused by
the subtraction of stronger emission in the chop-off position.} 
\begin{tabular}{lccc}
\hline
\hline
Star                        &\Oone                &\Oone              & \CII     \\ 
$\lambda$ (\mic) &63.18              & 145.53           & 157.75  \\
\hline
\hline
AB Aur           &851.2 (21.5)   & 44.6 (14.7)  & 51.0 (8.3)  \\
HD 31648      &94.9 (3.5)       &$<$ 7.8         &$<$ 9.7      \\
HD 35187      &32.8 (4.8)       &$<$ 4.6         &$<$ 5.9      \\
HD 36112      &37.3 (2.4)       &$<$ 6.3        &$<$ 9.6       \\
CQ Tau         &47.9 (4.0)       &$<$ 3.8        &$<$ 11.3      \\
HD 97048     &1592.5 (4.3)   &65.6 (2.6)     &106.8 (6.3) \\
HD 100453   &61.6 (7.5)        &$<$ 6.1        &$<$ 15.4     \\
HD 100546   &6043.4 (13.4)&194.7 (9.9)  &203.8 (8.6)  \\
HD 104237   &79.1  (3.5)      &$<$ 5.9        &$<$ 7.6        \\
HD 135344B&47.9 (5.8)       &$<$ 4.6        &$<$ 6.4        \\
HD 139614    &44.5 (6.1)      &$<$ 4.7        &$<$8.3         \\
HD 141569A &245.3 (4.8)   &24.9 (1.4)     &11.4 (2.1)     \\
HD 142527    &52.3 (3.8)    &$<$ 11.6      &$<$ 28.7       \\
HD 142666    &18.9 (3.1)      &$<$ 4.7        &$<$ 9.0        \\
HD 144668    &140.2 (4.3)    &$<$ 5.8        &$<$ 5.4        \\
HD 150193    &24.6 (2.8)      &$<$ 6.6        &$<$ 5.3        \\
KK Oph          &172.8 (5.0)    &6.2 (1.3)       &9.2 (1.2)       \\
51 Oph           &53.3 (2.5)      &$<$ 5.1        &$<$ 6.8         \\
HD 163296   &208.4 (4.2)  &$<$ 4.0          &ABS$^a$      \\ 
HD 169142    &91.5 (4.4)      &$<$ 3.6        &$<$ 8.0         \\
\hline
49 Cet             & $<$ 10.0        &$< $ 6.3       &$<$ 9.1       \\
HD 32297      &$<$7.4            &$<$ 4.2        &$<$7.1         \\
HR 1998        &$<$7.0            &$<$ 5.1        &$<$ 6.6         \\
HR 4796A     &$<$ 6.3           &$<$ 3.8        &$<$ 5.4         \\
HD 158352   &$<$ 8.2           &n.a.               &$<$ 3.8         \\ 
\hline
\end{tabular}
\label{t_atomflux}
\end{table}

\begin{figure*}[t!]
  \resizebox{\hsize}{!}{\includegraphics{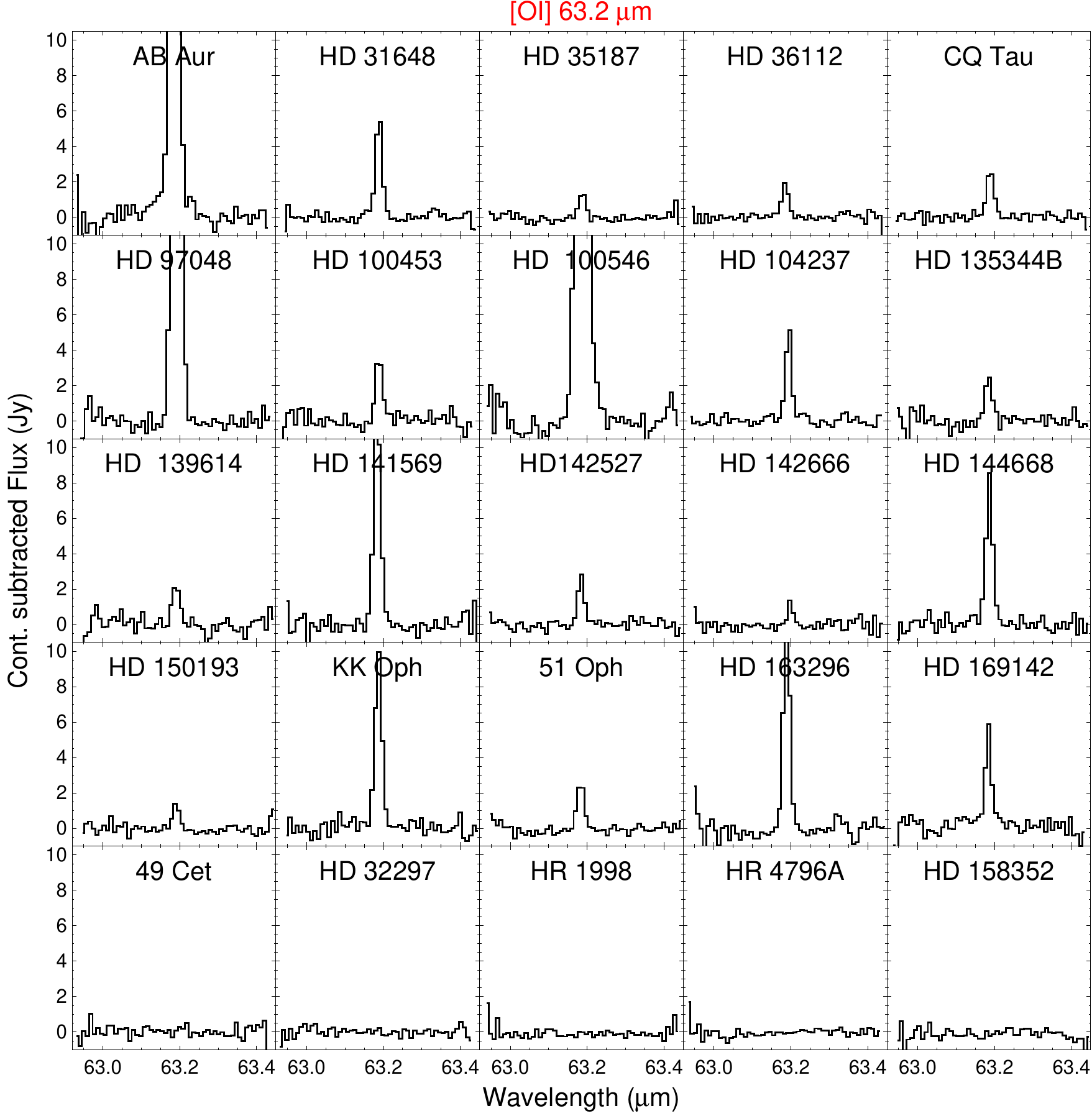}}
\caption{The \Odrie \,\mic \ lines for the entire sample. The line is seen in emission in all the 
HAEBEs, while absent in the more evolved debris disc objects 49 Cet, HD 32297, HR 1998,  
HR 4796A and HD 158352 (bottom row).}
\label{f_allstars_OI63}
\end{figure*}

\begin{figure}[t!]
  \resizebox{\hsize}{!}{\includegraphics{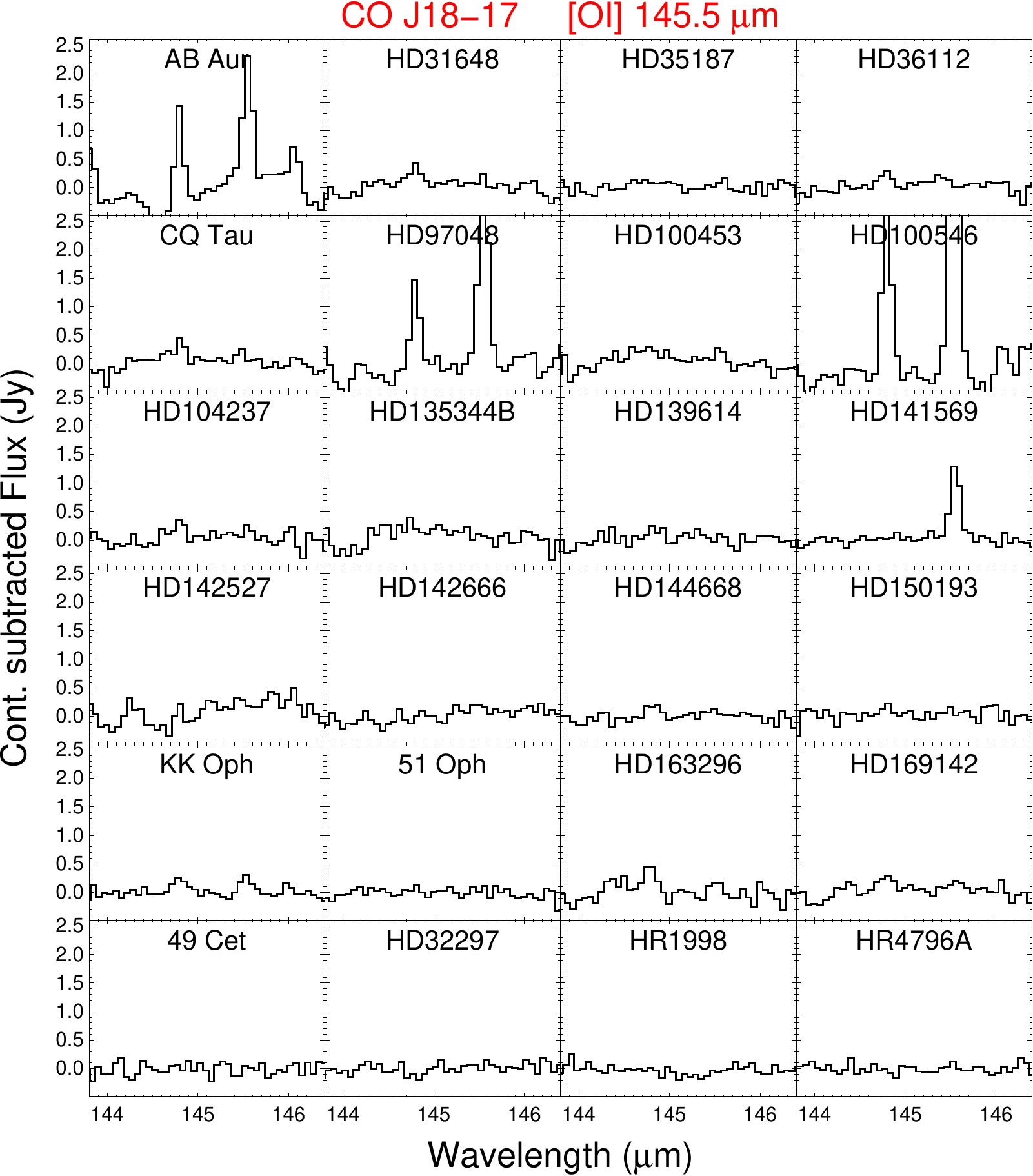}}
\caption{The settings around 145\,\mic. The lines of CO J=18-17 at 144.8~\mic \,\ and \Ohon \,\mic \,\ 
are only clearly detected in a three objects. }
\label{f_allstars_OI145}
\end{figure}

\begin{figure}[t!]
  \resizebox{\hsize}{!}{\includegraphics{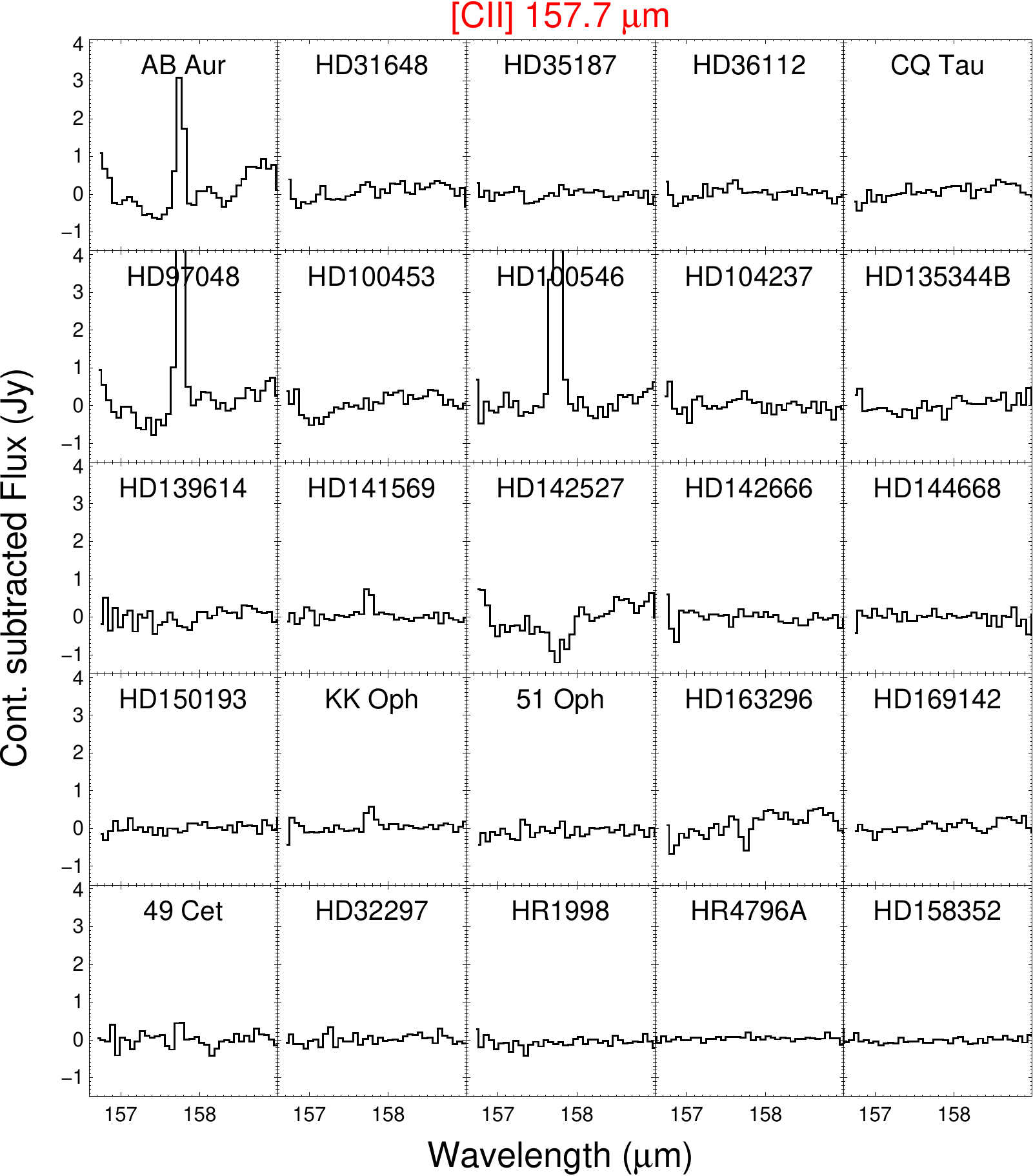}}
\caption{The \CII \ line at 157.7~\mic. The line is only clearly detected in three objects, more weakly
 in three others and is appears in absorption in HD 142527 and HD 163296, due to subtraction
 of the chop-off spectrum that has stronger emission.} 
\label{f_allstars_CII}
\end{figure}

%++++++++++++++++++++++++++++++++++++++++++++++++++++++++++++++++++++++
\subsection{Oxygen fine structure lines}
The third most abundant element in the interstellar medium is oxygen. Its fine structure line at 63.2 \mic \ 
is by far the strongest line observed in our spectra (see Table~\ref{t_atomflux}). It is detected in all Herbig 
Ae/Be stars, and absent in the debris discs.  For five of our objects, the line flux is larger than 200 $\times$ 
10$^{-18}$ W/m$^2$, while our faintest detection is  20 $\times$ 10$^{-18}$ W/m$^2$. The other fine 
structure line \Ohon \,\mic \ is also one of the strongest lines in our spectra, however it is only detected in 
five objects (25\% of the HAEBEs). In Figs.\,\ref{f_allstars_OI63} and \ref{f_allstars_OI145}, we show the 
spectra centered on \Odrie \,\ and \Ohon \ \mic \ for the whole sample.

%++++++++++++++++++++++++++++++++++++++++++++++++++++++++++++++++++++++
\subsection{Carbon fine structure line}
When detected, the \CII \ 157.7 \mic \ line can be strong - more than 100 $\times$ 10$^{-18}$ W/m$^2$. 
However, it is only seen in six objects (30\% of the HAEBEs; see Fig.~\ref{f_allstars_CII}) - these are the 
same objects for which \Ohon \ \mic \ was also detected (see Table~\ref{t_atomflux}), apart from HD 36112.
The low detection rate for \CII \ is surprising, given the high detection rate (83\%) in {\em ISO}/LWS spectra 
reported by Lorenzetti al. (\cite{lorenzetti2002}). This can be attributed to a difference in aperture: 80 arcsec 
for LWS versus 9 arcsec for PACS.  Moreover, we noticed that the line can also be present in the off-source 
chop positions (in a spatially variable amount), contaminating our spectra. In two cases, the dominating 
emission is present in all on and off-source spaxels, so that our chop-off subtracted spectra even show the 
feature in absorption as the chop-off position shows stronger emission (HD 142527 and HD 163296, see 
Fig.~\ref{f_allstars_CII}). The interpretation of the \CII \ emission line is complex. Besides originating in the 
disc, it could also form in the remaining envelope, or simply in cloud material in the line of sight. A detailed 
analysis of background \CII \ emission is beyond the scope of this paper, but a dedicated study is underway 
(Pantin et al., {\em in preparation}).

%++++++++++++++++++++++++++++++++++++++++++++++++++++++++++++++++++++++

\begin{table}
\caption{CO line strengths in units of 10$^{-18}$ W/m$^2$, 1$\sigma$ continuum RMS 
between brackets in case of a detection or, in case of a non-detection, 3$\sigma$  upper limits. 
With ':' we indicate that the feature is between 2 and 3 $\sigma$.  For HD 158352, not all settings 
were observed, therefore we write 'n.a.' (not available) when the spectrum is lacking.}
\label{t_molflux}
\begin{center}
\begin{tabular}{lccccc}
\hline
\hline
Star                        &CO J36-35  &CO J33-32 & CO J29-28&CO J18-17 \\
$\lambda$ (\mic) &72.85       & 79.36         & 90.16          &144.78 \\
\hline
AB Aur              &$<$ 37.9      & $<$ 33.4    &31.0 (4.3)   &26.6 (8.8)  \\
HD 31648         &$<$9.6        &$<$16.3      &$<$9.6        &6.9 (2.3)    \\
HD 35187         &$<$ 21.0     &$<$ 28.9    &$<$ 8.4      &$<$ 5.5       \\
HD 36112        &$<$ 30.2     &$<$ 17.6   &$<$ 17.7       &6.6 (1.1)    \\
CQ Tau           &$<$ 15.4     &$<$ 25.4    &$<$ 9.9      &5.9 (1.5)        \\
HD 97048        &$<$ 21.3     &$<$ 16.6    &15.3: (6.0) &28.5 (2.6)    \\
HD 100453     &$<$ 13.9      &$<$ 19.5    &$<$ 9.6      &$<$ 8.3        \\
HD 100546     &$<$ 28.7       &$<$ 38.8    &49.3 (16.1)&58.0 (3.3)   \\
HD 104237     &$<$ 19.6       &$<$ 16.1   &$<$ 9.8       &$<$ 5.6       \\
HD 135344B &$<$ 13.0        &$<$ 21.7    &$<$ 8.5        &$<$ 8.4     \\
HD 139614     &$<$ 16.4      &$<$ 19.9     &$<$ 9.0       &$<$ 6.9      \\
HD 141569A &$<$ 12.3       &$<$ 10.6     &$<$ 6.6      &$<$ 3.4       \\
HD 142527     &$<$ 30.9      &$<$ 23.7    &$<$ 12.4    &$<$ 9.8       \\
HD 142666    &$<$ 16.3       &$<$ 18.8    &$<$ 12.6     &$<$ 6.2      \\
HD 144668    &$<$ 11.9       &$<$ 12.1    &$<$ 8.2       &5.6 (1.3)     \\
HD 150193    &$<$ 14.8       &$<$ 13.1    &$<$ 7.2       &$<$ 6.2      \\
KK Oph          &$<$ 6.5          &$<$ 9.4     &$<$ 7.7       &6.9 (1.3)      \\
51 Oph           &$<$ 9.0         &$<$ 12.4     &$<$ 6.6     &$<$ 4.0        \\
HD 163296    &$<$ 8.0        &$<$10.7      &$<$ 7.9      &11.6 (1.5)    \\
HD 169142    &$<$ 10.0      &$<$ 10.9      &$<$ 10.0   &5.1: (1.7)     \\
\hline
49 Cet              &$<$ 13.6      & $<$ 18.6   &$<$ 9.8       &$<$ 6.7      \\
HD 32297         &$<$ 11.4    &$<$ 16.9     &$<$ 9.7       &$<$ 4.3      \\
HR 1998          &$<$ 12.6     &$<$ 13.5    &$<$ 6.6      &$<$ 4.8         \\
HR 4796A       &$<$ 12.8     &$<$ 17.1  &$<$ 7.6       &$<$ 4.1          \\
HD 158352    & n.a.              &$<$ 9.2       &n.a.             &n.a.                \\
\hline
\end{tabular}
\end{center}
\end{table}

\begin{table*}
\caption{Molecular line strengths of \water, hydroxyl and CH$^+$ for the sources with a least one detection in these lines. 
Line flux in units of 10$^{-18}$ W/m$^2$, 1$\sigma$ continuum RMS between brackets in case of a detection 
or, in case of a non-detection, 3$\sigma$  upper limits. With ':' we indicate that the feature is between 2 and 
3$\sigma$. At 90 and 179.5 \mic, there is a possible blend of CH$^+$ and \water. For the water lines, we also 
give the upper energy level in K.}
\label{t_water}
\begin{center}
\begin{tabular}{lccccccccc}
\hline
\hline
Star                         & o-\water   & o-\water        &CH$^+$     & o-\water       & OH              &  OH        &CH$^+$/ p-\water &CH$^+$/ o-\water& o-\water\\ 
 $\lambda$ (\mic)  & 63.32        & 71.946          &72.14          &  78.74        & 79.11         &79.18       &90.00      &179.52    & 180.42 \\
$E_{\rm{up}}$ (K)  &1071.0       & 843.5            &--                 &432.2           &--                  &--               &296.8      &114.4      &194.1 \\
\hline
AB Aur         &$<$29.8    &$<$19.0   &$<$36.7     &$<$26.5     &24.6: (7.9)  &$<$24.8  &$<$ 18.4      & $<$ 13.5    & $<$ 8.2\\
HD 31648   &9.8 (2.2)    &$<$12.4   &$<$21.6     &$<$17.7     &$<$9.9       &$<$10.3   &$<$9.1         &$<$7.5        &$<$5.9\\
HD 97048   &$<$16.4    &$<$16.8   &$<$18.6     &$<$18.2     &$<$17.0      &$<$18.9 &17.4 (4.9)     &$<$18.3     &$<$13.8\\
HD 100546  &$<$25.9    &$<$50.9  &127.5 (17.0)&$<$ 57.0   &$<$ 41.9    &$<$ 42.1 &116.0 (16.1)&30.5 (4.8)   &$<$ 17.1\\
HD 163296 &14.2:  (4.4)&16.5 (5.3)&$<$19.9     &10.2: (4.1) &11.4 (3.4)   &$<9.1$  &4.7: (1.9)      &$<$ 10.4    &$<$ 6.2\\
\hline
\end{tabular}
\end{center}
\end{table*}

%++++++++++++++++++++++++++++++++++++++++++++++++++++++++++++++++++++++
\subsection{Carbon monoxide}

\begin{figure}[t!]
   \resizebox{\hsize}{!}{\includegraphics{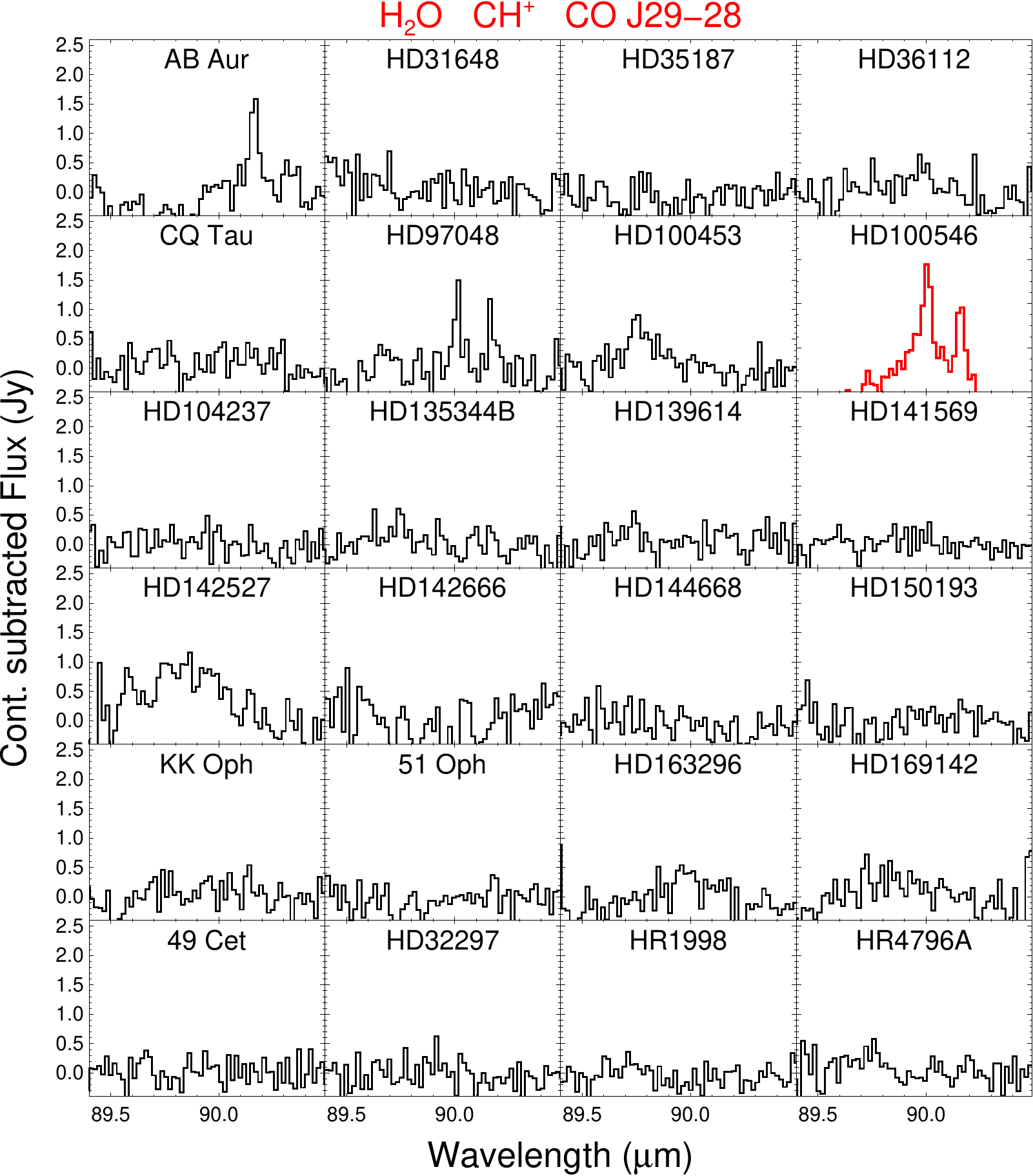}}
\caption{The spectra around 90 micron, covering p-\water \ at 89.988, CH$^+$ at 90.02 and CO at 
90.163\,\mic. The only sources with clearly detected lines are AB Aur (CO), HD 97048 and HD 100546 
(CH$^+$ and CO). HD 100546 is scaled by a factor 1/4, so that the emission lines would still fit in the 
plot window.} 
\label{f_allstars_90}
\end{figure}

In our PACS ranges, we cover four transitions of the CO molecule: $J$=36-35 (72.85\,\mic ), 
$J$=33-32 (79.36\,\mic), $J$=29-28 (90.16\,\mic) and $J$=18-17 (144.78\,\mic). These are all 
mid to higher $J$ transitions, with E$_{up}$ between $\sim$ 950 and 3700 K. The highest $J$ 
transitions in our settings ($J$=36-35 and $J$=33-32) are not detected in any star of our sample. 
In Fig.~\ref{f_allstars_90}, we show the region around 90\,\mic, covering the CO $J$=29-28 transition. 
This CO line is only clearly seen in AB Aur and HD 100546, with a tentative detection for HD 97048. 
For the lowest $J$ observable (18-17, at 144.78 \mic) we see many more detections in our spectra 
(see Fig.\,\ref{f_allstars_OI145}): it is detected in nine objects, and there is one tentative detection, 
for HD 169142. The strongest CO lines are observed in AB Aur, HD 97048 and HD 100546.
HD 141569A is the only star for which we have a clear (more than 5$\sigma$) detection of \Ohon \,\mic, 
but no detection of CO at 144.78 \mic, showing that both species trace different excitation conditions 
and chemistry (atomic versus molecular). We will discuss this further in Sect.\,\ref{s_co}.

%++++++++++++++++++++++++++++++++++++++++++++++++++++++++++++++++++++++
\subsection{Hydroxyl}

With our settings, we only cover the OH doublet at 79.11/79.18\,\mic. In Fig.\,\ref{f_water}, we can
see evidence for the doublet in HD 100546 and HD 163296; however, the only 3 $\sigma$ detection 
(of one line of this doublet) is found in AB Aur and HD 163296 (see Table~\ref{t_water}). Sturm et al. 
(\cite{sturm2010}) detected several OH lines at far-IR wavelengths (53-200 \mic) in the SED-mode 
PACS spectra of HD 100546.

%++++++++++++++++++++++++++++++++++++++++++++++++++++++++++++++++++++++
\subsection{Water}

\begin{figure*}
\begin{center}
   \resizebox{9cm}{6cm}{\includegraphics{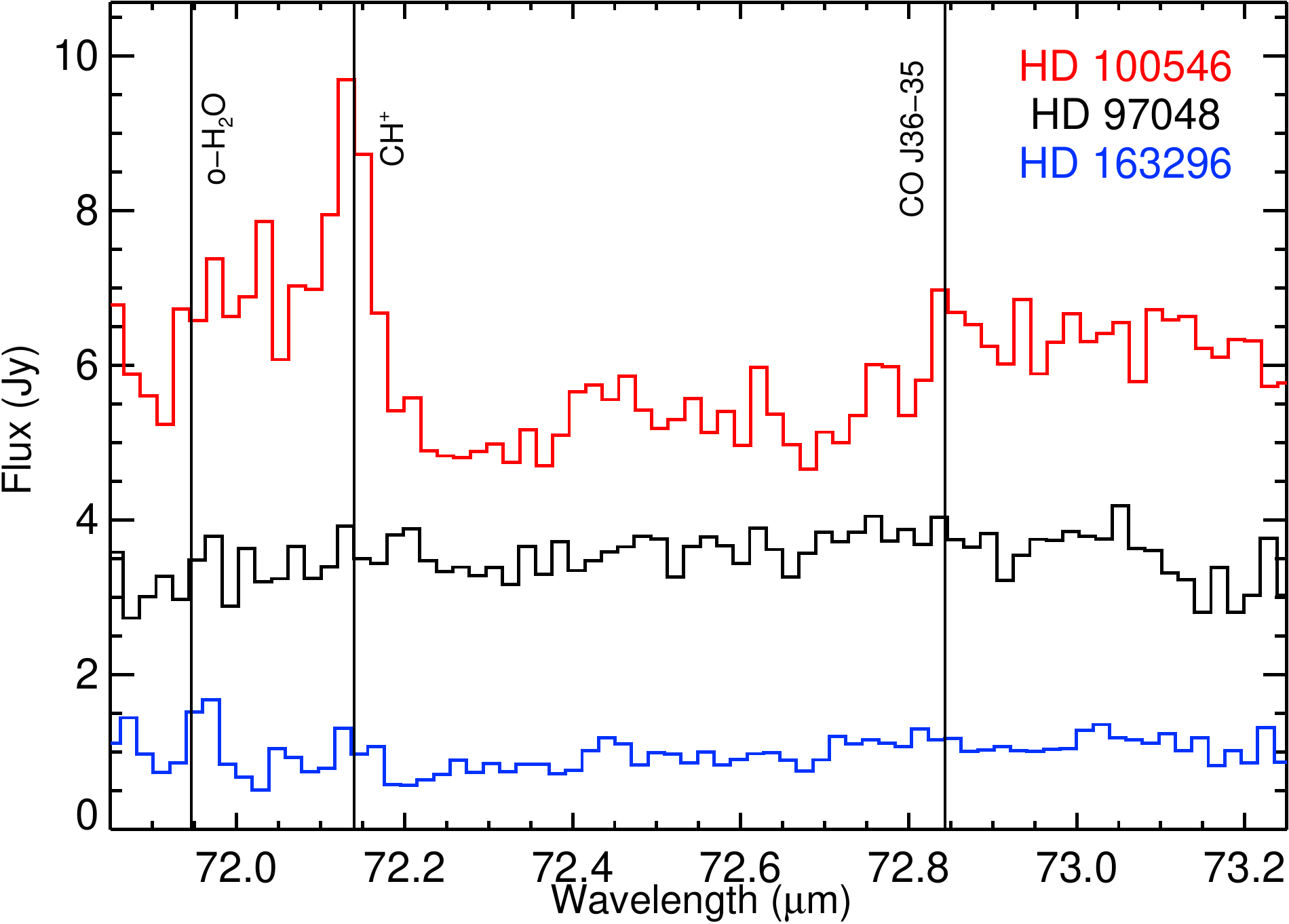}} 
   \resizebox{9cm}{6cm}{\includegraphics{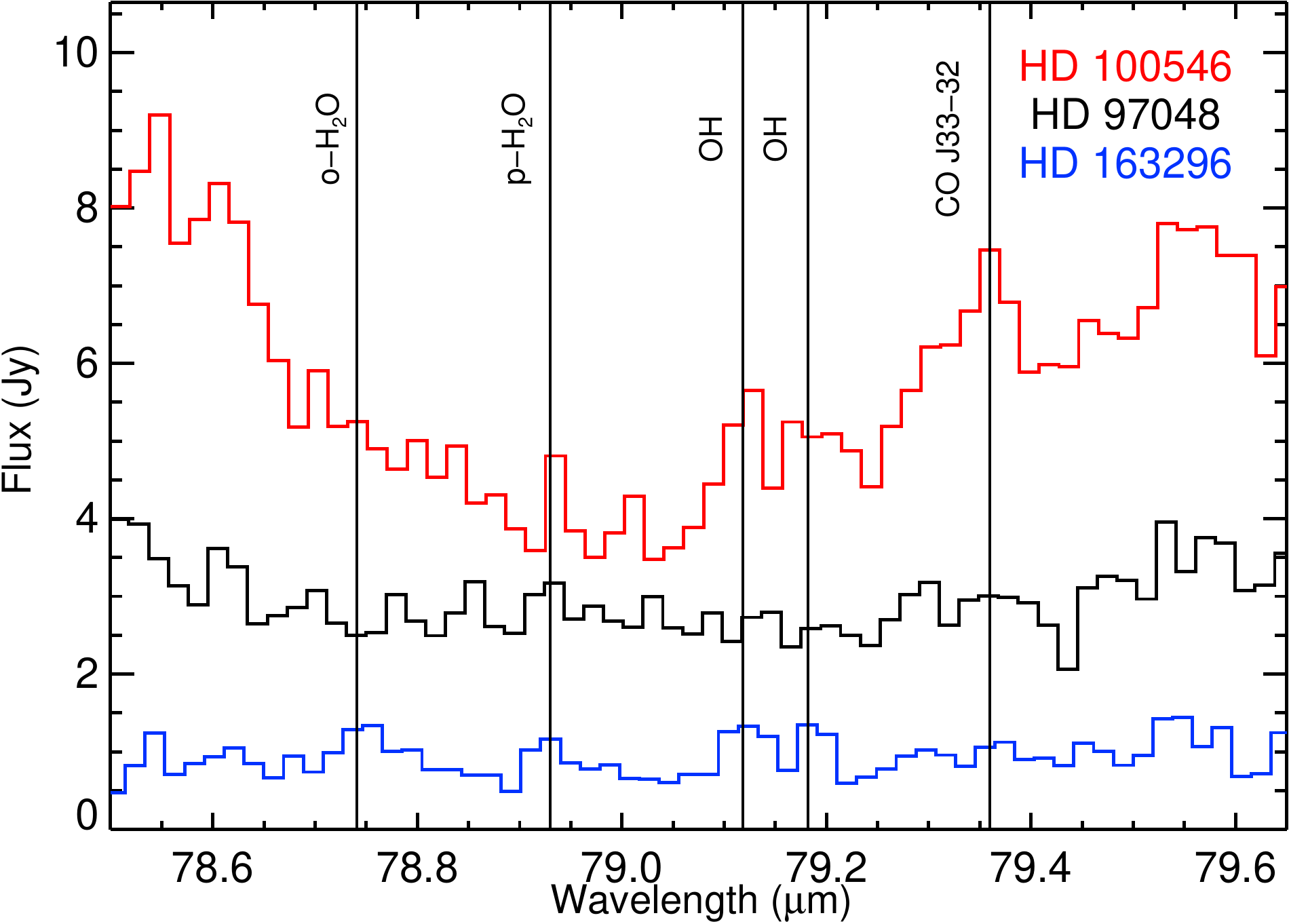}} 
   \resizebox{9.0cm}{6cm}{\includegraphics{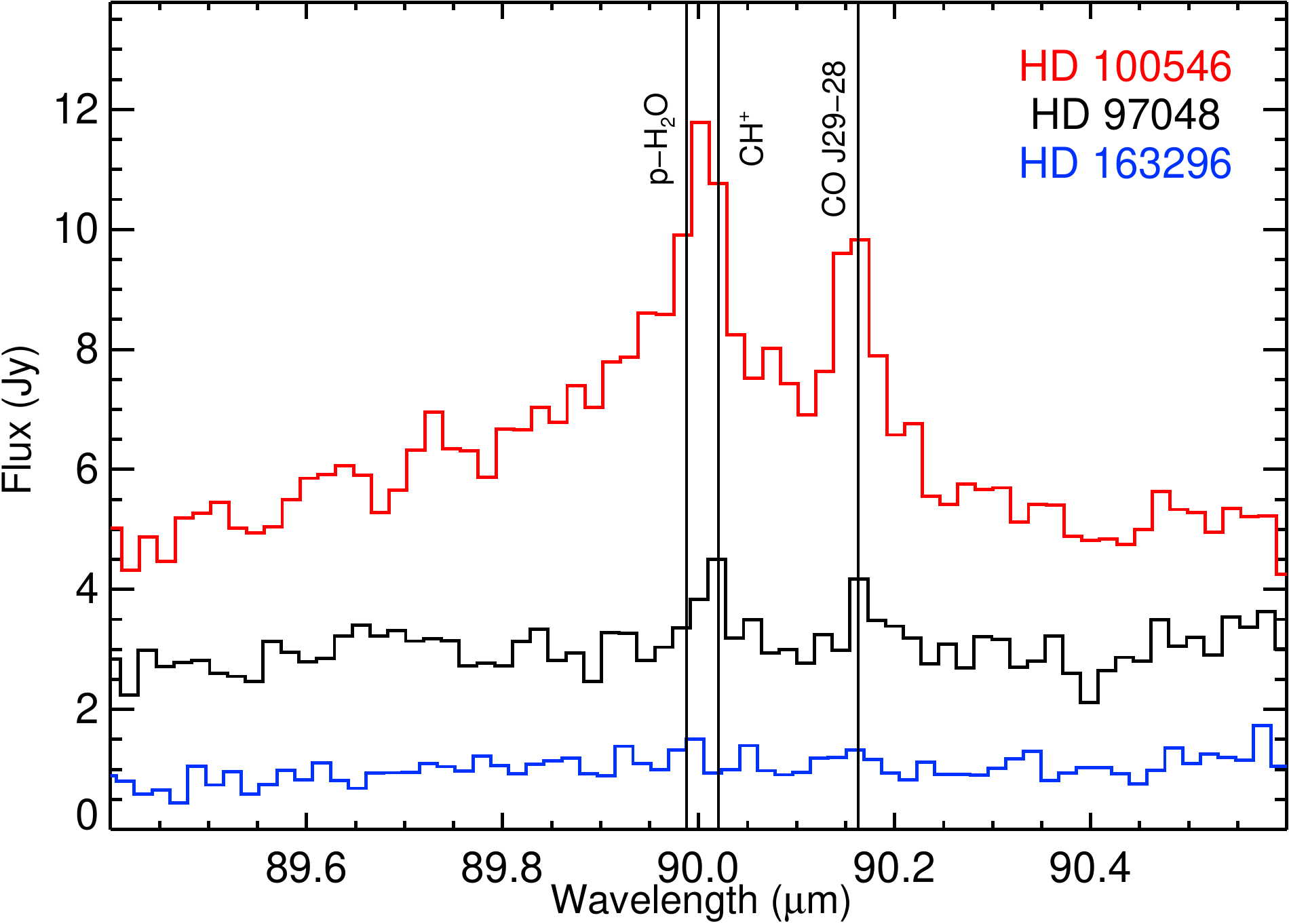}} 
   \resizebox{9.0cm}{6cm}{\includegraphics{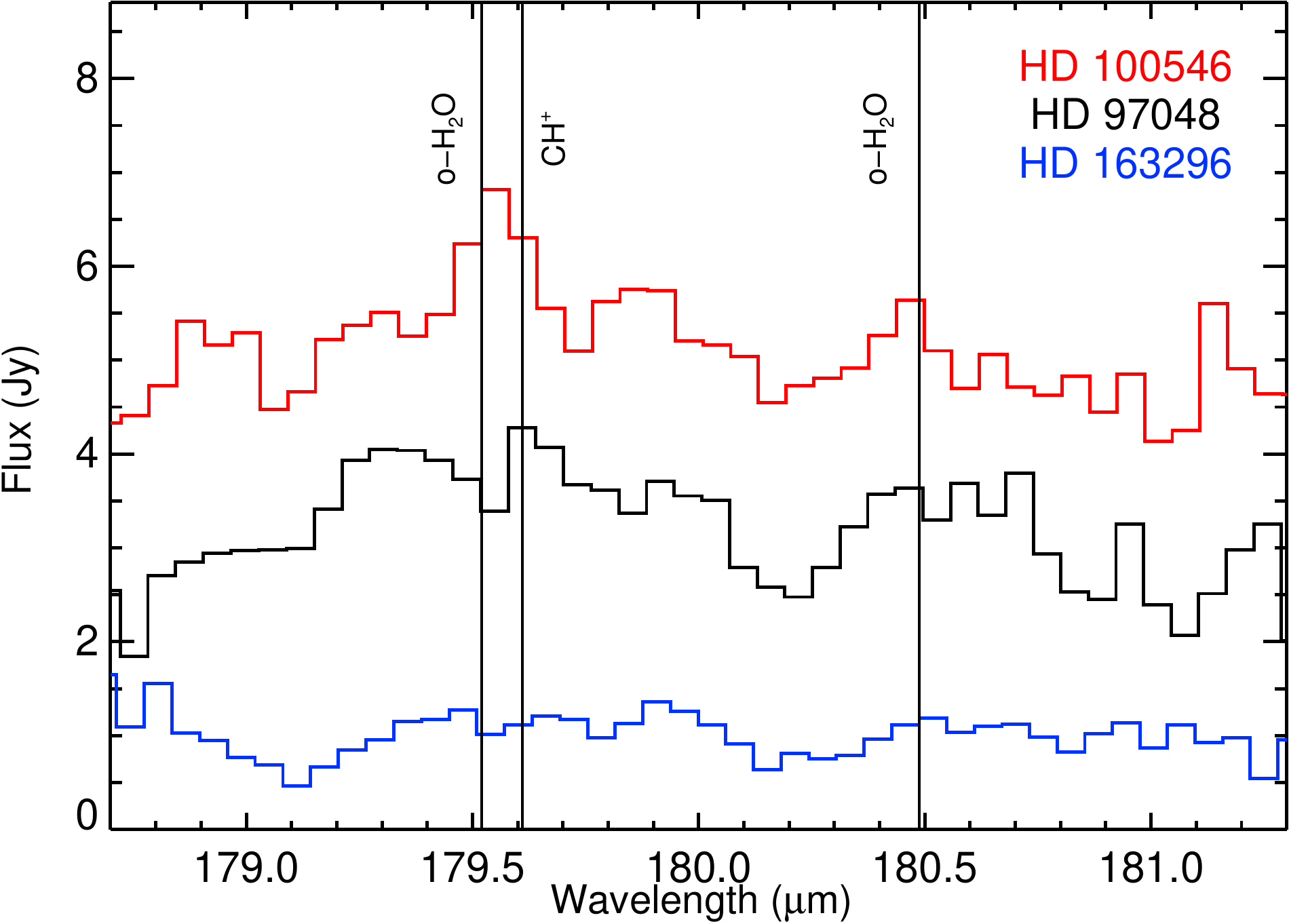}} 
\caption{Comparison of HD 100546 (top spectrum), HD 97048 (middle) and HD 163296 (bottom) at 
72.6 and 79.0, 90.0 and 180.0 \mic. We indicate the position of the lines of CO, CH$^+$, OH and \water.}
\label{f_water}
\end{center}
\end{figure*}

The only star in our sample with convincing evidence for water is HD 163296. 
The measured line fluxes are listed in Table\,\ref{t_water}. While the feature at 63.32 \mic \ is seen in 
HD 163296 with 3$\sigma$ confidence, we list it as a tentative detection given the spurious absorption 
feature next to it (in HD 31648, a potential feature of water at 63.32 \mic \ is redshifted by 0.009 \mic \, so 
we also consider it a tentative detection). Also in our deeper range scans at 71.946 and 78.74 \mic \ we 
see evidence for water in HD 163296 (see Fig.\,\ref{f_water}). The features are close to 3$\sigma$, 
however, the fact that we do see emission lines at the positions where water lines are predicted to be 
present further strengthens the detection of water in HD 163296. The analysis of the water in 
HD 163296 will be presented in a separate paper (Meeus et al. \cite{meeus2012}).

In two other objects,  HD 97048 and HD 100546, we do see several emission lines at the position
of water. In Fig.\,\ref{f_allstars_90} we show the region around 90 micron, where the transition 
of para-\water \ can appear. However, when we zoom in on these stars in the regions where other 
water lines are expected to be present, we get a different picture; in Fig.\,\ref{f_water} we show the 
spectra covering water lines. At 90.00\,\mic \ (see Fig.\,\ref{f_water}) we do see an emission peak 
for both stars, which could be the para-\water \ line (at 89.988\,\mic), but it is a blend with CH$^+$ 
at 90.02\,\mic. At other positions of water, which are not blended with CH$^+$, we do not see any 
emission line: 71.948, 78.741 and 78.93, 180.488\,\mic). 

To summarise, the only times we detect an emission feature at the position of a water line in these 
two objects, is in a blend with CH$^+$. We can conclude that there are no \water \ emission lines 
detected in these two objects at the sensitivity of our observations. Therefore, we do not confirm 
the detection of water in HD 100546 reported by Sturm et al. (\cite{sturm2010}) as also seen in a later
improved reduction of the original data (Sturm, priv. comm.).

\subsection{CH$^+$}

Thi et al (\cite{thi2011}) reported the first detection of CH$^+$ in a Herbig Ae/Be disc, HD 100546.
We detect the feature of CH$^+$ in both HD 97048 and HD100546 at 90.00\,\mic. For HD100546 
we have detections of two more lines, at 72.14 and 179.61\,\mic. The features at 90.0 and 
179.6\,\mic \ are also at the position of water features, but given the lack of other water features, 
they can be attributed to CH$^+$.

%++++++++++++++++++++++++++++++++++++++++++++++++++++++++++++++++++++++
%++++++++++++++++++++++++++++++++++++++++++++++++++++++++++++++++++++++
\section{Gas lines as tracers of the conditions in the disc}
\label{s_conditions}

The far-IR lines observed with PACS form in different regions of the disc. The exact locations 
vary with geometry, i.e. flaring, inner holes, gaps, but we will discuss here for simplicity the 
general case of a continuous flaring disc around a Herbig Ae star. 

The \CII \ line depends strongly on the irradiation of the star, especially UV photons 
shortward of 1200~\AA \ (Pinte et al. \cite{pinte2010}, Kamp et al. \cite{kamp2011}). 
The emission originates foremost in the upper tenuous layers of the disc (low critical density) 
where UV photons can penetrate. The \Odrie \ and \Ohon ~\mic \ form deeper in the 
disc where the atomic oxygen abundance is still high. Most of their emission comes from 
10-100~AU as the temperatures beyond 100~AU are generally too low to excite those
lines (Kamp et al. \cite{kamp2010}). The high excitation water lines ($E_{\rm upper}
\geq 400$~K) form mostly in the surface layers of the hot water reservoir inside the snow line
(15 AU for an effective temperature of 10500 K, moving inwards for cooler stars while
keeping the disc structure constant). The low excitation lines ($E_{\rm upper} \leq 200$~K) 
form in a thin layer beyond the snow line where water can be photodesorbed from the icy grains
into the gas phase (Cernicharo et al. \cite{cernicharo2009}, Woitke et al. \cite{woitke2009b}). 
The exception is the $89.988~\mu$m water line with an upper level energy of $\sim 300$~K
which forms across the snow line (for TW Hya, a T Tauri disc - Kamp et al. in preparation).
Bruderer et al. (\cite{bruderer2012}) modelled the CO ladder in HD 100546. They found that the 
high $J$ lines of CO can only be reproduced by a warm atmosphere in which T$_{\mathrm{gas}}$ 
is much higher than T$_{\mathrm{dust}}$. The low $J$ lines of CO (observed in the mm) trace 
the outer disc (at several 100 AU radial distance), while the mid to high $J$ lines observed in 
the far-IR originate at distance of several tens of AU. The highest $J$ lines of CO form mostly 
in the very inner disk (typically inside a few AU, or at the rim of transition discs). 

The fundamental ro-vibrational CO band ($\Delta v$ = 1) at 4.7 \mic \ band, tracing the terrestrial 
planet-forming region is routinely observed in HAEBEs (e.g. Brittain et al. \cite{brittain2007}). The 
bands are rotationally excited up to high $J \,  ( >$ 30), with \trot \ between 900 and 2500 K 
(van der Plas et al. \cite{plas2011}). If the gas is not in LTE, then the vibrational temperature, \tvib, 
can depart from \trot, when UV fluorescence causes super-thermal level populations. This is observed 
in several UV bright HAEBEs where \tvib \ $>$ 5000 K: HD 97048 and HD 100546 have \tvib \ $> $ 
6000 K, while \trot \ $\sim$ 1000 K (Brittain et al. \cite{brittain2007}, van der Plas et al. \cite{plas2011}). 
Also, in group I discs \trot \ $<$ \tvib, while in group II discs, \tvib \ $\lessapprox$ \trot (van der Plas et al. 
\cite{plas2011}). 

Furthermore, the line profile suggests CO depletion in the innermost regions of HAEBE 
discs (van der Plas et al. \cite{plas2009}), with group I clearing a larger radius ($r_{in}\sim$ 10 AU) than 
group II discs ($r_{in} \sim$ 1 AU; van der Plas et al. \cite{plas2011}). 
The {\em transitional disc}  HD 141569A stands out for having a low \trot \ $\sim$ 250 K, while its \tvib \  
($\sim$ 5600 K) is in a similar range of the hottest CO observed in HAEBEs, attributed to UV fluorescence 
(Brittain et al. \cite{brittain2007}). Furthermore, Goto et al. (\cite{goto2006}) showed that this disc has an 
inner clearing in CO up to a radius of 11 AU, comparable to the group I discs. 

$^{12}$CO lines observed in the millimeter come from low $J$ transitions of optically thick CO located in the 
outer disc surface. These pure rotational transitions of cold CO ($\Delta v$ = 0) are routinely detected in 
HAEBE discs (e.g. Pi\'etu et al. \cite{pietu2003}, Dent et al. \cite{dent2005}).  Earlier, the existence of 
Keplerian rotation in discs was confirmed with mm interferometry of CO lines (Koerner et al. \cite{koerner1993}). 
Furthermore, as the lines are optically thick, a simple model of the line profile allows for an estimate of the outer 
disc radius and even inclination (e.g.\ Dent et al. \cite{dent2005}, Panic et  al. \cite{panic2008}). 

Woitke et al. (\cite{woitke2010}) calculated a grid of disc models with the thermo-chemical radiation 
code \prodimo. This model grid, called "Disc Evolution with Neat Theory"  is a useful tool to 
derive statistically meaningful dependencies on stellar and disc properties. Kamp et al. (\cite{kamp2011}) 
used the model grid to derive line diagnostics that are relevant for the PACS observations. We will refer to 
these diagnostics in our discussion below. Our sample includes several objects that stood out in earlier 
papers, in terms of detections of \hh, CO and/or OH, which can be attributed to a high level of UV 
fluorescence; we will relate these results also to our new observations. We will now discuss the results 
presented in the previous sections in the context of our current understanding of these discs. The following 
sections present our interpretation of observational correlations and their meaning in the context of disc 
structure and evolution.

\begin{table*}
\begin{center}
\caption{Probability $p$ (in percentage) that the two parameters (x,y) under consideration are not 
correlated, calculated with several statistical methods: Spearman's, Kendall's and Cox-Hazard's. 
Under 'Result' we indicate what we can derive from these statistical tests.}
\begin{tabular}{llcccccc}
\hline
\hline
\multicolumn{2}{c}{Parameters}                           & $p$ Spearman & $p$ Kendall& $p$ Cox-Hazard &Result                 &\multicolumn{2}{c}{Linear fit: y = a $\times$ x + b}\\
\multicolumn{1}{c}{x}     &\multicolumn{1}{c}{y}  &                            &                        &                                &                            &a                       &b\\
\hline
log \Odrie \mic                            &log \Ohon  \mic         &0.07             &0.06               &$<$0.1                   & Correlated        &1.2 $\pm$ 0.21& 1.8 $\pm$ 3.3 \\
log \Odrie \mic                            &log CO $J$18-17     &0.06             &0.03               &$< $0.1                  & Correlated        &0.64 $\pm$ 0.07&-7.1 $\pm$ 1.1\\
log L$_{\mathrm{UV }}$           &log \Odrie \mic         &0.33              &0.32               &--                             & Correlated        &0.72 $\pm$ 0.20& -16.1 $\pm$ 0.1\\
log \teff                                         &log L$_{\mathrm{UV}}$&0.05       &$<$0.1          &$<$0.1                   & Correlated        &9.3 $\pm$ 0.2&-36.5 $\pm$ 4.4\\
log L$_{\mathrm{[OI]6300 \AA}}$&log \Odrie \mic &0.88                 &0.45               &--                             & Correlated        &0.83 $\pm$ 0.21 & -12.4 $\pm$ 1=0.9\\
log F$_{63 \mu \mathrm{m}}$ &log F$_{\mathrm{1.3 mm}}$&0.19&0.09               &0.01                       & Correlated        &0.72 $\pm$ 0.1&-0.20 $\pm$ 0.2\\

\hline
log F$_{63 \mu \mathrm{m}}$   &log \Odrie \mic           & 4.6               &4.4                  &2.7                       & Correlated?     &&\\
log CO $J$ 3-2                             &log \Odrie \mic          &3.7                &2.4                  &--                          & Correlated?     &&\\
log \teff                                           &log \Odrie \mic          &4.7                &3.5                  &1.6                       & Correlated?     &&\\
log L$_{\mathrm{PAH6.2 \mu m}}$ &log \Odrie \mic   &6.6                &4.4                  &--                          & Correlated?    &&\\
log L$_{\mathrm{Br \gamma}}$ &log \Odrie \mic         &4.6                &4.4                  &--                          & Correlated?     &&\\

\hline
log CO $J$ 2-1                  &log \Odrie \mic             &54                       &31                  &--                             & Not correlated   &&\\
log L$_{*}$                          &log \Odrie \mic             &7.2                     &9.2                 &18                           & Not correlated   &&\\
log L$_{\mathrm{X}}$       &log \Odrie \mic             &84                      &71                   &--                             & Not correlated  &&\\
log F$_{\mathrm{1.3 mm}}$&log \Odrie \mic          &34                      &27                  &44                           & Not correlated  &&\\
log L$_{\mathrm{Acc}}$ (BaDis)&log \Odrie \mic  &84                     &83                   &--                             & Not correlated  &&\\
log F$_{\mathrm{IR}}$/F$_{*}$&log \Odrie \mic    &21                      &27                   &47                          & Not correlated   &&\\
slope b                                &log \Odrie \mic             &43                       &38                  &72                          & Not correlated   &&\\
\hline
\end{tabular}
\label{t_stats}
\end{center}
\end{table*}

In order to remove the bias that can be introduced by the distance of the stars, we scaled our data to 
a distance of 140 pc (to ease comparison with objects in Taurus and predictions of the model grid). The 
scaled data are all the PACS line and continuum fluxes, the $^{12}$CO $J$ = 3-2 and 2-1 line fluxes, and 
the mm continuum fluxes. 

The relations between parameters are analysed with their corresponding '$p$-values' (see Table~\ref{t_stats}), 
which gives the probability that the two variables considered are not correlated. Two parameters will be 
classified as 'correlated' if one or more of the three (Spearman, Kendall and Cox-Hazard) differently obtained 
$p$-values are not larger than 1\%, and as 'tentatively correlated' when 1$ < p <$ 5\% (e.g. Bross \cite{bross1971}). 
When $p > 5$ \%, the parameters are classified as 'not correlated', as their $p$-values are similar to $p$-values 
derived for randomly generated samples. The $p$-values and linear fits provided in 
Table~\ref{t_stats} take into account that several of our datasets include upper/lower limits instead of detections, as
they were derived with the ASURV package (Feigelson \& Nelson \cite{feigelson1985}, Isobe et al. \cite{isobe1986}, 
and Lavalley et al. \cite{lavalley1992}), that was specifically designed to deal with censored data.
We used the Spearman's partial correlation technique to quantify the influence of the common distance parameter 
on the probability of false correlation and found that this is negligible for our sample -- the $p$-values considering 
the distances, or random values instead, are practically equal. The absence of influence of the distances on the 
correlations most probably comes from the relatively narrow range covered by this parameter in our sample. We 
excluded 51 Oph in the correlation test with L$_{\mathrm{UV}}$ and \Odrie, as it is an outlier when comparing its
extremely high \lUV \ to the rest of the sample (see Table~\ref{t_excess} and Fig.~\ref{f_starpara}), and an
enigmatic object (e.g. van den Ancker et al. \cite{ancker2001}).

%++++++++++++++++++++++++++++++++++++++++++++++++++++++++++++++++++++++
\subsection{Oxygen fine structure lines}

\begin{figure}[t!]
 \resizebox{\hsize}{!}{\includegraphics{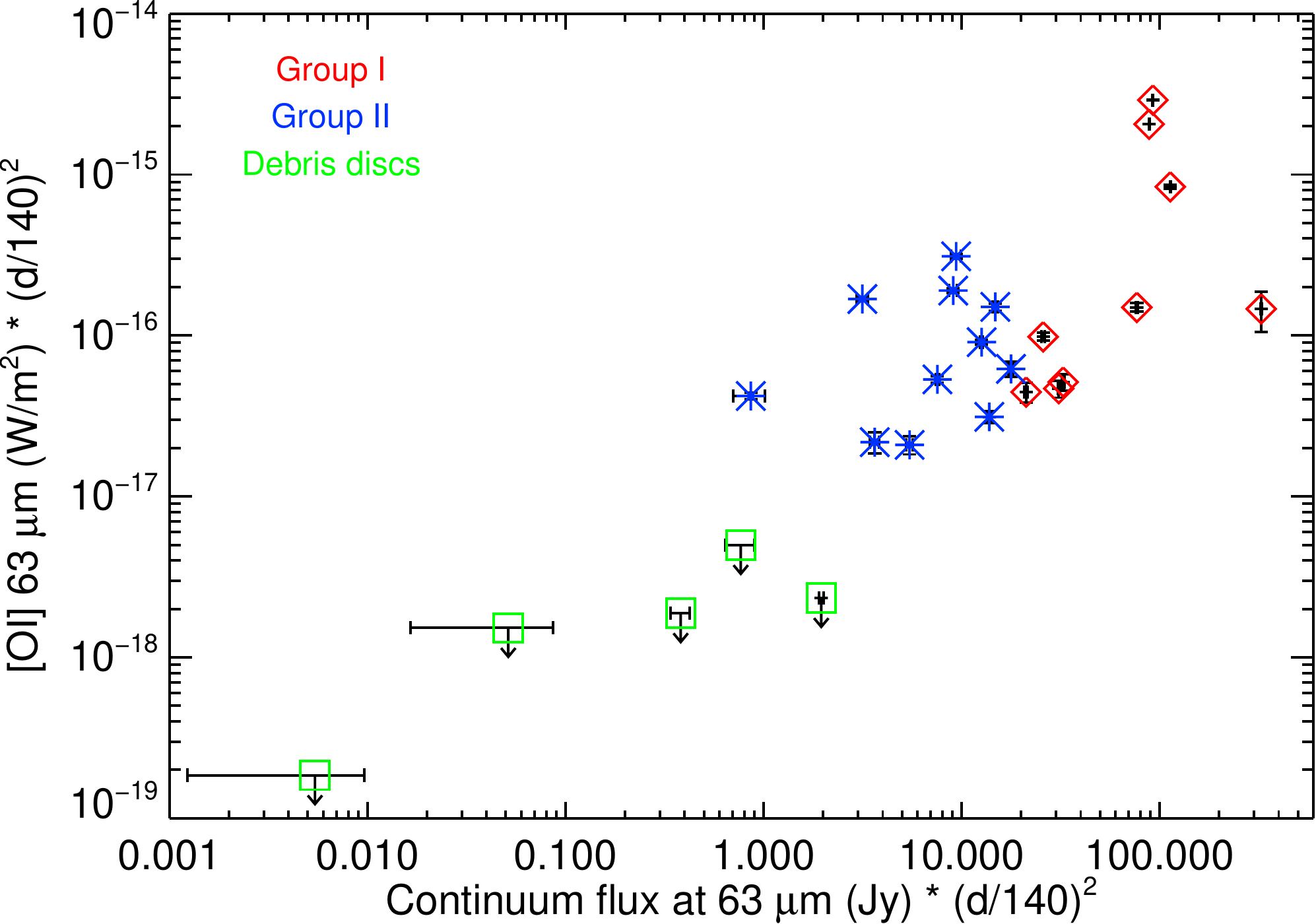}}
 \resizebox{\hsize}{!}{\includegraphics{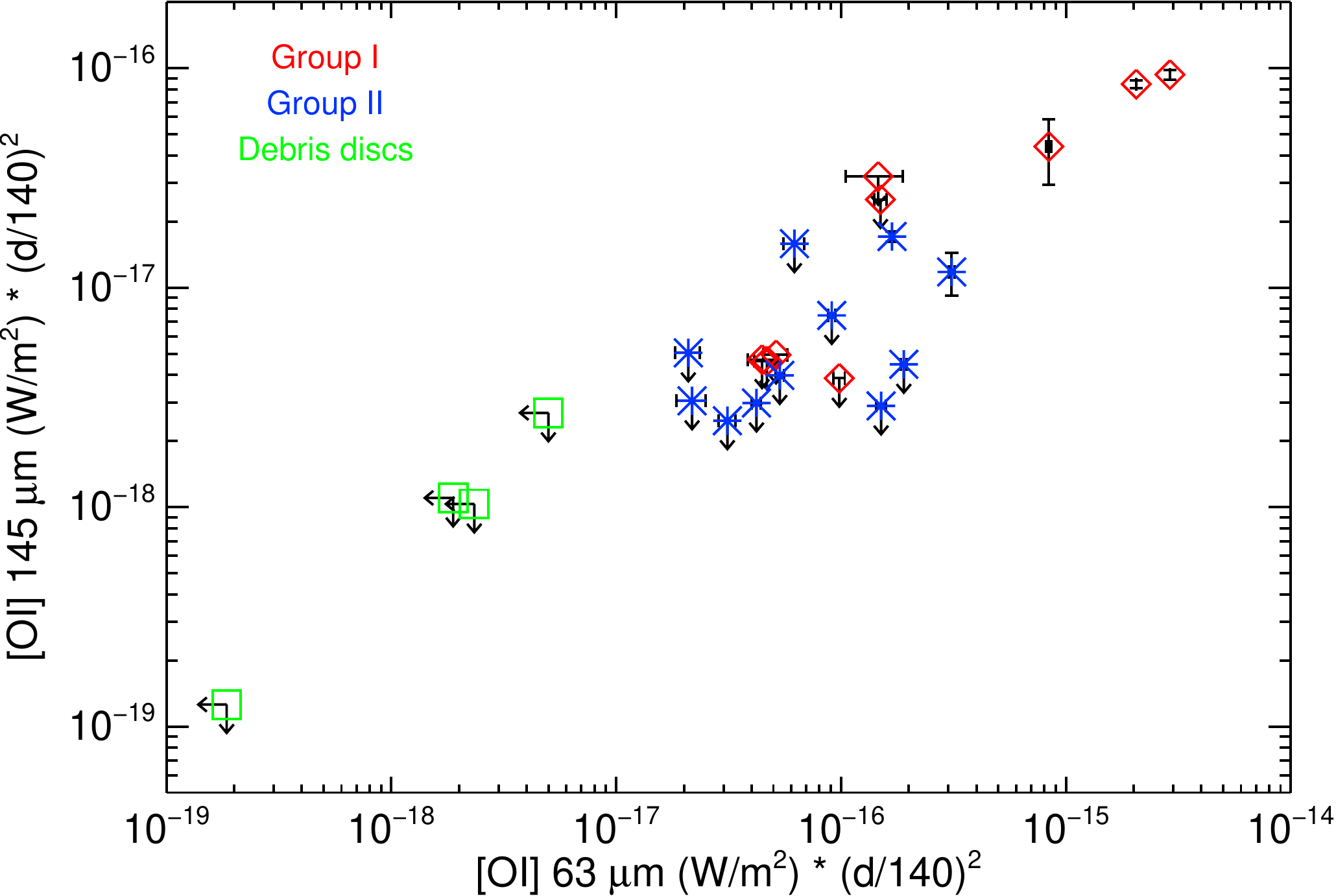}} 
\resizebox{\hsize}{!}{\includegraphics{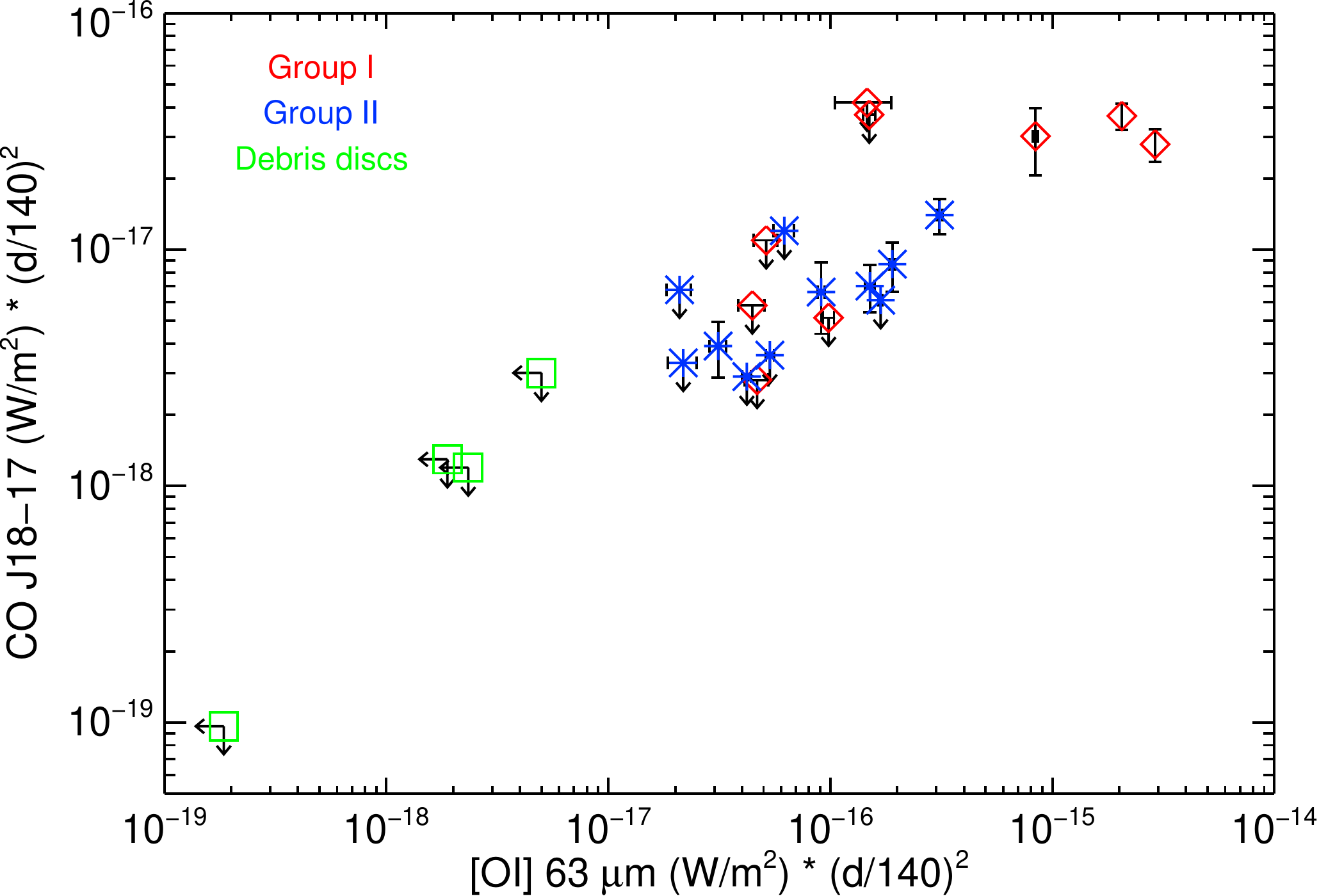}}
\caption{Top to bottom: \Odrie \ \mic \ as a function of the continuum flux at 63\,\mic; \Ohon 
\,\mic \ as a function of  \Odrie \,\mic \, and CO J=18-17 as a function of the \Odrie \ \mic \ flux.
All fluxes are normalised to a distance of 140 pc. Diamonds: group I sources, asterix: group II
sources, squares: debris discs.}
\label{f_linesratio}
\end{figure}

In Fig.~\ref{f_linesratio}, we plot the strength of the \Odrie~\mic \  line as a function of the 
continuum flux at 63~\mic. The variables are weakly correlated (see Table~\ref{t_stats}). The three 
sources with the highest line fluxes also have the highest continuum fluxes: AB Aur, HD 97048 
and HD 100546, to which we will refer to as 'the bright three'. These are also the only HAEBEs 
in which \hh \ emission has been detected in the IR (see Sect.~\ref{s_intro}).

The ratios of the fine structure line fluxes of \Odrie, \Ohon \ and \CII \ 157 \mic \ are diagnostics of the 
excitation mechanism (e.g. Kaufman et al. \cite{kaufman1999}). Unfortunately, for most of the sources 
we only obtained upper limits for one or more of these lines. We show the line flux of  \Odrie \ as a 
function of \Ohon \ in Fig.~\ref{f_linesratio}, and find a clear correlation (see Table~\ref{t_stats}).  We find 
line ratios of \Odrie /\Ohon \ between 10 and 30. These ratios are not compatible with predictions of the 
PDR model in Tielens \& Hollenbach (\cite{tielens1985}) for optically thick lines with T$_{\rm{gas}} <$ 
200 K. Our line ratios (median of 24) are in agreement with predictions from the model grid, which gives a 
median line ratio of 25 (Kamp et al. \cite{kamp2011}). These authors conclude, based on those disc models,
that the line ratio is not sensitive to the average oxygen gas temperature (for 50 $< T_{\rm{gas}} < $ 500 K), 
but instead correlates with the gas to dust ratio.

\subsection{Ionised carbon fine structure line}

The \CII \ line flux is very sensitive to the UV radiation field, and the line is mostly optically thin.
Unfortunately, most of our sources are background contaminated, as \CII \ is also detected
in off-source positions, in a variable amount, depending on the location. For the few sources with 
solid, non-\CII \ background contaminated detections, we find line ratios of \Odrie/\CII \ 157 between 10 and 30. 

\subsection{Carbon monoxide}
\label{s_co}

\begin{figure}[t!]
  \resizebox{\hsize}{!}{\includegraphics{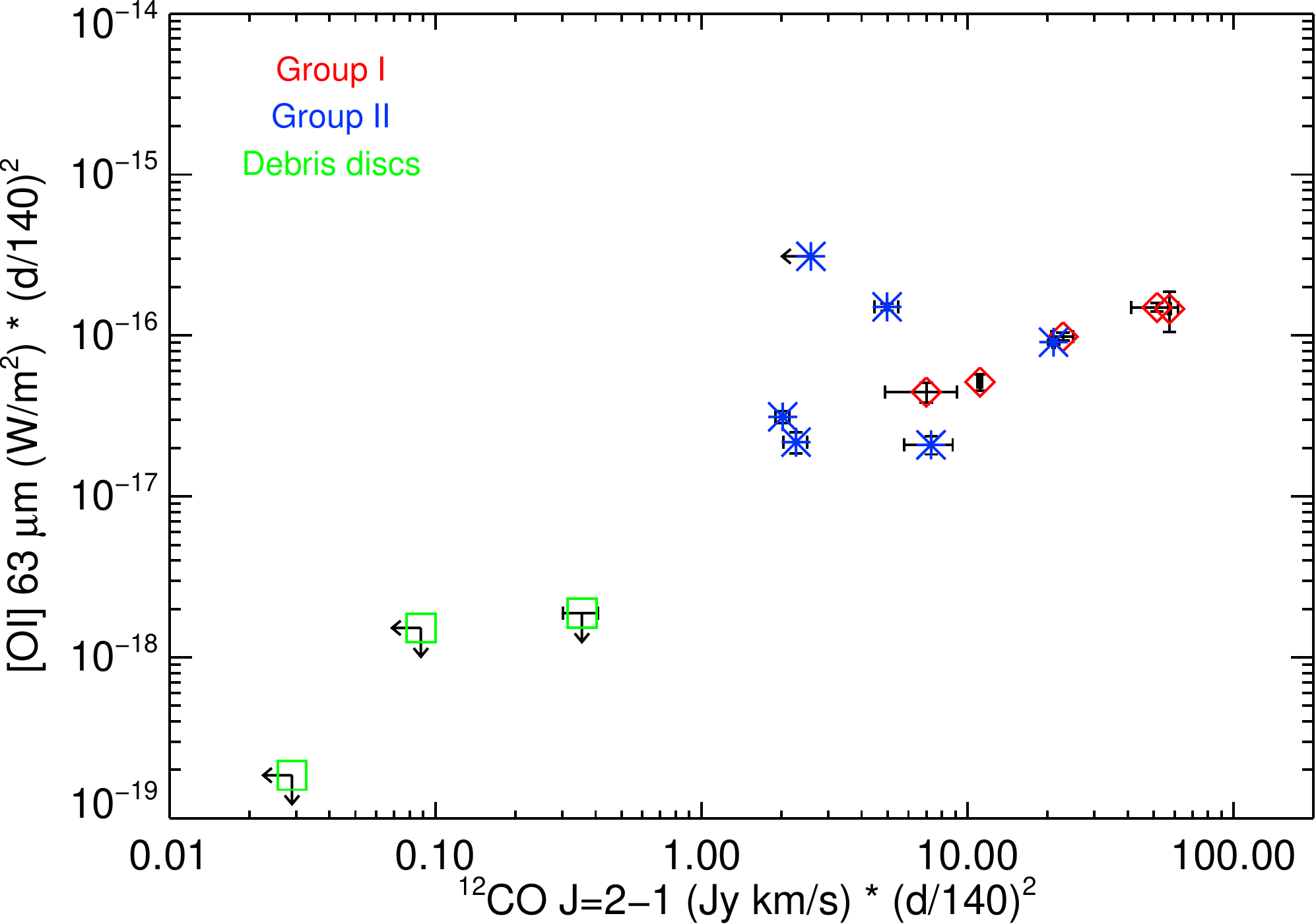}} 
  \resizebox{\hsize}{!}{\includegraphics{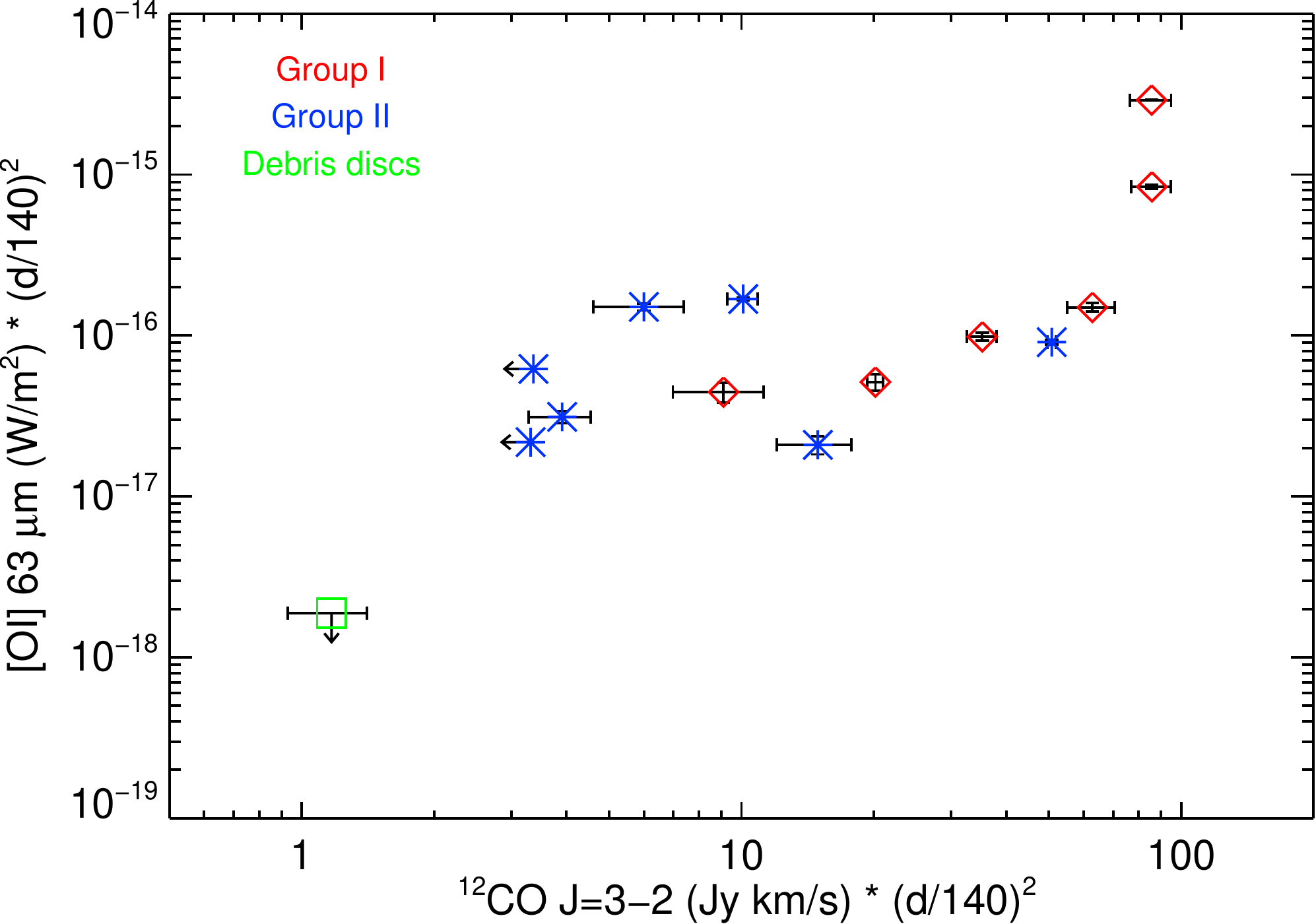}} 
\caption{ \Odrie \,\mic \ line flux versus strength of the $^{12}$CO $J$ = 2-1 (top)
and $J$=3-2 transitions (bottom). All fluxes are normalised to a distance of 140 pc. 
Diamonds: group I sources, asterix: group II sources, squares: debris discs.}
\label{f_mmco}
\end{figure}

Freeze-out of CO on grains is not expected in the disc of an A-type star (e.g. Panic et al. 
\cite{panic2009}), so that the strength of the low-$J$ \ $^{12}$CO lines can be used to obtain a lower 
limit on the cold gas mass. The disc size can be derived from the $^{12}$CO flux and profile.
In Fig.~\ref{f_mmco}, we plot the \Odrie \ line flux as a function of the $^{12}$CO $J$=2-1 and $J$=3-2 
line strengths (data from Dent et al. \cite{dent2005}, Panic et al. \cite{panic2009}, Isella et al. 
\cite{isella2010}, \"Oberg et al. \cite{oberg2010}, \cite{oberg2011} and our own data, see 
Appendix~\ref{a_sma}). We do not find a clear correlation with the $J$=2-1 transition, but we do find 
a weak correlation with the $J$=3-2 transition (see Table~\ref{t_stats}). 

Kamp et al. (\cite{kamp2011}) showed that the ratio of \Odrie /$^{12}$CO $J$=2-1 can be used
to derive the gas mass in the disc to an order of magnitude. The idea is that this ratio is determined
by the average gas temperature in the disc. If the \Odrie \ line is optically thin, the line flux will mainly
depend on the gas mass and average \Oone \ temperature (Woitke et al. \cite{woitke2010}). Once the 
temperature is known, the line flux of \Odrie \ can thus be related to 
the disc gas mass. In our sample, we have nine sources for which the $^{12}$CO $J$=2-1 line flux 
is known. We calculated the ratio for  those sources, and found that log  (\Odrie /$^{12}$CO $J$=2-1) 
falls between 2.5 and 3.5. This means that we have a similar average gas temperature in all cases. 
We apply the relation for this ratio range between log \Odrie \ and the gas mass, derived by Kamp et 
al. (\cite{kamp2011}) to obtain an estimate of the disc mass. The results are shown in Table~\ref{t_mgas}. 
We derive M$_{\rm{gas}}$ between 0.24 and 25 $\times \ 10^{-3}$ \Msun. These values are of course only 
indicative; for a more accurate estimate, a full model of all the available observations needs to be done
for each disc. The masses are consistent with the estimates derived from a detailed modelling of  
HD 163296 (M$_{\mathrm{gas}} \sim 15-120 \times 10^{-3}$ \Msun; Tilling et al. \cite{tilling2012}) and 
HD 169142 (M$_{\mathrm{gas}} \sim 3-7 \times 10^{-3}$ \Msun; Meeus et al. \cite{meeus2010}).

\begin{table}
\begin{center}
\caption{Line fluxes of $^{12}$CO $J$=2-1, log of the line ratios and derived gas masses.}
\begin{tabular}{lccr}
\hline
\hline
Object  & $^{12}$CO $J$=2-1 & log (\Odrie / &\multicolumn{1}{c}{M$_{\rm{gas}}$}\\
             & (10$^{-18}$ W/m$^2$)& CO $J$=2-1)&\multicolumn{1}{c}{(\Msun)}   \\
\hline
HD 31648       &  0.169      &  2.75    &  6.5 $\times 10^{-3}$  \\%-2.187
HD 35187       &  0.026      &  3.10    &  2.5 $\times 10^{-3}$ \\%-2.597
HD 36112       &  0.099      &  2.58    &  0.24 $\times 10^{-3}$ \\% -3.624  
CQ Tau            &  0.024      &  3.30    &  4.4 $\times 10^{-3}$ \\%-2.361   
HD 135344 B  &  0.080      &  2.78   & 2.2 $\times 10^{-3}$ \\%-2.658  
HD 139614     &  0.054      &  2.92    & 2.2 $\times 10^{-3}$ \\%-2.660   
HD 142527     &  0.160      &  2.52    &  0.66 $\times 10^{-3}$ \\%-3.180
HD 142666    &   0.052      &  2.57    &  0.62 $\times 10^{-3}$ \\%-3.21   
HD 163296     &  0.053      &  3.59    &  25.4 $\times 10^{-3}$ \\%-1.595
HD 169142    &  0.164       &  2.75    &  5.3 $\times 10^{-3}$ \\%-2.274
49 Cet             &   0.015      &  2.84    & $<$ 3.3 $\times 10^{-3}$ \\%-2.484 
\hline
\end{tabular}
\label{t_mgas}
\end{center}
\end{table}

\subsection{Hydroxyl}

Although less abundant than \hh \ and CO, hydroxyl (OH) is also an important molecule, as it 
plays a central role in the formation/destruction of \water, \hh \ and [O\,{\sc i}]. 
Mandell et al. (\cite{mandell2008}) were the first to detect ro-vibrational transitions of warm OH 
(at 3.0-3.7 \mic) in two HAEBEs, AB Aur and HD 36112. They derived a rotational 
temperature of 650-800 K, and argue that fluorescent excitation is responsible for the emission 
of OH located in the disc surface layer. Fedele et al. (\cite{fedele2011}) also searched for OH 
in 11 HAEBEs with CRIRES, and detected it in 4 sources with spectral types between B5 and A1; 
none of those objects are in our sample. They find that objects with an OH detection tend to be Meeus 
group I sources.  Recently, several transitions of OH around 3 \mic \ were detected in HD 100546 
(Liskowsky et al. \cite{liskowsky2012}). In our PACS spectra, we only detect OH in HD 163296 (and a 
tentative detection in AB Aur) - but we only cover one doublet, which is not the strongest in the far-IR,
so it might be we are not sensitive enough. Indeed, Sturm et al. (\cite{sturm2010}) detected several 
OH lines in their full SED range mode PACS spectra of HD 100546, from which the 84 \mic \ doublet
is the strongest. In our spectrum at 79 \mic, we do see a indications for the OH doublet, but it is
not a 3$\sigma$ detection (see Fig.~\ref{f_water}).

\subsection{Water}

Water was not yet reported to be detected in a Herbig Ae/Be disc despite several searches in the 
near- and mid-IR (e.g. Pontoppidan et al. \cite{pontoppidan2010}, Fedele et al. \cite{fedele2011}). 
However, in the study by Pontoppidan et al. (\cite{pontoppidan2010}), HD163296 shows a \water \
emission line at 29.85 \mic, but in that paper, water was only confirmed to be present when detected 
both at 15.17 and 17.22 \mic, with at least 3.5$\sigma$ confidence.

Thi \& Bik (\cite{thi2005}) showed that the ratio \water/OH declines when the ratio of the intensity of 
the UV field over the density increases. Thus in lower density regions with a lot of UV radiation,
the amount of water expected is low, compared to OH. Indeed, Fedele et al. (\cite{fedele2011}) conclude 
that, if water vapour is present, it must be located in deeper, colder layers of the disc than where OH is 
found; the disc atmosphere is depleted in water molecules.

We detect at least one water line and have evidence for several others in HD 163296, a group II
source. Tilling et al. (\cite{tilling2012}) modelled the disc of HD 163296 based on our earlier, 
shallower, range scans, and showed that the disc is mostly settled, which results in slightly warmer 
dust and increased line flux. This fact, together with the rather high UV luminosity, can probably 
explain the water detections in this disc. 

Kamp et al. (\cite{kamp2011}) found that strong dust settling will increase the water abundance
in the disc surface. The reason is complex (we refer the reader to Sect.~5.3 of Kamp et al.),
but the main idea is that there is an efficient cold water formation route in these discs. 
It is interesting to note that HD 100546, the source that has the highest UV flux and is richest in 
other, strong lines (\Odrie, CO and CH$^+$) does not show evidence of warm water. HD 100546 is a 
group I source, thought to have its inner disc cleared (e.g. Bouwman et al. \cite{bouwman2003}, 
Benisty et al. \cite{benisty2010}). Woitke et al. (\cite{woitke2009b}) showed that water lines 
originate in 3 distinct regions: 1) a deep midplane behind the inner disc wall, up to 10 AU, hosting
most of the water vapour; 2) a midplane region between 20 and 150 AU where water freezes out and
there is a small amount of cold water vapour; and 3) a warm water layer between 1 and 50 AU
higher up in the disc. In HD 100546, region 1 and part of region 3 are missing, so that the amount
of gas phase water is much smaller. 

Furthermore, while in the inner disc of HD 163296 the density is too low for water to form (from OH + \hh) 
to balance the fast photodissociation (see e.g. Thi \& Bik \cite{thi2005}), water can survive in the inner 
10 AU of the warm atmosphere. In contrast, in the UV strong star HD 100546, even at 30 AU the UV field 
is too strong for water to survive. This, in combination with the greater amount of settling in the 
HD 163296 disc, might explain the absence of detectable \water \ emission in the disc of HD 100546 
while it is detected in HD 163296. 

\subsection{CH$^+$}

The formation of CH$^+$ is controlled by the gas-phase reaction C$^+$ + \hh \ $\rightarrow$ 
CH$^+$ + H, which has an activation energy of 4500 K. Therefore, CH$^+$ not only 
traces the presence of \hh \ but also the presence of hot gas. In our sample, we have IR 
detections of molecular hydrogen in 3 targets: AB Aur, HD 97048 and HD 100546. In two of 
these objects, we also detect CH$^+$, suggesting that their formation and excitation 
mechanisms are indeed related. For a more in-depth discussion of CH$^+$ in HD 100546
we refer to Thi et al. (\cite{thi2011}).

%++++++++++++++++++++++++++++++++++++++++++++++++++++++++++++++++++++++
%++++++++++++++++++++++++++++++++++++++++++++++++++++++++++++++++++++++
\section{Correlations with stellar and disc parameters}
\label{s_ana}

In the next paragraphs, we will look for correlations between the observed \Odrie \ line
fluxes and the properties of the objects. For this purpose, we did not include the debris
discs, as their discs have a very different nature. Besides, in the debris discs, we did not 
detect any \Odrie \ line, so that we would only be able to compare upper limits. 

\subsection{The influence of \teff, UV and X-ray luminosity}

\begin{figure}[t!]
  \resizebox{\hsize}{!}{\includegraphics{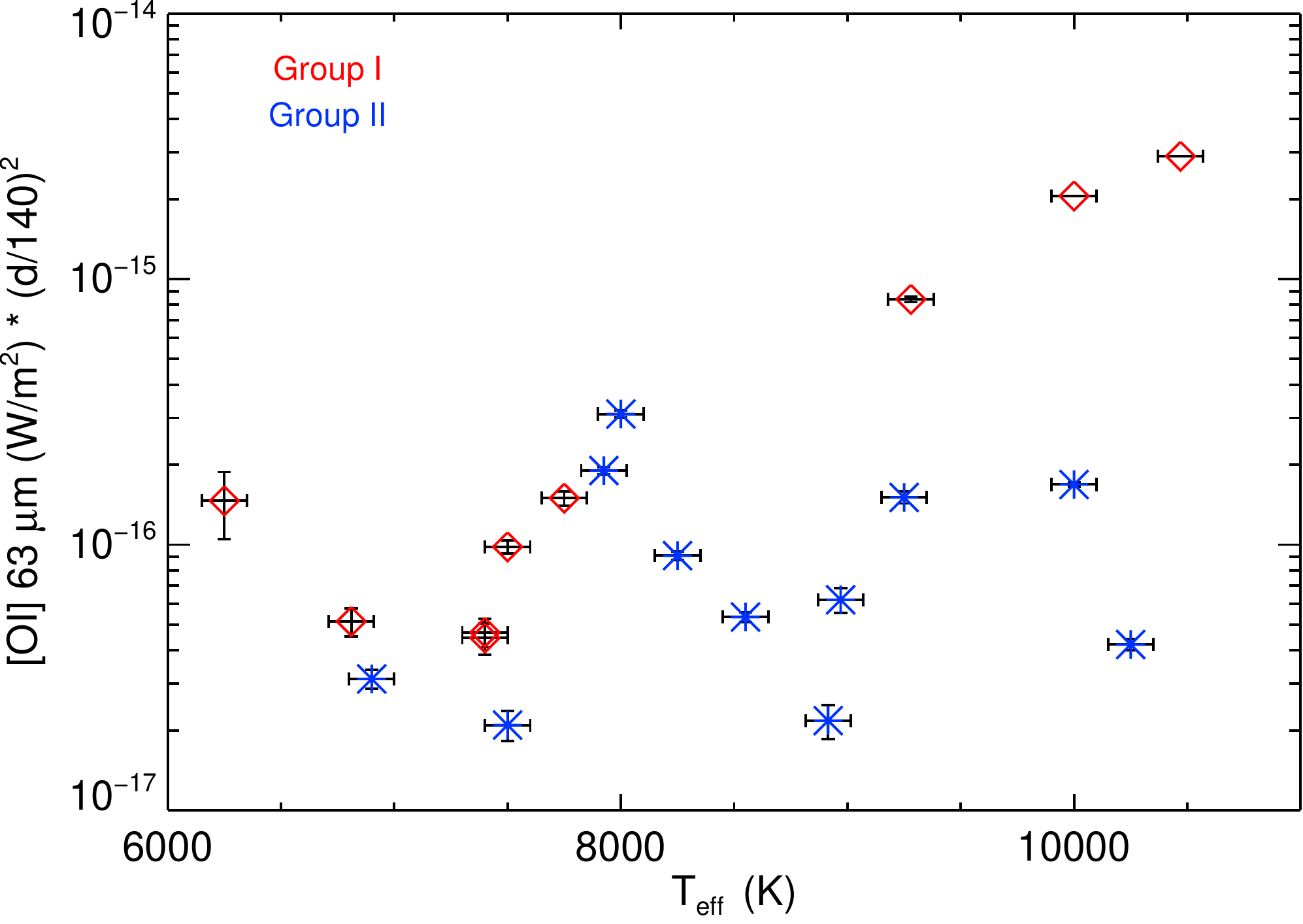}} 
  \resizebox{\hsize}{!}{\includegraphics{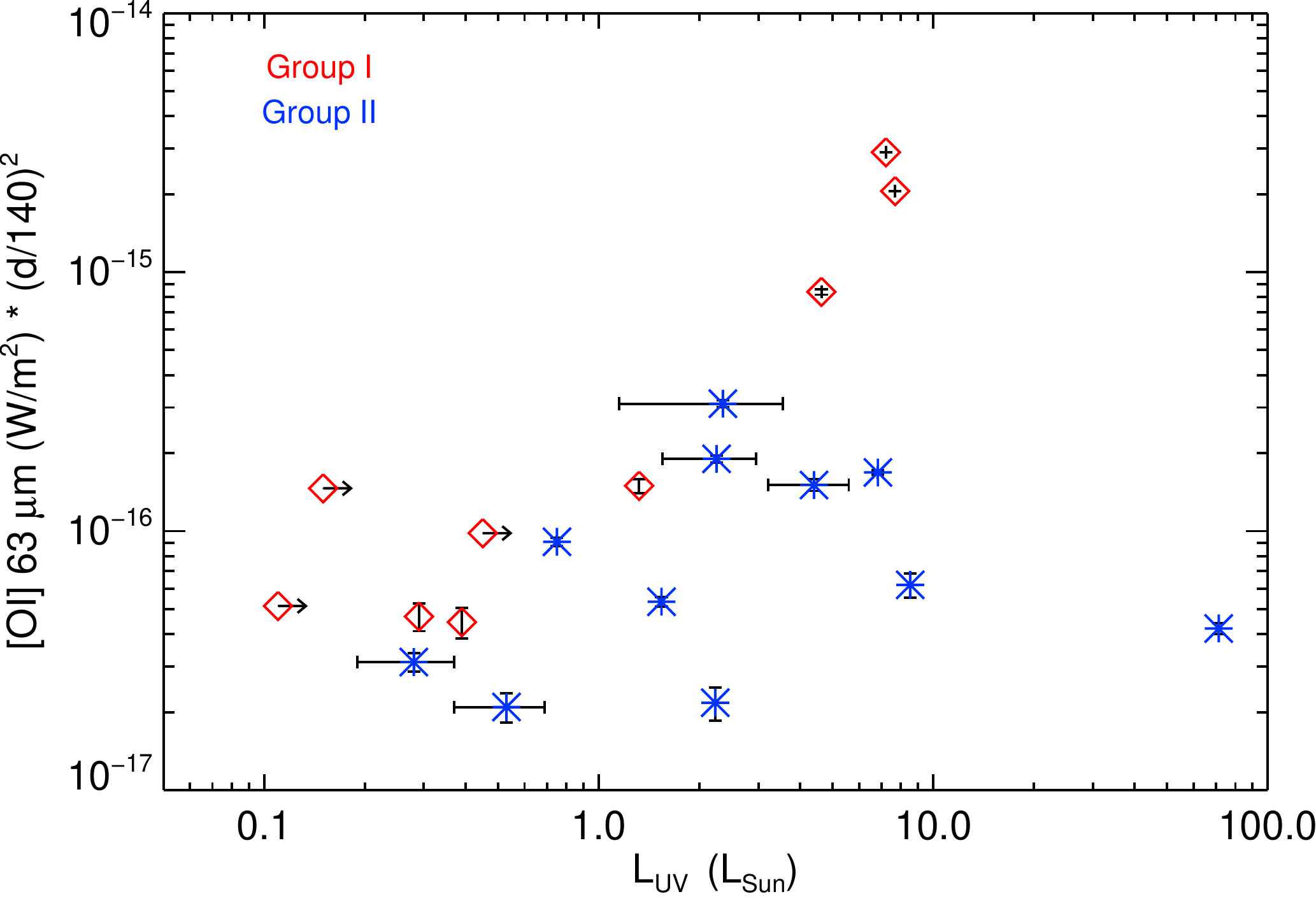}} 
  \resizebox{\hsize}{!}{\includegraphics{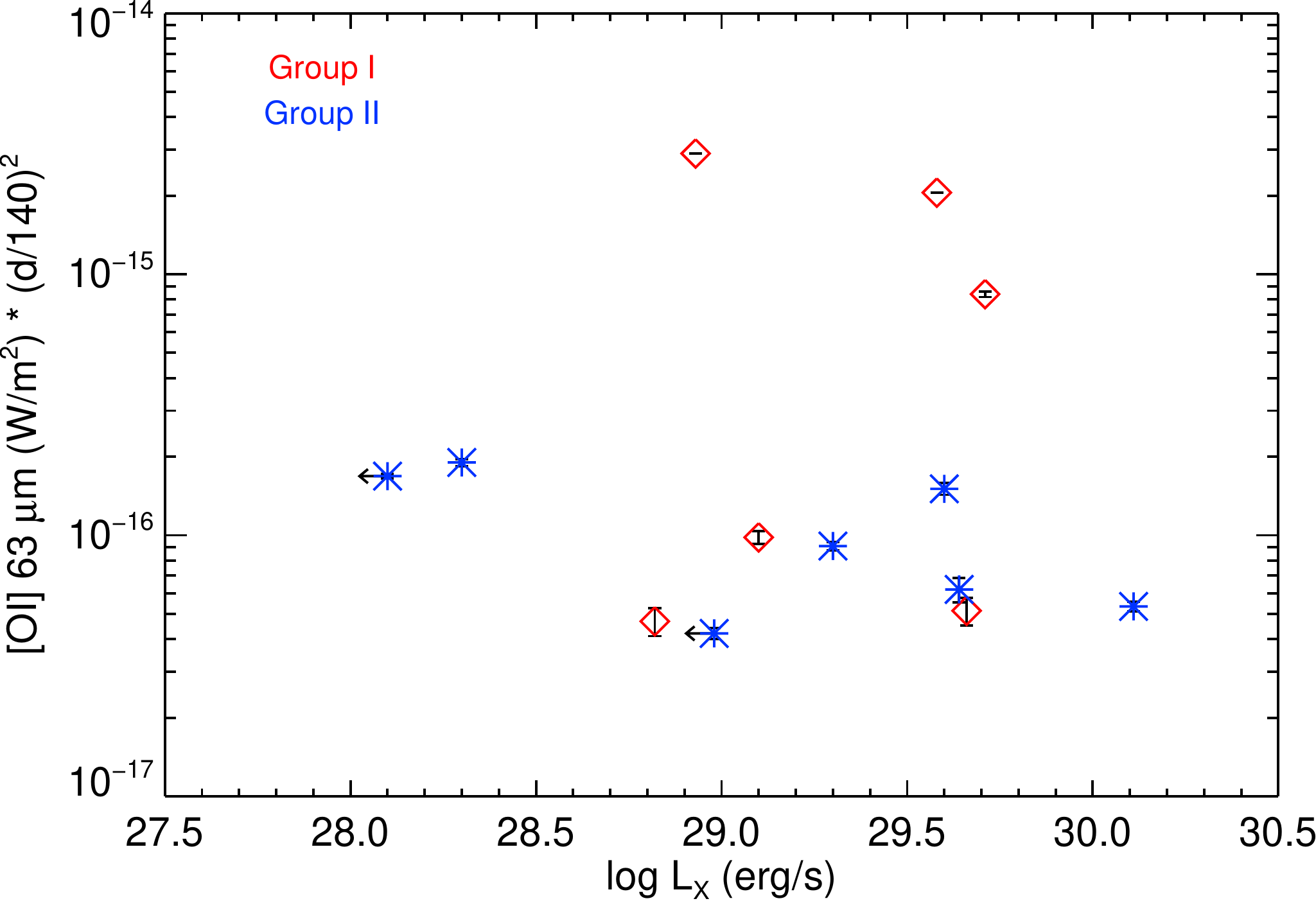}} 
\caption{\Odrie \ \mic \ versus effective temperature, UV luminosity and X-ray 
luminosity. Diamonds: group I sources, asterix: group II sources.}
\label{f_starpara}
\end{figure}

We searched for a correlation between stellar parameters and the \Odrie \
line flux. We did not find a trend with age nor with stellar luminosity. In Fig.~\ref{f_starpara}, 
we show the relation between the line flux of \Odrie \ and the effective temperature 
of the stars. Both appear uncorrelated, until \teff \ reaches 10000 K, when the \Odrie \ 
flux increases dramatically for a few sources. Our $p$-values (see Table~\ref{t_stats}) are 
inconclusive, therefore, the significance of a possible correlation cannot be established 
from our statistical analysis.

The UV and X-ray photons play an important role in the chemistry and temperature 
balance of protoplanetary discs. For HAEBEs, the UV photons are important in heating 
the disc, through their absorption by PAHs and the subsequent photoelectric effect. 
In Fig.\,\ref{f_starpara}, we show the relation between the line flux of  \Odrie \,\mic, the UV 
luminosity calculated from IUE spectra (see Sect.\,\ref{s_targets}), and the X-ray luminosity 
(data mainly found in Hubrig et al. \cite{hubrig2009}, see Table~\ref{t_xray} for a full list). 
There is a clear correlation between the \Odrie \ flux and the UV luminosity (see 
Table~\ref{t_stats}), as reported earlier for a limited sample in Pinte et al. (\cite{pinte2010}). 
This could be related to an increase in OH photo-dissocation in the disc surface and/or to a 
more efficient photoelectric heating of the gas by PAHs in those sources with a higher UV 
luminosity. 

X-ray photons can ionise atoms and molecules. The X-ray fluxes observed in HAEBEs ($\log$ 
L$_\mathrm{X}$ = 28-30) are on average lower when compared to those of the lower-mass 
T Tauri stars (TTS; $\log$ L$_\mathrm{X}$ = 29-32). Aresu et al. (\cite{aresu2011}) found from
theoretical modeling of discs a correlation between the X-ray luminosity and the \Odrie \ line
flux for X-ray luminosities above 10$^{30}$ erg/s. Below that value, the gas temperature in the
region where the \Odrie \ line forms is dominated by UV heating while above that value, X-rays
provide an additional heating source, thereby increasing the total line flux. Since all objects in
our sample are below this L$_\mathrm{X}$ threshold, it is not surprising that we do not see a 
correlation with the X-ray luminosity. Furthermore, X-rays in HAEBEs are softer than in TTS, so 
that they cannot penetrate as deep in the discs as in T Tauri discs. A dedicated study will use spectral 
X-ray properties to interpret the observed PACS spectra (G\"udel et al., private communication). 

%++++++++++++++++++++++++++++++++++++++++++++++++++++++++++++++++++++++
\subsection{Relation with accretion rate}

\begin{figure}[t!]
   \resizebox{\hsize}{!}{\includegraphics{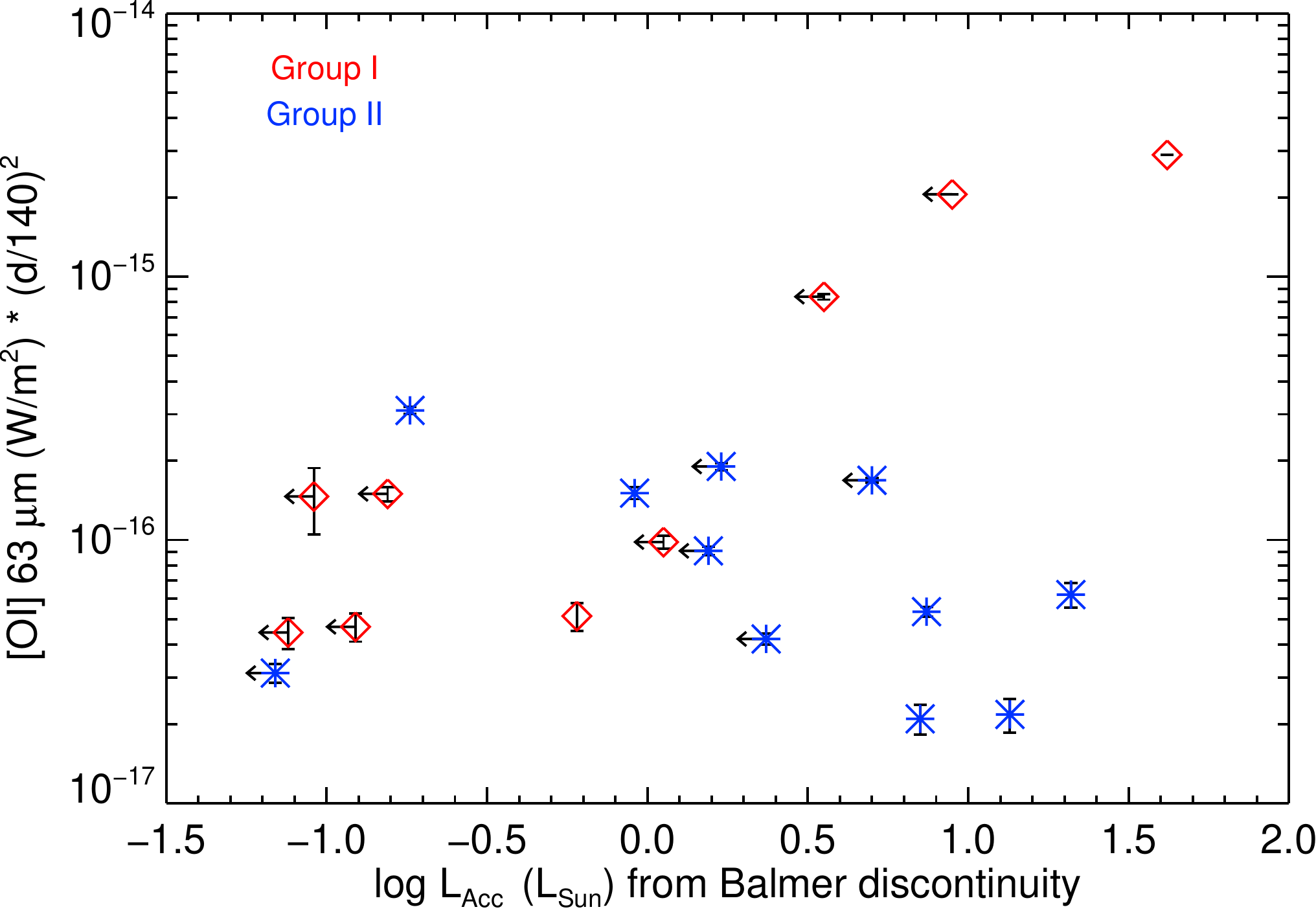}} 
   \resizebox{\hsize}{!}{\includegraphics{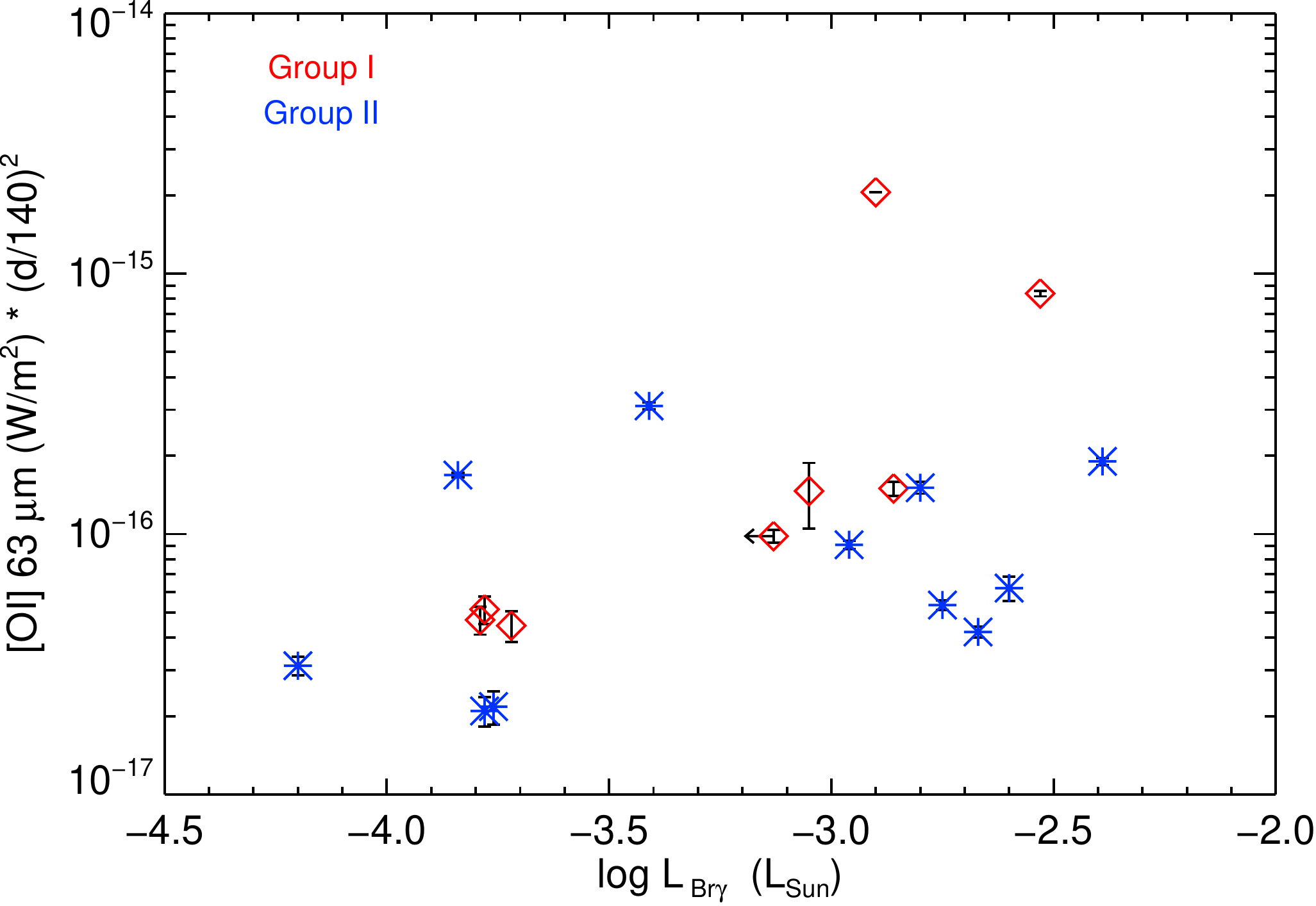}} 
   \resizebox{\hsize}{!}{\includegraphics{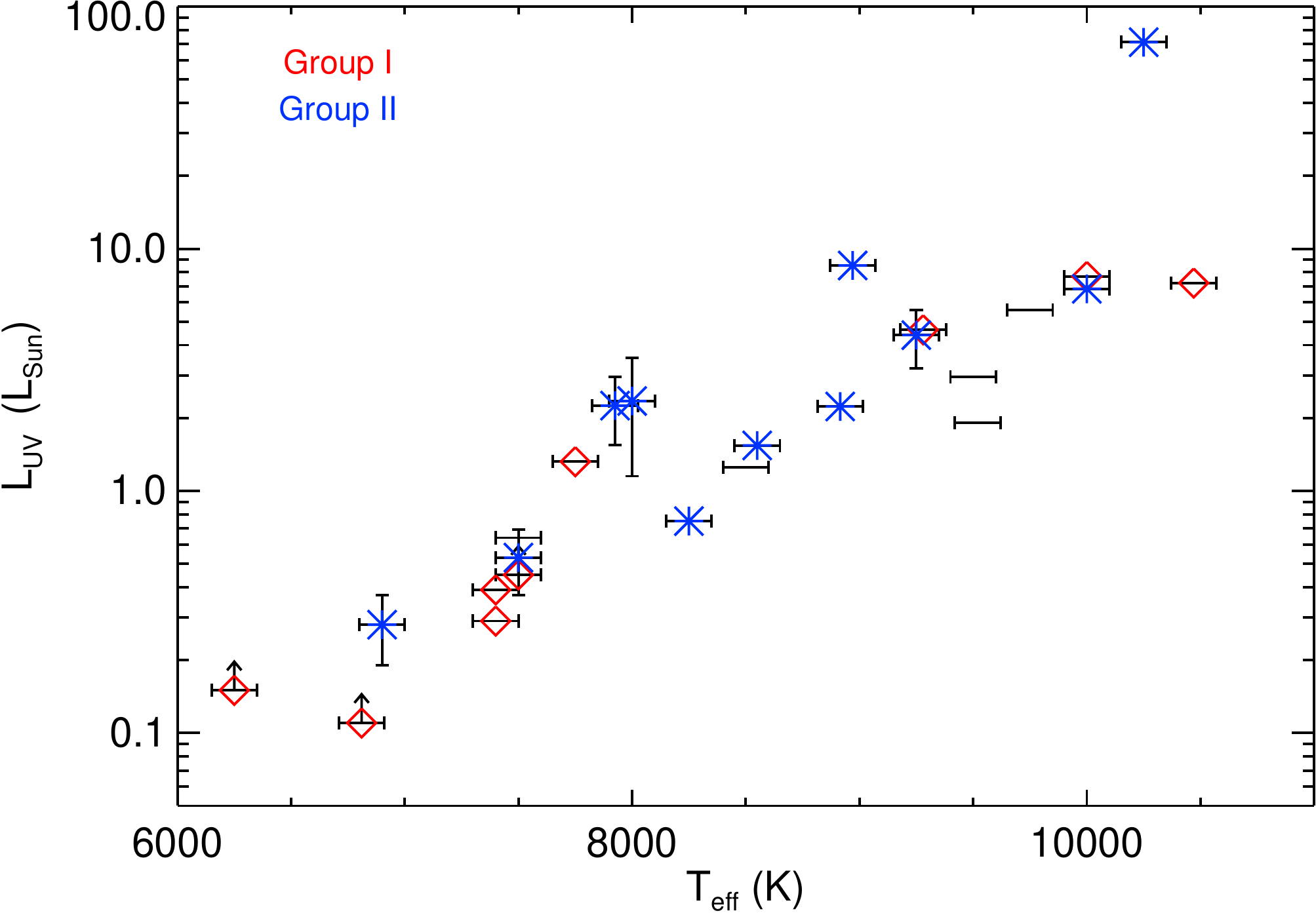}} 
\caption{Top: \Odrie \ \mic \ versus the accretion rate derived from the excess in the Balmer discontinuity.
Middle: \Odrie \ versus the luminosity of the Br$_\gamma$ line. Bottom: L$_{\rm{UV}}$ vs. \teff.
Diamonds: group I sources, asterix: group II sources.}
\label{f_OIacc}
\end{figure}

In Fig.~\ref{f_OIacc}, we plot the line flux of \Odrie \, \mic \, as a function of  L$_{\mathrm{acc}}$ 
derived from the excess in the Balmer discontinuity. There is no trend visible, just more scattering 
at higher L$_{\mathrm{acc}}$. However, we should point to the difficulties that  lie in an accurate 
determination of the accretion rate, that will introduce additional scatter in the values. We also plot the 
\Odrie \,\ line flux against the Br$\gamma$ luminosity (data from Garc\'ia-Lopez et al.\ \cite{garcia2006} 
and Donehew et al.\ \cite{donehew2011}). Here we see a tentative correlation with 
L$_{\mathrm{Br\gamma}}$.
Given that we do not see a clear correlation between the accretion rate and the \Odrie \ line flux, we 
can conclude that the accretion is not an important contributor to the excitation of \Odrie \  in HAEBEs. 
This confirms the findings of an earlier study of a few HAEBEs by Pinte et al. (\cite{pinte2010}), that 
the emission from HAEBE discs can be explained by photospheric heating alone. Indeed, in 
Fig.~\ref{f_OIacc}, bottom, we plot L$_{\rm{UV}}$ vs. \teff, and see that both are very well correlated. 
This means that the bulk of the UV luminosity in HAEBEs is photospheric, rather than originating in 
accretion, as is often observed in the cooler T Tauri stars (e.g. Yang et al. \cite{yang2012}).

%++++++++++++++++++++++++++++++++++++++++++++++++++++++++++++++++++++++
\subsection{Relation with PAH and [OI]6300 \AA}

\begin{figure}[t!]
  \resizebox{\hsize}{!}{\includegraphics{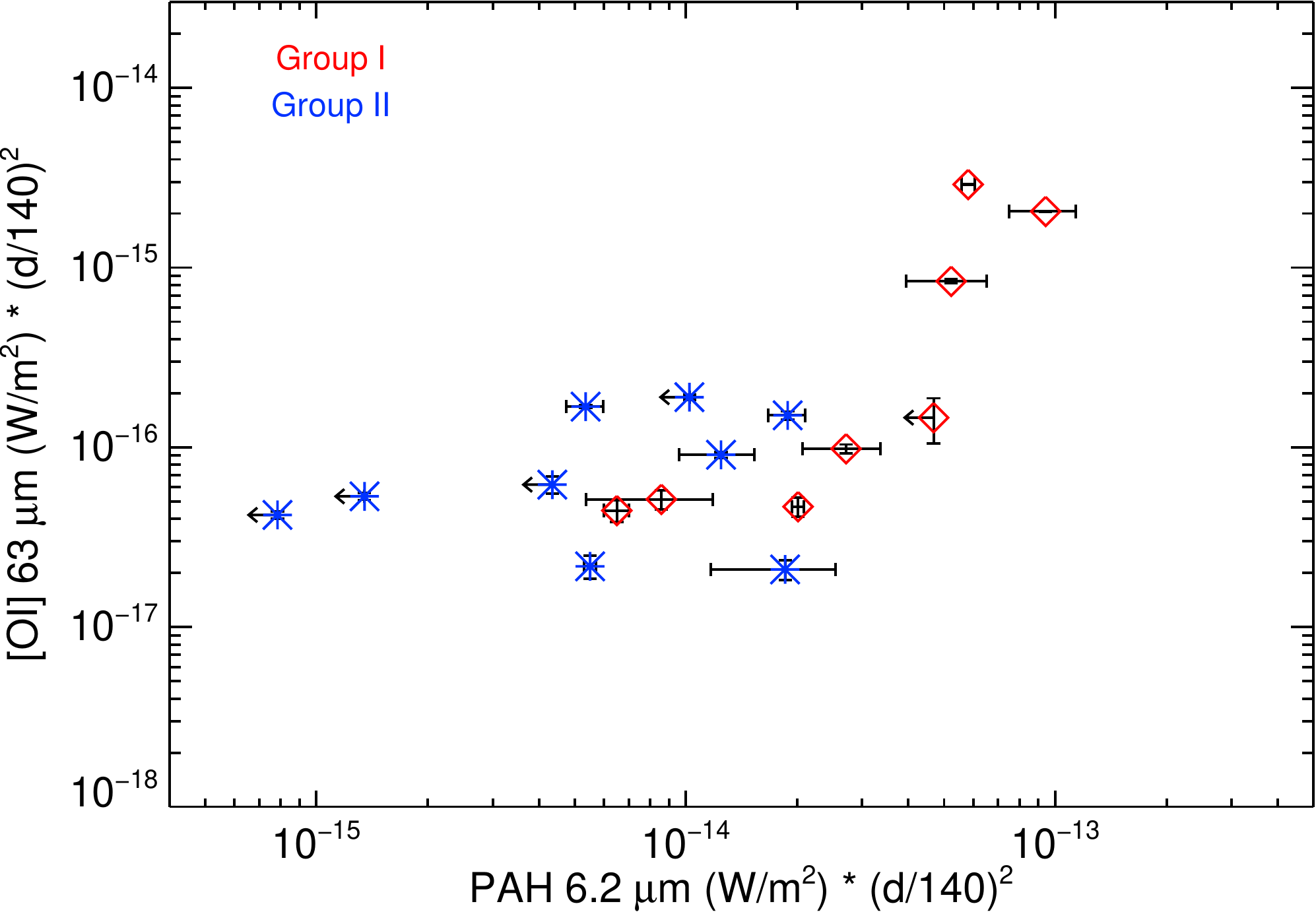}} 
  \resizebox{\hsize}{!}{\includegraphics{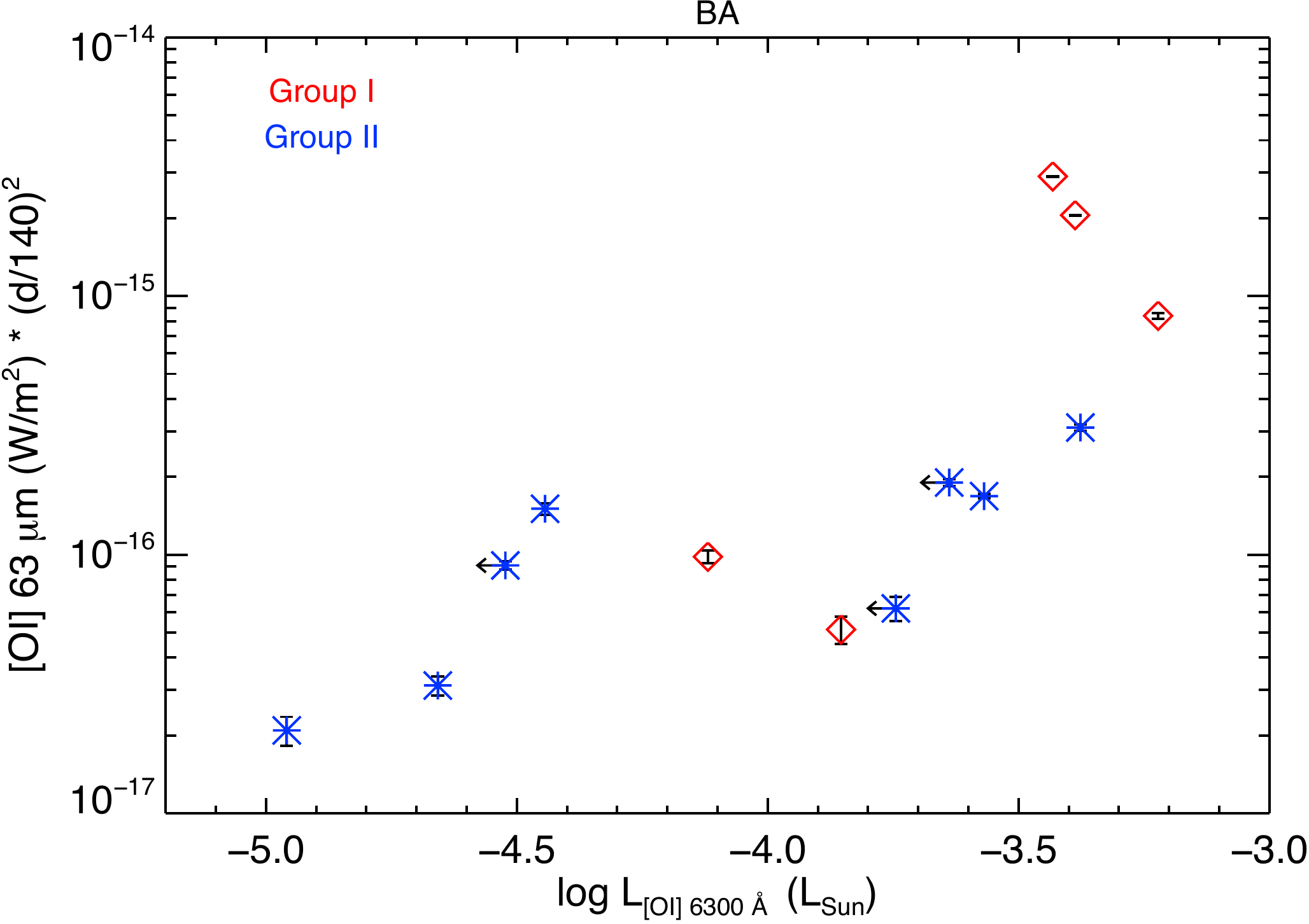}} 
  \caption{Top: \Odrie \ \mic \ line flux versus PAH 6.2 \mic \ line flux.  Bottom: \Odrie \ versus the
  luminosity of  \Oopt. Diamonds: group I sources, asterix: group II sources.}
\label{f_OIpah}
\end{figure}

PAHs are important for the heating of the disc through the photo-electric effect. In HAEBEs, 
the PAH luminosity, L$_\mathrm{PAH}$/L$_*$  is observed to reach up to 9 $\times 10^{-3}$. 
In Fig.~\ref{f_OIpah}, we show the relation between the PAH luminosity and the \Odrie \ 
line fluxes (data from Acke et al. \cite{acke2004b} and Keller et al. \cite{keller2008}). The 
\Odrie \ flux and the PAH flux weakly correlate with each other  (see Table~\ref{t_stats}). 

HD 141569A is the only star for which we saw a firm detection of \Ohon \,\mic, but no detection 
of CO at 144.8 \mic. The \Odrie \,\ over \Ohon \,\ ratio is smaller than 10, while it is around 20 in 
AB Aur, HD 97048 and HD 100546. Moreover, Brittain et al. (\cite{brittain2007}) showed from 
CO 4.7\,\mic \ modelling that \trot \,\ is of the order of 250 K, while \trot \ in HD 97048 and 
HD 100546 is much higher, $\geq$ 1000 K. This difference cannot be attributed to a lower UV 
luminosity as it is rather similar (6.83 in HD 141569A vs. 7.69 in HD 97048 and 7.22 \Lsun \ in 
HD 100546). However, a lower PAH luminosity is observed in HD 141569A (a factor 10 less; 
Acke et al. \cite{acke2010}), so there is less heat contribution to the disc - there might also be an 
intrinsic difference in PAH abundance. 

The excitation mechanism of \Oopt \  is still not well determined, it could be either thermal
or non-thermal (e.g. fluorescence or OH photodissociation). The emission line is 
commonly found in HAEBEs (e.g. Corcoran \& Ray  \cite{corcoran1998}). Acke et al. 
(\cite{acke2005}) derived from a simple model of the spectroscopically resolved \Oopt \ line 
that the emission originates from the disc surface, assuming that it is non-thermally excited by 
the UV photo-dissociation of OH molecules in the disc surface. It is more often found in flared 
than in self-shadowed discs (Acke et al. \cite{acke2005}) and traces the disc between 0.1 and 
50 AU (van der Plas et al. \cite{plas2009}). 
In the bottom plot of Fig.~\ref{f_OIpah}, we plot \Odrie \ against \Oopt \ (data from Acke et al. 
\cite{acke2005}, van der Plas et al. \cite{plas2008} and Mendigut\'ia et al. \cite{mendi2011a}). Here 
we see a correlation that could be explained by the excitation mechanism for both lines 
being related to UV photons.  We already showed that there is no obvious correlation between the 
accretion rate and the \Odrie \ line strength. The special properties of the 'bright three', which have a 
high UV luminosity as witnessed by the high line flux of \Odrie \ and PAH luminosity, thus must 
mainly be because of their high effective temperature, when compared to the rest of the sample.
The only other objects in our sample with a temperature around 10000 K, are HD 141569A (a 
transitional disc) and 51 Oph, a special case with a sharp drop in the SED at longer ($>$ 20 micron) 
wavelengths, which can be attributed to a compact disc in which the dust has settled towards the 
midplane. 

%++++++++++++++++++++++++++++++++++++++++++++++++++++++++++++++++++++++
\subsection{Relation with disc properties} 

\begin{figure}[t!]
  \resizebox{\hsize}{!}{\includegraphics{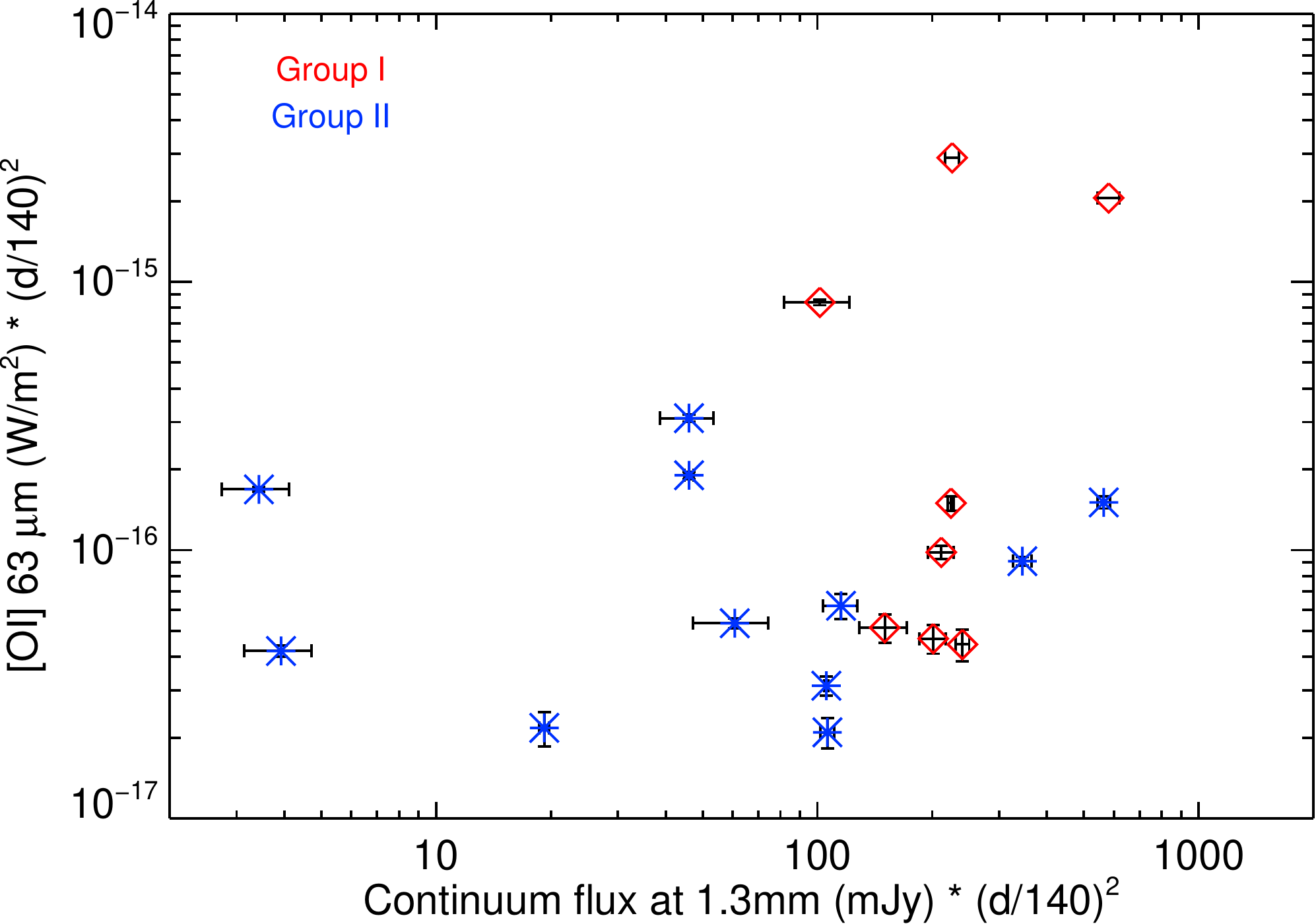}} 
  \resizebox{\hsize}{!}{\includegraphics{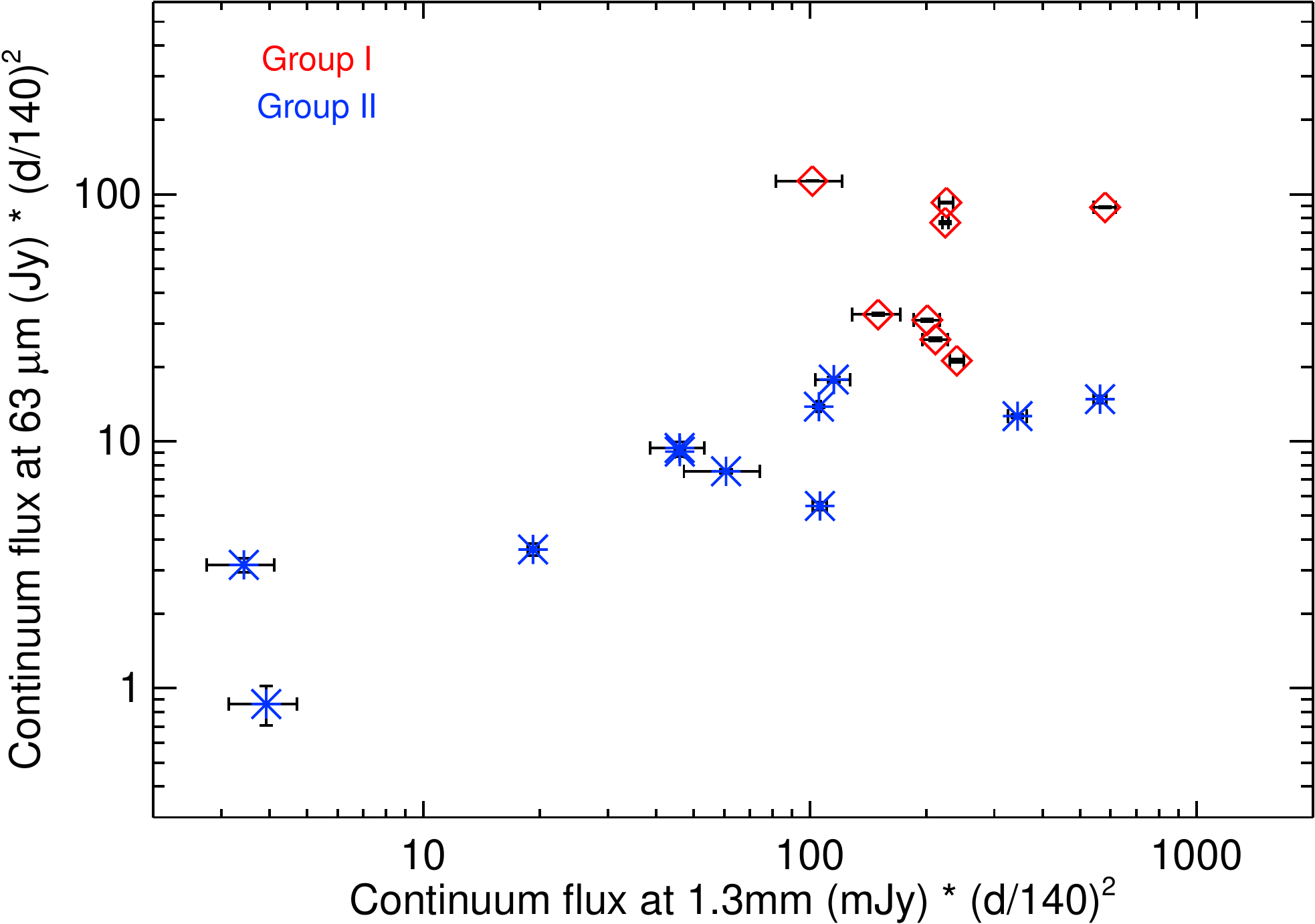}} 
\caption{Top: \Odrie \ versus the continuum flux at 1.3~mm, there is a weak trend of stronger 
line flux with higher continuum flux. Bottom: The continuum flux at 63~\mic \ versus the 
continuum flux at 1.3~mm, where we see a clear correlation. Diamonds: group I sources, asterix: 
group II sources.}
\label{f_mm}
\end{figure}

\begin{figure}[t!]
  \resizebox{\hsize}{!}{\includegraphics{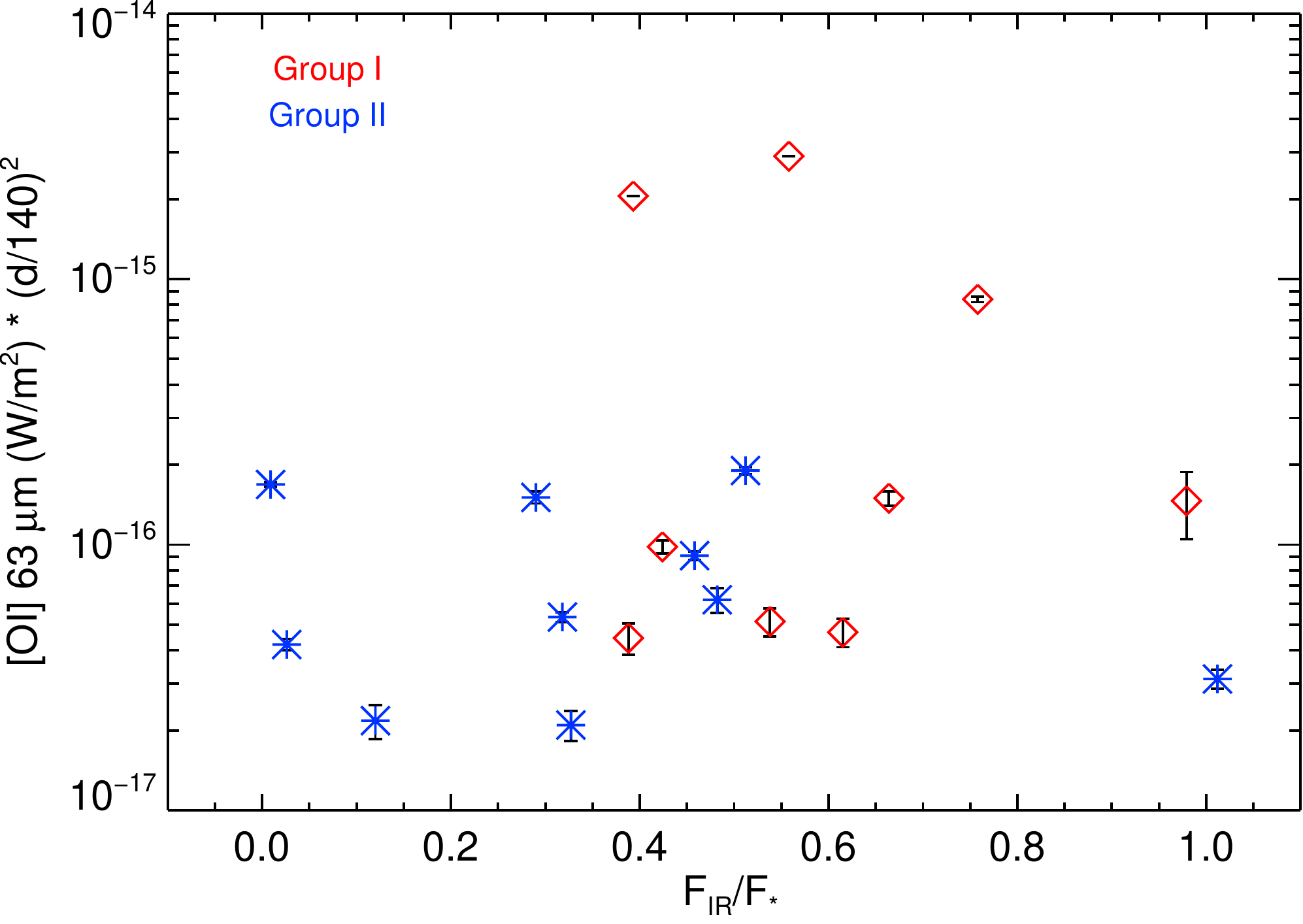}} 
  \resizebox{\hsize}{!}{\includegraphics{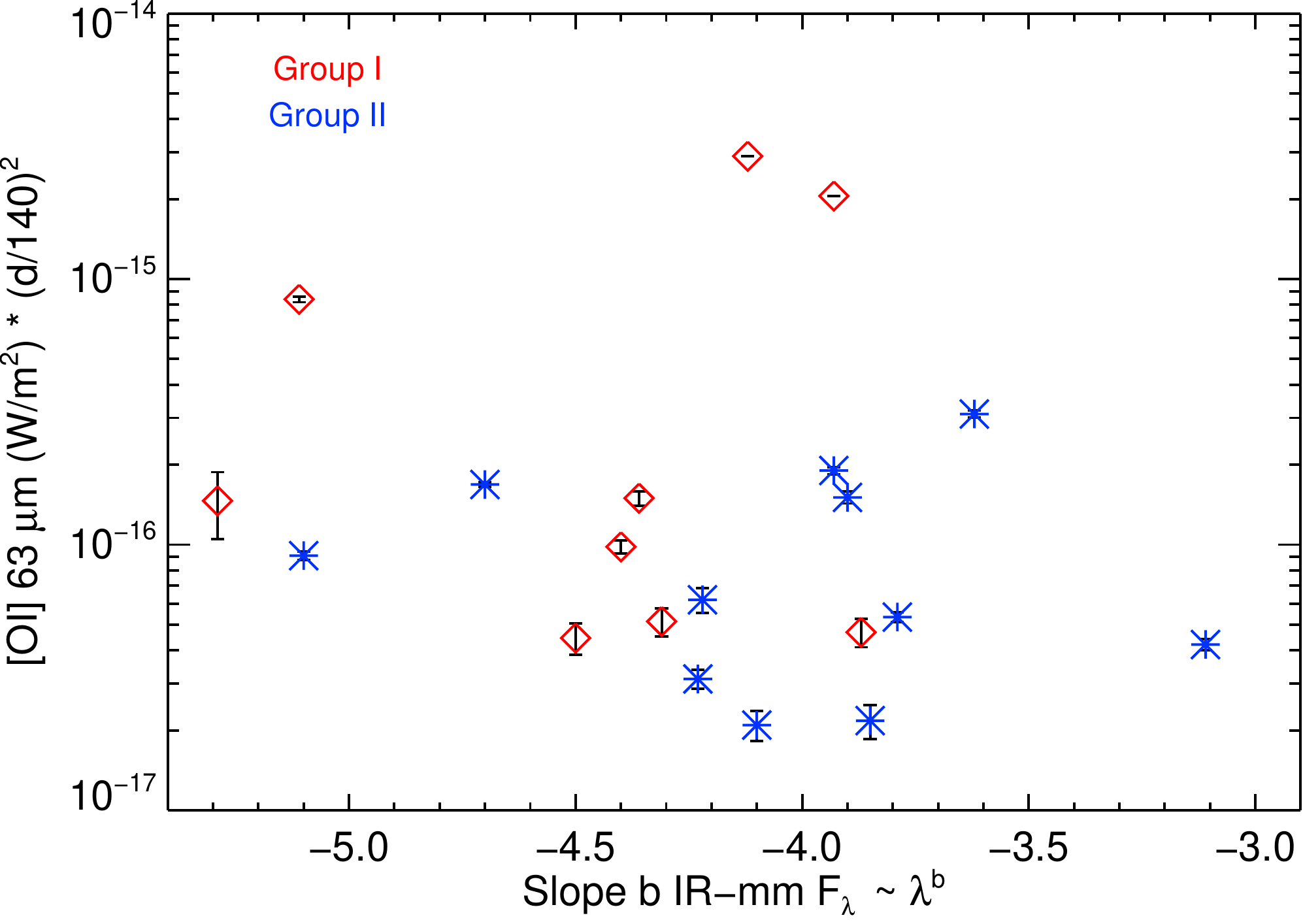}} 
\caption{Top: \Odrie \  versus  amount of IR excess. We do not find a correlation between these 
variables. The 2 objects to the left are HD 141569A and 51 Oph, a transitional and a compact
disc, respectively. Bottom:  \Odrie \ versus the slope b of the SED at far-IR to mm 
wavelengths. The \Odrie \ flux also does not correlate with the SED slope. Diamonds: group I 
sources, asterix: group II sources.}
\label{f_IR}
\end{figure}

The continuum flux at 1.3\,mm is often used to derive a minimum dust mass of the disc, under the 
assumption that the dust emission is optically thin (Beckwith et al. \cite{beckwith1990}).  In 
Fig.~\ref{f_mm}, we plot the \Odrie \ flux as a function of the continuum flux at 1.3~mm (our mm 
data collected in the literature are listed in  Appendix~\ref{t_mm13}; we also use our own 
unpublished SMA data, see Appendix~\ref{a_sma}). We did not find that these variables are
correlated (see Table~\ref{t_stats}). On the other hand, the continuum flux at 63 \mic \ and at 
1.3 mm are strongly correlated (see Fig.~\ref{f_mm}).

In Fig.\,\ref{f_IR}, we plot the line flux of \Odrie \, \mic \, as a function of the total IR excess, a proxy 
for the amount of dust continuum observed. We do not find a correlation. This is likely because the 
IR continuum is rather a tracer of the dust disc, and the scale height of the gas disc can be higher 
than that of the dust, as is already observed in a few HAEBEs (e.g. van der Plas et al. \cite{plas2009}). 
HD 141569A has a transitional disc, with a much lower \irratio \ than the rest of the sample (see 
Table~\ref{t_excess}), indicating that the disc is barely flaring and perhaps has already dissipated 
much of its disc material. All these diagnostics lead to confirm that the disc of HD 141569A is much 
different (inner disc mostly cleared out from dust) from that of HD 97048 and HD 100546, which still 
have flaring gas-rich discs. Modelling of the HD 141569A disc with \prodimo \ will help us to better 
understand these differences in terms of excitation mechanisms, abundance of gas and disc 
structure (Thi, {\em in preparation}). Our data suggest that, in HD 141569A, the CO is located deeper 
in the disc (closer to the dust) where it can be thermalised and/or shielded from photodissociation by 
direct UV photons. 

The slope b of the far-IR to mm SED, where F$_{\lambda} \sim \lambda^{\mathrm{b}}$, can be 
related to the size of the dust grains radiating at mm wavelengths. However, grain size is not the 
only factor influencing the slope, also the composition (e.g. amount of carbon) and grain shape can 
be important factors. Acke et al. (\cite{acke2004a}) showed that the SED far-IR to mm slope is related 
to the SED group: self-shadowed discs (group II) have on average shallower slopes than their flaring 
counterparts (group I). In our sample, we do not see a correlation between the \Odrie \ line strength 
and the SED slope (see Fig.~\ref{f_mm}).
 
{\subsection{Non-detections in debris discs}

In our sample we have five debris discs, for which the \Odrie \ line was not detected. We obtain 3$\sigma$ 
upper limits for the  \Odrie \ line flux (at their respective distances) $\sim$ 6-10 $\times 10^{-18}$ W/m$^2$. 
This contrasts with the young debris disc  $\beta$ Pic, where \Odrie \ \mic \ and \CII \ 158 \mic \  emission lines 
were detected with {\it Herschel}/PACS (Brandeker et al. \cite{brandeker2012}). These authors give an \Odrie \ 
\mic \ line flux = 13.2  $\times 10^{-18}$ W/m$^2$ for $\beta$ Pic (at a distance of 19 pc), what would not have 
been detected at the distance of our debris discs. The only exception is HR 1998, at a distance of 22 pc, for which 
we have an upper limit  $\sim$ 5 $\times 10^{-18}$ W/m$^2$ when scaled to 19 pc, almost a factor 3 lower than 
the $\beta$ Pic detection. $\beta$ Pic appears to be a special debris disc that is relatively rich in gas, originating 
from the ongoing vaporisation of dust through grain-grain collisions, comet evaporation, and/or photodesorption 
of grain surfaces (Lagrange et al. \cite{lagrange1998}; Czechowski \& Mann \cite{czechowski2007}; Chen et al. 
\cite{chen2007}).

%++++++++++++++++++++++++++++++++++++++++++++++++++++++++++++++++++++++
\section{Conclusions}
\label{s_conc}

In this paper, we studied the gas content with {\em Herschel} PACS spectroscopy for a 
sample of 20 Herbig Ae stars and five A-type debris discs, that can be summarised as follows:

\begin{enumerate}
\item{We detect the \Odrie \ \mic \ line in all the Herbig Ae stars of our sample, while it is absent 
in the debris discs, confirming the lack of a large amount of gas in these discs. The \Odrie \ line 
is by far the strongest line observed in our spectra, next in strength (if detected) are \Ohon \ 
and \CII; they are only detected in 5 (25\%) and 6 (30\%) sources, respectively. }

\item{The CO mid to high $J$ transitions (18-17 and 29-28) are only detected in 9 (45\%) and 
2  (10\%) objects, respectively. %High $J$ lines are formed in the inner disc, within a few AU. 
The highest $J$ (33-32 and 36-35) CO lines covered in our spectra are not seen
at all in our sample. The three detections of CO $J$=29-28 are in the three strongest UV emitting 
objects, AB Aur, HD 97048 and HD 100546, revealing the need for a large amount of UV
photons for this line to become visible. Interesting in this respect is the transitional disc of 
HD 141569A, where we did not detect CO $J$=18-17, but did detect a strong line of \Ohon. This 
cannot be attributed to a difference in UV luminosity but rather to significant inner disc clearing,  
and to a more tenous disc.}

\item{We detect two lines of CH$^+$ in HD100546, and also detect CH$^+$ for the first time in 
HD 97048, only the second Herbig Ae star in which it is detected.}

\item{Hydroxyl and \water \ are important ingredients of the disc chemistry. However, we found 
water and OH in only one object, HD 163296, which has a settled disc. The previous detection of 
\water, announced by Sturm et al. (\cite{sturm2010}) in HD 100546 cannot be confirmed. The 
misidentification was caused by a blend with the CH$^+$ line, often present at the 
same wavelength as \water.  The non-detection of \water \ in most sources is 
in agreement with findings of Pontoppidan et al. (\cite{pontoppidan2010}) and Fedele et al. 
(\cite{fedele2011}), who also did not detect water at IR wavelengths, despite dedicated surveys.}
\end{enumerate}

We correlated the strength of the \Odrie \,\mic \,\ line with stellar parameters, as well 
as disc properties. We can summarise our findings as follows:

\begin{enumerate}
\item{The \Odrie \ line flux correlates weakly with the continuum flux at 63 \,\mic. The line flux ratios 
of  \Odrie / \Ohon \ and \Odrie / \CII \ are between 10 and 30.}

\item{We found that three of our sources, AB Aur, HD 97048 and HD 100546, have very strong \Odrie 
\ line fluxes, when compared to the rest of the sample. These three sources have group I discs and
have the highest \teff \ values in the sample, and thus have more stellar UV flux. Indeed, we see a 
correlation between the {\em total} (stellar + accretion) UV luminosity and the strength of the \Odrie \ line. 
We do not see a correlation with the X-ray luminosity, which is rather low in our sample of  HAEBE stars. }

\item{We did not find a correlation between the accretion rate estimated from the Balmer discontinuity, 
and a tentative one with the Br$\gamma$ line. This shows that accretion is not the main driver of the 
\Odrie \ excitation in HAEBEs. The bulk of the UV luminosity is photospheric rather than from accretion.}

\item{Sources with high \Odrie \ fluxes also have high PAH luminosity, which can both be related to 
their high UV fluxes. We also see a correlation with the luminosity of the \Oopt \ line. }

\item{The disc geometry (flat versus flared) does not uniquely determine the strength of the \Odrie \,\ 
line flux. The three strongest lines are observed in flared discs, but once these sources are excluded, 
there is no significant difference in line strength observed between the group I and II discs.}

\item{We found a strong correlation between the continuum at 63\,\mic \ and at 1.3\,mm. There is no 
correlation between the \Odrie \,\ line strength and the strength of the dust continuum at 1.3\,mm. We 
also did not  find a correlation with the slope of the far-IR to mm SED, nor with the IR excess. }

\item{We see a weak correlation with the strength of $^{12}$CO $J$ = 3-2 line. Based on the line 
ratio \Odrie /$^{12}$CO $J$=2-1, we can derive an estimate of the gas mass present in the disc. 
We found M$_{\rm{gas}}$ between 0.25 and 25 $\times \ 10^{-3}$ \Msun, consistent with
the estimates derived from a detailed modelling of  HD 163296 (M$_{\mathrm{gas}} \sim 15-120 \times 
10^{-3}$ \Msun; Tilling et al. \cite{tilling2012}) and HD 169142 (M$_{\mathrm{gas}} \sim 3-6.5 \times 
10^{-3}$ \Msun; Meeus et al. \cite{meeus2010}).}
\end{enumerate}

A picture emerges for the protoplanetary discs around Herbig Ae/Be stars where the stellar UV flux is 
the main parameter controlling the strength of the \Odrie \ line, which is formed just below the disc surface. 
An increased amount of settling can enhance the line flux for those species (such as water or OH) that are 
formed deeper in the disc, where the density is higher. We plan to follow-up on this study with detailed 
modelling of a few key objects: AB Aur \& HD 97048 and HD 135344 B \& HD 142527 (group I, high and 
low UV, respectively), HD163296 (group II), HD141569 A (transitional disc), and finally the enigmatic 
compact disc of 51 Oph. Our modelling results will further aid in the understanding of the chemistry and 
physical processes present in Herbig Ae/Be discs.

%++++++++++++++++++++++++++++++++++++++++++++++++++++++++++++++++++++++
\begin{acknowledgement}

We would like to thank the PACS instrument team for their dedicated support and A. Carmona for
discussions about gas line diagnostics. G. Meeus, C. Eiroa, 
I. Mendigut\'ia and B. Montesinos are partly supported by AYA-2008-01727 and AYA-2011-26202. 
G. Meeus is supported by RYC-2011-07920. CAG and SDB acknowledge NASA/JPL for funding support. 
WFT thanks CNES for financial support. FM thanks the Millennium Science Initiative (ICM) of the Chilean 
ministry of Economy (Nucleus P10-022-F). FM, IK and WFT acknowledge support from the EU FP7-2011 
under Grant Agreement No. 284405. CP acknowledges funding from the EU FP7 under contract 
PERG06-GA-2009-256513 and from ANR of France under contract ANR-2010-JCJC-0504-01.
PACS has been developed by a consortium of institutes led by MPE (Germany) and including UVIE (Austria); 
KUL, CSL, IMEC (Belgium); CEA,  OAMP (France); MPIA (Germany); IFSI, OAP/AOT, OAA/CAISMI, LENS, 
SISSA (Italy); IAC (Spain). This development has been supported by the funding agencies BMVIT (Austria), 
ESA-PRODEX  (Belgium), CEA/CNES (France), DLR (Germany), ASI (Italy), and CICT/MCT (Spain).
This research has made use of the SIMBAD database, operated at CDS , Strasbourg, France.

\end{acknowledgement}
%++++++++++++++++++++++++++++++++++++++++++++++++++++++++++++++++++++++
{}

%+++++++++++++++++++++++++++++++++++++++++++++++++++++++++++++++++++++++++++++++++
\begin{appendix}

\section{PACS observation identifications}

In Table~\ref{t_obsid}, we show the {\it Herschel} observation identification numbers (obsids) 
of our observations. Several stars were observed twice in the range mode, to obtain a deeper 
observation. These are also indicated in the table. In Figs.~\ref{f_allstars_72}, \ref{f_allstars_79} 
and \ref{f_allstars_180}, we show all the stars at 72, 79 and 180~\mic. 

\begin{table}
\begin{center}
\label{t_obsid}
\caption{Overview of the obsids that were  observed. (D) means that it
was a deeper observation than our regular settings; (D1) for settings at
79/158~\mic, (D2) for settings at 72/145~\mic \ and 79/158~\mic \ and (D3) 
for all 3 range settings. }
\begin{tabular}{lll}
\hline
\hline
Star                & Line Spec      & Range Spec \\
\hline
HD 9672         & 1342188424 & 1342188423 \\
AB Aur            & 1342191355 & 1342191354 \\ 
                         &                         & 1342226000 (D2)\\
HD 31648      &1342226002 &1342226003\\				
HD 32297      & 1342217849 & 1342217850 \\	
HD 35187      &1342226900 &1342226901 \\ %digit forsterite	
HD 36112      & 1342227635 & 1342227636\\	  
HD 36910      &1342227638 &1342227637  \\
HR 1998        & 1342226192 & 1342226193  \\
HD 97048      & 1342188436 & 1342188435 \\ %golofsson spirespec, evans sed red &blue
HD 100453    & 1342203059 & 1342203058 \\ %evans sed red &blue 	
                         & 1342212228 & 1342203442 \\  %deep 
HD 100546    & 1342188438 & 1342188437 \\ %golofsson spirespec, evans sed red &blue      
HD 104237    & 1342212234 & 1342212233  \\ %golofsson spirespec, evans sed red &blue      	 
HR 4796A      & 1342199242 & 1342199243 \\
                         &                         & 1342234488 (D1)\\
HD 135344B & 1342190370 & 1342190369 \\ %evans sed red &blue 	
HD 139614    & 1342191300 & 1342191299 \\ %evans sed red &blue 
HD 141569    & 1342190376 & 1342190375 \\ %digit forsterite
                         &                         & 1342204340 (D3) \\ %deep
HD 142527    &1342216173  & 1342216172     \\  %golofsson spirespec, evans sed red &blue
HD 142666    & 1342214224 & 1342214225 \\ %digit forsterite
HD 144668   & 1342192146  & 1342216200 \\ %evans sed red &blue 	
HD 150193   & 1342216625  &1342216626  \\
KK Oph          & 1342192148  & 1342192149  \\
                        & 	                         &  1342228205 (D2)\\
HD 158352   & 1342190377  & 1342227800 (D1)\\ %who is that??
HD 158643    & 1342217821 & 1342217822 \\
HD 163296    & 1342192161  & 1342192160 \\ %evandish HIFI
                         &                          & 1342229742 (D3) \\ %evandish HIFI
 HD 169142   & 1342186310  & 1342186309 \\ %evans sed red &blue 
                         &                          & 1342215676  (D2)\\
\hline
\end{tabular}
\end{center}
\end{table}

\begin{figure}[t!]
   \resizebox{\hsize}{!}{\includegraphics{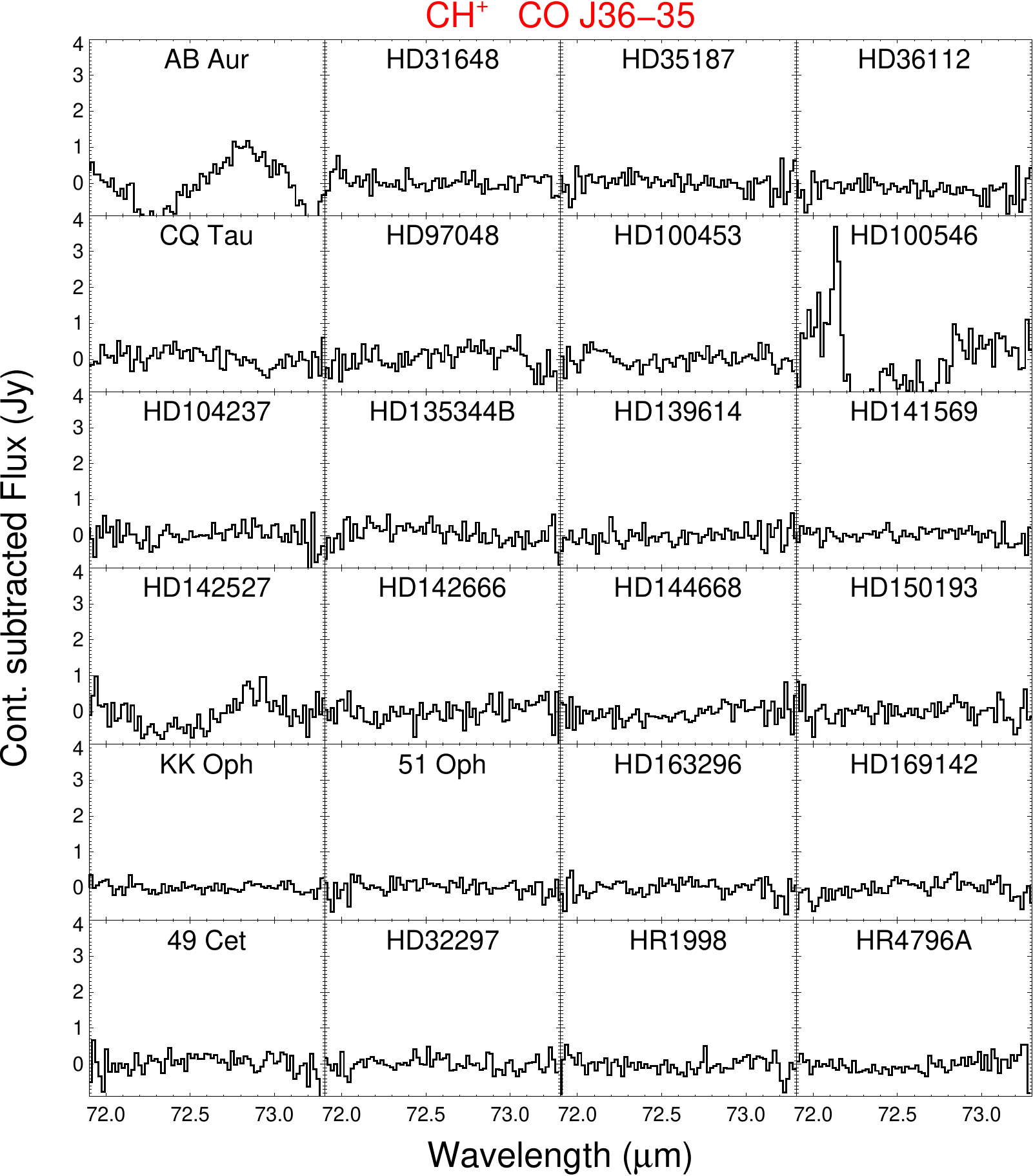}}
\caption{The region around 72 micron.}
\label{f_allstars_72}
\end{figure}

\begin{figure}[t!]
   \resizebox{\hsize}{!}{\includegraphics{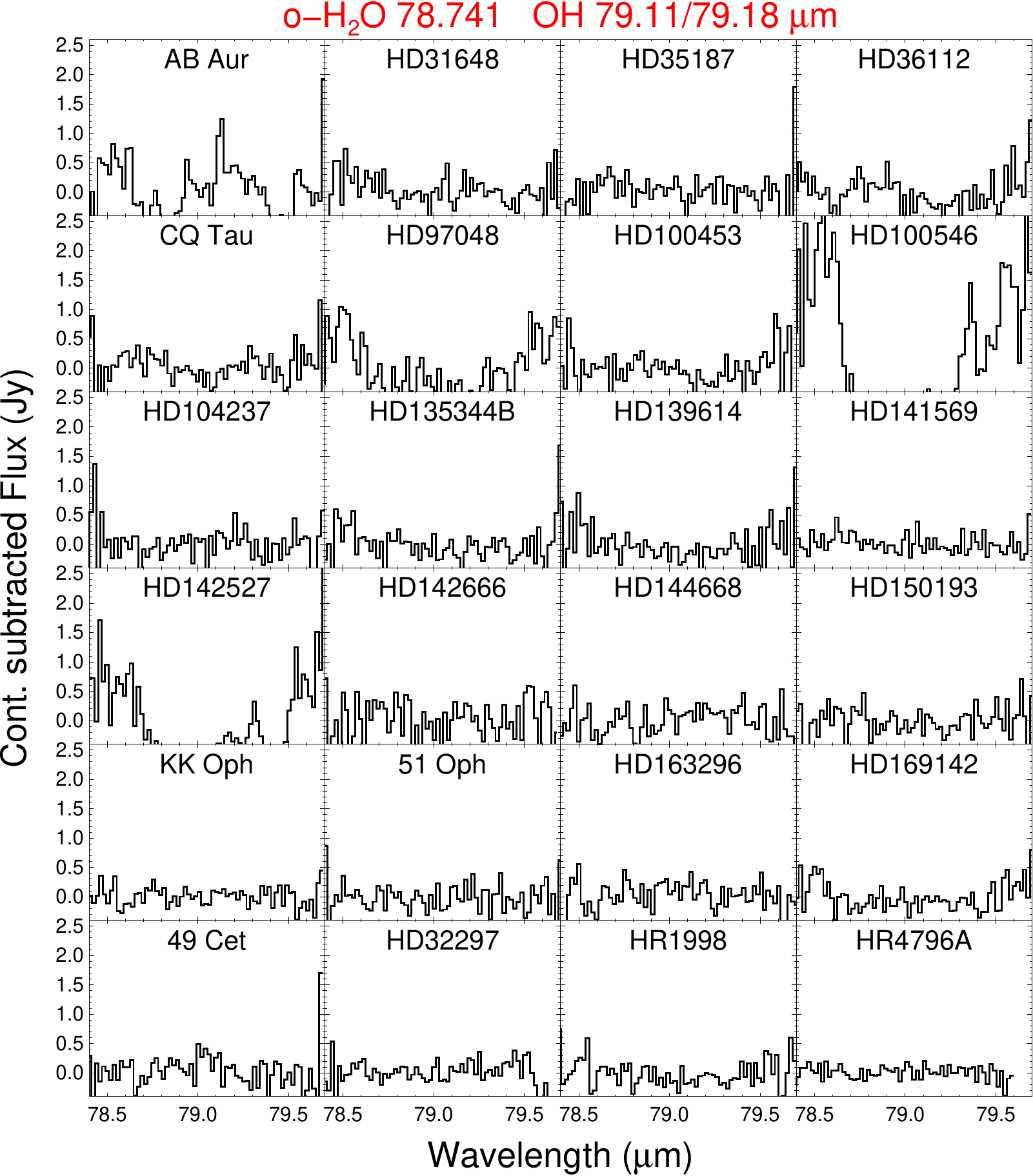}}
\caption{The region around 79 micron.}
\label{f_allstars_79}
\end{figure}

\begin{figure}[t!]
   \resizebox{\hsize}{!}{\includegraphics{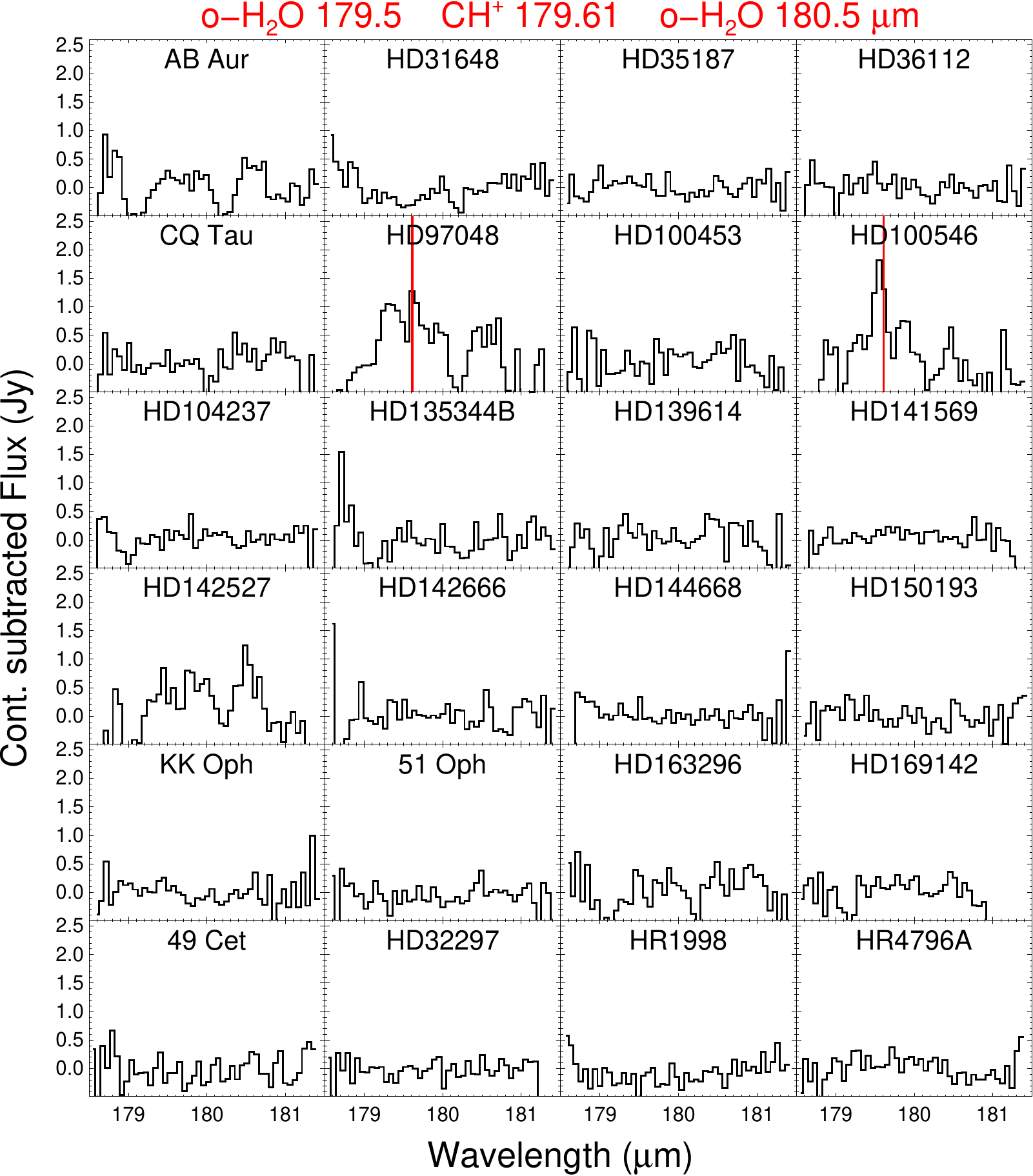}}
\caption{The region around 180 micron. With a vertical line we indicate the position of CH$^+$.}
\label{f_allstars_180}
\end{figure}

\end{appendix}

%+++++++++++++++++++++++++++++++++++++++++++++++++++++++++++++++++++++++++++++++++
\begin{appendix}
\section{Additional millimeter observations}
\label{a_sma}

\begin{table}
\begin{center}
\caption{$^{12}$CO 2-1 and $^{12}$CO 3-2 line fluxes observed with SMA.} 
\begin{tabular}{llllll}
\hline
\hline
Star                &$^{12}$CO 2-1 & Error    &$^{12}$CO 3-2 & Error \\
                       &\multicolumn{2}{l}{(Jy km/s)}    &\multicolumn{2}{l}{(Jy km/s) } \\
\hline
HD 35187    &$<$ 0.55            &--           &--                &--         \\
HD 142666 &$<$ 0.62            &              &--                &--               \\
HD 144668 &\multicolumn{2}{c}{confused}              & \multicolumn{2}{c}{confused}  \\
KK Oph         &$<$ 1.08            &--           &--               &--                 \\
HR 1998      &$<$ 1.17             &--           &--               &--                 \\
HD158352  &$<$ 0.48             &--           &--               &--                  \\
\hline
\end{tabular}
\label{t_sma}
\end{center}
\end{table}

\begin{table}
\begin{center}
\caption{Continuum fluxes at  1.2 mm, observed with MAMBO and at 1.3 mm, observed with SMA.} 
\begin{tabular}{llllll}
\hline
\hline
Star                &F$_{\mathrm{1.2mm}}$ & Error &F$_{\mathrm{1.3mm}}$ & Error   \\
                       &\multicolumn{2}{l}{(mJy)}&\multicolumn{2}{l}{(mJy)}\\
\hline
HD 32297    &3.14       &0.822&  3.10      &0.74\\
HD 35187    &33.960&1.001   & 28.95   &0.85\\
HD 141569 A&4.785  &0.507&--&--\\
HD 142666 &102.500&2.022  &99.0       &4.3\\
HD 144668 &--             &--           &34.3     &0.9\\
HD 151093 &47.860   &2.091&--&--\\
KK Oph       &36.160&2.260    &24.5        &4.3\\
51 Oph        &5.007     &0.599 &--&--\\
HR 4796 A  &   --          &--         &$<$ 9.3 &--\\
HD 158352 &  --           &--        &$<$ 0.7 5&--&\\
\hline
\end{tabular}
\label{t_smacont}
\end{center}
\end{table}

Interferometric observations of HD 32297, HD 35187, HD 142666, HD 144668, HD 158352, HR 1998, and KK Oph 
were carried out in five tracks with the Sub-Millimeter Array (SMA; Ho et al. \cite{ho2004}) in the compact configuration 
from November 2010 to October 2011.  The observations were carried out at a central frequency of 223.9 GHz, with 
upper and lower bandpasses of 4 GHz bandwidth and the center of each bandpass offset from the central frequency 
by 2.5 GHz.  The continuum was sampled at 3.25 MHz, and a 104 MHz regions was set aside for 0.203 MHz 
resolution observations of the CO J=2-1 line (230.5379700 GHz) line.
The compact array observations included seven or eight antennas, with baselines from 10--70 m.  Each object was 
observed for $\sim$2.5 hours total integration time in good conditions (zenith $\tau_{225 \mathrm{GHz}}$ 0.06 - 0.25) 
with system temperatures of $\sim$100-230 K .  

For gain calibration, we interleaved 5 minute observations of close quasars between 20 minute observations of the 
targets.  We combined observations of two or three targets per track.  We used 60--90 minute observations of bright 
quasars for bandpass calibration, and 20 minute observations of available planets for flux amplitude calibration.  
Observations of flux and bandpass calibrators were carried out before or after our targets were available.  
We flagged and calibrated the data using standard routines in the facility IDL package MIR.  We carried out baseline 
based phase calibration, finding rms phase errors of $\sim$10--20\degr.  Based on variations in the measured fluxes 
of our gain calibrators, the flux calibration has an uncertainty of $\sim$15\%. 

To measure the 1.3 mm continuum flux, we combined wideband continuum channels from all observations and both sidebands 
for each object.   We used the MIRIAD command uvfit to the fit a point source to the observations. In order to confirm that 
sources are unresolved in our observations, we carried out standard Fourier inversion, CLEAN deconvolution and image 
reconstruction using natural weighting with the facility reduction tool MIRIAD.  The typical synthesized beam of the 
combined observations is $\sim3\farcs5 \times 2\farcs8$, with an rms of 0.3 -- 1.5 mJy/beam in the continuum.  We 
separately carried out the same imaging process on the higher spectral resolution line data, binning in 1.0 km/s velocity 
channels centered on the CO J=2$\rightarrow$1 line.  The beamsize is slightly smaller than in the continuum, with rms in 
each channel of 0.4 to 2 Jy/beam.  We list the line fluxes in Table~\ref{t_sma} and the 1.3 mm continuum fluxes in 
Table~\ref{t_smacont}.

In addition, HD 32297, HD 35187, HD 142666, KK Oph, HD 141569, HD 150193, and HD 158643 were observed for 
continuum emission at 1.2 mm using the MAMBO2 bolometer array (Kreysa et al. \cite{kreysa1998}) on the IRAM 30m telescope 
at Pico Veleta, Spain.   Observations were conducted during the Nov. 2008 bolometer pool.  Zenith opacity for our observations 
was typically $\sim$0.2 -- 0.3, and observations were carried out to a target 1$\sigma$ sensitivity of 1 mJy, typically 20 minutes on 
source, in an ON-OFF pattern of 1 minute on target followed by 1 minute on sky, with a throw of 32\arcsec.  Flux calibration was 
carried out using Mars, and local pointing and secondary flux calibration was carried out using nearby bright quasars.  The data 
were reduced using the facility reduction software, MOPSIC\footnote{http://www.iram.es/IRAMES/mainWiki/CookbookMopsic}.  
We list the 1.2 mm continuum fluxes in Table~\ref{t_smacont}.

\end{appendix}

%+++++++++++++++++++++++++++++++++++++++++++++++++++++++++++++++++++++++++++++++++
\begin{appendix}
\section{Data collected from the literature}
\label{a_lit}

In the Tables~\ref{t_xray}, \ref{t_mm13}, \ref{t_co21} and \ref{t_co32}, we list the X-ray luminosities, millimeter
continuum fluxes and CO line fluxes that we collected from the literature, as well as their references. 

\begin{table}
\begin{center}
\caption{X-ray luminosities collected from the literature, references listed: 
(1) Telleschi et al. (\cite{telleschi2007}),
(2) Grady et al. (\cite{grady2010}),
(3) Stelzer et al. (\cite{stelzer2004}),
(4) Collins et al. (\cite{collins2009}), 
(5) Stelzer et al. (\cite{stelzer2006}), 
(6) Grady et al. (\cite{grady2009}) ,
(7) Stelzer et al. (\cite{stelzer2009}) ,
(8) Berghoefer et al. (\cite{berghoefer1996}), 
(9) Grady et al. (\cite{grady2007}) and
 (10) Stelzer \& Neuh\"auser  (\cite{stelzer2000}).}
\begin{tabular}{llll}
\hline
\hline
Star                &log L$_{\mathrm{Xray}}$    &Instrument  &ref. \\
                       &(erg/s)                                     &                      &\\
\hline
AB Aur          &29.71                                       &XMM&(1)  \\
HD 31648     &29.30                                      &Chandra&(2)\\
HD 97048     &29.58                                      &XMM&(3)\\
HD 100453   &28.82                                      &Chandra&(4)\\
HD 100546   &28.93                                      &Chandra&(5)\\
HD 104237   &30.11                                      &Chandra&(5) \\
HD 135344 B&29.66                                      &ROSAT&(6) \\
HD 141569A &$<$ 28.1                                 &Chandra&(5)\\
HD 144668   &28.3                                          &Chandra&(7)\\
HD 150193   &29.64                                       &Chandra&(5) \\
51 Oph          &$<$ 28.98                                &ROSAT&(8)\\
HD 163296   &29.6                                          &Chandra&(5)\\
HD 169142   &29.1                                          &Chandra&(9)\\
HR 4796 A   &29.38                                         &ROSAT&(10)\\

\hline
\end{tabular}
\label{t_xray}
\end{center}
\end{table}

\begin{table}
\begin{center}
\caption{Continuum fluxes at 1.3mm collected from the literature; references listed: 
(1) Henning et al. (\cite{henning1994}),
(2) Mannings et al. (\cite{mannings1997}),
(3) Chapillon et al. (\cite{chapillon2008}),
(4) Meeus et al. (\cite{meeus2002}, \cite{meeus2003})),
(5) Sylvester et al. (\cite{sylvester1996}),
(6) Acke et al. (\cite{acke2004a}),
(7) Hughes et al. (\cite{hughes2008}).
} 
\begin{tabular}{lllll}
\hline
\hline
Star                &F$_{\mathrm{1.3mm}}$ & Error         &Instrument     &ref. \\
                       &(mJy)             &(mJy)  &&\\
\hline
AB Aur           &103 &20  &IRAM&(1)\\
HD 31648     &360 &20  &OVRO&(2)\\
HD 36112     &56   &1    &IRAM&(3)\\
CQ Tau          &162 &2    &IRAM&(3)\\
HD 97048     &450 &30  &IRAM&(1)\\
HD 100453  &265  &21  &SIMBA/SEST&(4)\\
HD 100546  &470  &20  &IRAM&(1)\\
HD 104237  &90    &20  &IRAM&(1)\\
HD 135344B&142 &19  &UKT14/JCMT&(5)\\
HD 139614  &242  &15  &UKT14/JCMT&(5)\\
HD 142527  &1190&30  &ATCA&(6)\\
HD 163296&780    &30   &OVRO&(2)\\
HD 169142&197    &15   &UKT14/JCMT&(5)\\
49 Cet           &2.3    &0.6  &SMA&(7)\\
\hline
\end{tabular}
\label{t_mm13}
\end{center}
\end{table}

\begin{table}
\begin{center}
\caption{$^{12}$CO 2-1 line fluxes collected from the literature; references listed: 
(1) \"Oberg et al. (\cite{oberg2010}),
(2) Isella et al. (\cite{isella2010}),
(3) Panic et al. (\cite{panic2009}),
(4) Hughes et al. (\cite{hughes2008}).
} 
\begin{tabular}{lllll}
\hline
\hline
Star                &$^{12}$CO 2-1 & Error         &Instrument     &ref. \\
                       &(Jy km/s)             &(Jy km/s)  &\\
\hline
HD 31648    &22.0                     &0.2             &SMA              &(1)\\
HD 36112    &12.9                     &2.58           &SMA              &(2)\\
CQ Tau         &3.10                    &0.18           &SMA              &(1)\\
HD 135344B&10.39                 &0.21           &SMA              &(1)\\
HD 139614  &7.00                     &2.11           &RxA3/JCMT &(3) \\
HD 142527  &20.76                 &0.23            &SMA              &(1)\\
HD 142666  &3.12                     &0.5              &SMA              &(22)\\
HD 163296  & 6.90                    &--                &RxA3/JCMT &(3) \\
HD 169142  & 21.30                  &--                &RxA3/JCMT &(3) \\
49 Ceti         &2.0                       &0.3              &SMA              &(4)\\
\hline
\end{tabular}
\label{t_co21}
\end{center}
\end{table}

\begin{table}
\begin{center}
\caption{$^{12}$CO 3-2 line fluxes collected from the literature; references  listed: 
(1) Dent et al. (\cite{dent2005}) and (2) Panic et al. (\cite{panic2009}). ($a)$: JCMT
flux dominated by emission from the dark cloud.
}
\begin{tabular}{lrrll}
\hline
\hline
Star                &$^{12}$CO 3-2 &\multicolumn{1}{c}{Error}&Instrument     &ref. \\
                       &(Jy km/s)             &(Jy km/s)  &\\
\hline
AB Aur          &143.20$^a$         &3.29         &RxB3/JCMT&(1)  \\
HD 31648    &53.00                    &1.16         &RxB3/JCMT&(1)  \\
HD 35187    &$<$ 5.00               &--             &RxB3/JCMT&(1)  \\
HD 36112    &15.70                     &1.93       &RxB3/JCMT&(1)  \\
CQ Tau         &6.00                      &0.97        &RxB3/JCMT&(1)  \\
HD 100546  &178.4                    &19.20     &APEX             &(2) \\
HD 135344B&18.80                   &0.77        &RxB3/JCMT&(1)  \\
HD 139614  &9.10                      &2.12        &RxB3/JCMT&(1)  \\
HD 141569A&14.70                   &1.16        &RxB3/JCMT&(1)  \\
HD 142666  &13.90                    &2.70       &RxB3/JCMT&(1)  \\
HD 150193  &$1.40$                 &--             &RxB3/JCMT&(1)  \\
HD 163296  &8.30                      &1.93       &RxB3/JCMT&(1)  \\
HD 169142  &32.9                     &2.51        &RxB3/JCMT&(1)  \\
49 Cet           & 6.58                     &1.35        &RxB3/JCMT&(1)  \\
\hline
\end{tabular}
\label{t_co32}
\end{center}
\end{table}

\end{appendix}

\end{document}